\pdfoutput=1

\documentclass[11pt,a4paper]{article}
\usepackage[utf8]{inputenc}
\usepackage{textgreek}
\usepackage{jheppub}
\hypersetup{unicode=true,bookmarksopen=true,pdfencoding=unicode}

\usepackage{mathtools}
\usepackage{lmodern}
\usepackage[T1]{fontenc}
\usepackage{bbm}
\DeclareMathAlphabet{\mathbfi}{OML}{cmm}{b}{it}
\usepackage{amssymb}

\usepackage{graphicx}
\usepackage{subcaption}

\let\originalleft\left
\let\originalright\right
\renewcommand{\left}{\mathopen{}\mathclose\bgroup\originalleft}
\renewcommand{\right}{\aftergroup\egroup\originalright}

\makeatletter
\newenvironment{equations}[1][]{\subequations\ifx\relax#1\relax\else\label{#1}\fi\align\ignorespaces}{\endalign\ignorespacesafterend\endsubequations}
\def\@spliteq#1{\begin{equation}\begin{aligned}#1\end{aligned}\end{equation}}
\def\splitequation{\collect@body\@spliteq}

\makeatother

\usepackage{accents}
\newlength{\dhatheight}

\renewcommand{\vec}[1]{{\ifnum9<1#1\mathbf{#1}\else\ifcat\noexpand#1\relax\boldsymbol{#1}\else\mathbfi{#1}\fi\fi}}
\newcommand{\mathe}{\mathrm{e}}
\newcommand{\mathi}{\mathrm{i}}
\let\oldre\Re
\let\oldim\Im
\renewcommand{\Re}{\oldre\mathfrak{e}\,}
\renewcommand{\Im}{\oldim\mathfrak{m}\,}
\newcommand{\total}{\mathop{}\!\mathrm{d}}

\newcommand{\eqend}[1]{\,#1}
\newcommand{\bigo}[1]{\mathcal{O}\left({#1}\right)}

\newcommand{\brst}{\mathop{}\!\mathsf{s}\hskip 0.05em\relax}

\newcommand{\hypergeom}[2]{\,{}_{#1}\mathrm{F}_{#2}}
\newcommand{\normord}[1]{\mathopen{:}{#1}\mathclose{:}}

\newcommand{\g}{\bar g}

\newcommand{\h}{\hat h}
\newcommand{\cc}{\hat c}
\newcommand{\ac}{\hat{\bar c}}

\allowdisplaybreaks

\makeatletter
\gdef\@fpheader{\strut}
\makeatother

\begin{document}

\title{De Sitter quantum gravity within the covariant Lorentzian approach to asymptotic safety}

\author[a]{Edoardo D'Angelo}
\author[b]{\!\!, Renata Ferrero}
\author[c]{and Markus B. Fr{\"o}b}

\affiliation[a]{Dipartimento di Matematica, Dipartimento di Eccellenza 2023-2027, Università di Milano,\\ Via Cesare Saldini 50, 20133 Milano, Italy}
\affiliation[b]{Institute for Quantum Gravity, Friedrich-Alexander-Universit{\"a}t Erlangen-N{\"u}rnberg,\\ Staudtstr. 7, 91058 Erlangen, Germany}

\affiliation[c]{Institut f{\"u}r Theoretische Physik, Universit{\"a}t Leipzig,\\ Br{\"u}derstra{\ss}e 16, 04103 Leipzig, Germany}

\emailAdd{edoardo.dangelo@unimi.it}
\emailAdd{renata.ferrero@fau.de}
\emailAdd{mfroeb@itp.uni-leipzig.de}

\abstract{Recent technical and conceptual advancements in the asymptotic safety approach to quantum gravity have enabled studies of the UV completion of Lorentzian Einstein gravity, emphasizing the role of the state dependence. We present here the first complete investigation of the flow equations of the Einstein-Hilbert action within a cosmological spacetime, namely de~Sitter spacetime. Using the newly derived graviton propagator for general gauges and masses in de~Sitter spacetime, we analyze the dependence on the gauge and on finite renormalization parameters. Our results provide evidence of a UV fixed point for the most commonly used gauges.}


\maketitle

\section{Introduction}
\label{sec_intro}

Models of quantum gravity in de~Sitter spacetime have a well-established history \cite{Bunch:1978yq,Bros:1994dn,Witten:2001kn}, as this spacetime plays an essential role in understanding the universe across various epochs. In the context of early cosmology, de~Sitter spacetime provides a robust framework for describing an inflationary universe, a scenario supported by evidence from the cosmic microwave background radiation (CMBR) \cite{Planck:2018vyg}. At late times, observations of distant supernovae indicate an accelerated expansion of the universe \cite{SupernovaSearchTeam:1998fmf,SupernovaCosmologyProject:1998vns}, which can also be modeled using de~Sitter spacetime. Within general relativity, this spacetime is uniquely characterized as a solution to Einstein's field equations with constant curvature and a positive cosmological constant.

The standard cosmological model fundamentally depends on the de~Sitter solution of general relativity. Crucially, theoretical evidence strongly suggests the necessity for a UV-complete theory of gravity, one that unifies Einstein gravity with the quantum nature of the universe's fundamental constituents. Such a unification would provide a full-fledged theory of quantum gravity, but as is well-known the union of those two theories by means of standard perturbative QFT methods leads to a non-renormalizable theory with infinitely many parameters (or coupling constants) that need to be determined experimentally \cite{tHooft:1974toh,Goroff:1985sz,Goroff:1985th}.

In the past decades there have been several attempts to tackle the problem of quantum gravity.
The asymptotic safety quantum gravity program, in a continuum-based approach, is based on the realization of the UV completion of quantum gravity through a non-trivial fixed point of the renormalization group (RG) flow \cite{Reuter:2001ag,Reuter:2005kb,Percacci:2017fkn, Reuter:2019byg}. In the language of the functional renormalization group (FRG), this conjecture can be expressed as follows~\cite{Reuter:1996cp}: a theory is non-perturbatively renormalizable if there exists a RG trajectory that reaches a non-trivial fixed point of the RG flow in the UV, the critical surface is finite-dimensional, and there is a complete trajectory connecting the UV to an IR fixed point. In this case, only finitely many parameters need to be determined experimentally.

The majority of the computations in this program have used methods from the RG toolbox, namely the ones usually applied to critical phenomena in Euclidean signature \cite{Weinberg:1976xy,Weinberg:1980gg,Donoghue:2019clr,Bonanno:2020bil}. In the FRG approach, the RG governs the flow of an effective action $\Gamma_k$, depending on an RG scale $k$, under changes of this scale. The flow of this effective action is described by a functional flow equation, and the most widely used flow equations are the Polchinski equation \cite{Polchinski:1983gv} (for which $k$ is a UV cutoff), the proper time equation \cite{Bonanno:2004sy} and, particularly in quantum gravity, the Wetterich equation \cite{Wetterich:1992yh,Morris:1993qb, Reuter:1996cp} (for which $k$ is an IR cutoff).

In Lorentzian signature, several issues arise within the FRG framework. Most prominent is the fact that the effective action must break Lorentz invariance if $k$ is a true momentum cutoff, since momenta can be spacelike, timelike or null. Even if one considers the RG trajectories obtained from the Euclidean approach without insisting on $k$ as a cutoff, there is no clear prescription for integrating out the momentum eigenmodes of kinetic operators derived from Lorentzian metrics~\cite{Ferrero:2022hor}. Moreover, properties inherently related to Lorentzian spacetimes, such as ambiguities related to the choice of a state, or of the observer, could not be addressed. Only recently, approaches to evaluate the RG flow in Lorentzian signature have been developed. These include the algebraic framework for QFT in curved spacetime~\cite{DAngelo:2022vsh,DAngelo:2023tis,DAngelo:2023wje}, the use of Lorentzian heat kernel methods~\cite{Thiemann:2024vjx, Ferrero:2024rvi,Banerjee:2024tap}, the flow of only spatial degrees of freedom~\cite{Kaya:2013bga,Serreau:2013eoa,Guilleux:2015pma,Banerjee:2022xvi}, the reconstruction of the graviton propagator by means of a double regularization~\cite{Pawlowski:2023gym} and analytic continuation via Wick rotation~\cite{Baldazzi:2018mtl} in an ADM formalism~\cite{Manrique:2011jc,Saueressig:2023tfy,Saueressig:2025ypi}.

In this paper we will exploit the techniques developed in Refs.~\cite{DAngelo:2022vsh, DAngelo:2023tis}, where the FRG formalism has been adapted to the algebraic approach to QFT in curved spacetimes. In this fully covariant approach, instead of employing a momentum cutoff which would break covariance, a mass-like term quadratic in the fields and local in position space is chosen as regulator. Such terms are also known as Callan--Symanzik cutoffs~\cite{Alexandre:2000eg}, and they have been used in the study of the RG flow of the graviton spectral function~\cite{Fehre:2021eob}. While they regulate the IR, they do not act as a UV cutoff and therefore an additional UV renormalization is required. In Refs.~\cite{DAngelo:2022vsh, DAngelo:2023tis}, the UV renormalization is performed using the Epstein--Glaser inductive procedure~\cite{EpsteinGlaser1973}, which has two main advantages: it works in position space, so it can be applied to arbitrary curved spacetimes in which Fourier space methods are not available, and, while it is perturbative in nature, it can be applied to theories that are power-counting non-renormalizable, namely effective field theories~\cite{Gomis:1995jp,Burgess:2003jk}. On a general curved spacetime, to ensure covariance under diffeomorphisms the Epstein--Glaser procedure is tightly constrained and implies the subtraction of UV divergences in the FRG equation in the form of a Hadamard parametrix~\cite{HollandsWald2001,HollandsWald2002}, as we will explain in Sec.~\ref{sec:FRGE}. Thus, we obtain a locally covariant RG flow, independent of the choice of a coordinate system on the spacetime.

Strictly speaking, the flow equation derived in this context is the flow of renormalized theories under changes in the mass parameter, with no Wilsonian interpretation. In fact, the running scale $k$ is a physical mass scale, since correlation functions (or scattering amplitudes in flat space) depend on the RG scale $k$ through an additional mass term. Thus, this approach suggests an alternative point of view to the controversy on whether the RG scale in the Wetterich equation describes the running under an artificial, unphysical cutoff scale, or the running under a physical variable described by the measurement of a scattering amplitude at some energy scale, see for example the recent works~\cite{Donoghue:2019clr,Bonanno:2020bil,Buccio:2023lzo,Buccio:2024hys}

Despite the fact that the regulator term acts as an IR cutoff only, it is still possible to prove that the effective average action $\Gamma_k$ reduces to the classical action $S$ in the UV limit $\Gamma_k \xrightarrow{k \to \infty} S$, while it reduces to the full quantum effective action in the IR limit of vanishing regulator, $\Gamma_k \xrightarrow{k \to 0} \Gamma$. Therefore, in this sense, solutions to the flow equation describe an RG flow between the UV and the IR limits, and we can talk of a RG flow equation in Lorentzian spacetimes.

The RG flow equation in Lorentzian spacetimes exhibits an explicit state dependence, so that the same theory (for example a scalar field on some fixed background) has different phase diagrams depending on the choice of a reference state (such as a vacuum or a thermal state~\cite{DAngelo:2022vsh}). This state dependence is often overlooked in the Euclidean context, where there is usually a unique choice of vacuum state, but it becomes an important physical effect in Lorentzian, curved spacetimes, where no unique vacuum may exist and different choices of ground states are possible. These Lorentzian RG techniques have been first applied to the case of quantum gravity in Ref.~\cite{DAngelo:2023wje}. In that first investigation, only state- (that is, present in any Hadamard state) and background-independent contributions were considered in the RG flow. These universal contributions give raise to a non-trivial fixed point in the flow, but state-dependent, non-universal contributions may significantly alter the phase diagram, even destroying the fixed point. Thus, taking into account the full state dependence of the RG flow is essential to understand if asymptotic safety is realized in specific examples and in particular states.

De~Sitter spacetime is the ideal background to evaluate the RG flow, both for its relative theoretical simplicity and its relevance for the physical universe. The investigation of the functional RG flow for quantum gravity in de~Sitter space is the topic of this paper. In particular, we will construct an effective description of a quantum gravitational theory on de~Sitter spacetime. The analysis of the RG flow in this background will relate the UV theory to the effective, IR description of the universe we need in cosmology. In light of this, we will also discuss the dependence of our results on the gauge choice.

Before addressing the computation, let us emphasize that several theoretical challenges appear for de~Sitter spacetime, including the choice of the vacuum, gauge-fixing, issues related to analytic continuation, IR divergences and secular effects, and the absence of asymptotic states. In particular, while for massive fields there is a unique de~Sitter invariant vacuum state of Hadamard form~\cite{Schomblond:1976xc,Bros:1995js,Allen:1985ux,Allen:1985wd,Frob:2013qsa}, the massless limit of the corresponding propagators either diverges or does not agree with the corresponding strictly massless propagator~\cite{Glavan:2022nrd}, which breaks de~Sitter invariance. Moreover, while fundamental solutions of the equations of motion for massive vector and tensor fields have been obtained recently~\cite{Farnsworth:2024iwc}, these are not propagators because they do not satisfy the Hadamard condition. In this work we therefore also introduce a missing ingredient, namely the propagators in de~Sitter spacetime for massive scalars, vector and tensors.

\medskip

This paper is structured as follows: In Sect.~\ref{sec:FRGE} we review the recently developed approach to asymptotic safety in Lorentzian signature. In Sec.~\ref{sec:propagators} we supply the missing piece to the use of regularized propagators in de~Sitter space: we present the Hadamard expansion of the Feynman propagator of massive scalars, vectors and tensors. Sec.~\ref{sec:FRGE-dS} is dedicated to a review of the important gauge choices done in the literature and to a discussion about renormalization in de~Sitter. Secs.~\ref{sec:zeta=1/2} and~\ref{sec:harmonic gauge} contain the main results of this paper, showing the RG flow, its fixed point structure, and its gauge and parameter dependence. Finally, in Sec.~\ref{sec:discussion} we conclude and give an outlook.

\medskip

Our conventions are the ``+++'' ones of MTW, which are a mostly plus metric, $\nabla_\mu \nabla_\nu v_\rho - \nabla_\nu \nabla_\mu v_\rho = R_{\mu\nu\rho\sigma} v^\sigma$, and $R_{\mu\nu} = R^\rho{}_{\mu\rho\nu}$.

\section{Flow equation in Lorentzian signature}
\label{sec:FRGE}

We start by setting up the functional renormalization group equation (FRGE) in Lorentzian signature in the algebraic approach to QFT on curved backgrounds~\cite{DAngelo:2022vsh,DAngelo:2023wje}, which are based on the first derivation of the Wetterich equation~\cite{Wetterich:1992yh} for quantum gravity by Reuter~\cite{Reuter:1996cp}.

We consider an arbitrary globally hyperbolic spacetime $(\mathcal{M}, \g)$ as our background, with a fixed background metric $\g$. We will later restrict to the case in which $\g$ is the de~Sitter metric. The dynamical quantum fields are the metric fluctuations $\h$, together with the standard (Faddeev--Popov) ghost and antighost fields $\cc$ and $\ac$ that arise in the standard BRST approach to gauge theories. We collectively denote the quantum fields by $\varphi^A = \{ \h_{\mu\nu}, \cc^\mu, \ac^\mu \}$, where the index $A$ includes field species as well as Lorentz indices. The dynamics is governed by the action
\begin{equation}
S(\varphi) \coloneqq S_\text{grav}(\g + \h) + S_\text{gf}(\g, \h) + S_\text{gh}(\g, \h, \cc, \ac) \eqend{,}
\end{equation}
where $S_\text{grav}$ is the gravitational action as a function of the full metric $\gamma \coloneqq \g + \h$, invariant under diffeomorphisms $\delta_v \gamma_{\mu\nu} = \mathcal{L}_v \gamma_{\mu\nu} = \nabla^\gamma_\mu v_\nu + \nabla^\gamma_\nu v_\mu$, where $\mathcal{L}$ is the Lie derivative and $\nabla^\gamma$ the Levi-Civita covariant derivative associated to $\gamma$. $S_\text{gf}$ is the gauge-fixing term defined by
\begin{equation}
\label{eq:action_gaugefixing}
S_\text{gf}(\g, \h) \coloneqq - \frac{1}{2 \xi} \int_\mathcal{M} \g_{\mu\nu} F^\mu[\g,\h,\zeta] F^\nu[\g,\h,\zeta] \eqend{,}
\end{equation}
where $F^\mu$ is a linear functional of the metric fluctuation $\h$, and we employ the notation
\begin{equation}
\int_\mathcal{M} f \coloneqq \int f \sqrt{-\det \g} \total^d x.
\end{equation}
The gauge-fixing term depends on the two gauge parameters $\xi$ and $\zeta$. It breaks the \emph{quantum} gauge symmetry of the metric fluctuation given by $\delta^\text{Q}_v \h = \mathcal{L}_v(\g + \h)$, $\delta^\text{Q}_v \g = 0$, but preserves the \emph{background} gauge symmetry $\delta^\text{B}_v \h = \mathcal{L}_v \h$, $\delta^\text{B}_v \g = \mathcal{L}_v \g$. This can be achieved if $F^\mu$ depends separately on the background metric $\g$ and the fluctuations $\h$ and not just on the full metric $\gamma = \g + \h$, thus breaking the split symmetry of the gravitational action. 

Lastly, $S_\text{gh}$ is the ghost term, obtained as usual by a gauge transformation of the gauge-fixing term in which the parameters $v$ are substituted by the ghost field $\cc$:
\begin{equation}
\label{eq:action_ghost}
S_\text{gh}(\g, \h, \cc, \ac) \coloneqq - \int_\mathcal{M} \ac^\mu \g_{\mu\nu} \, \delta^\text{Q}_{\cc} F^\nu \eqend{.}
\end{equation}
The classical action $S$ is now invariant under BRST transformations, $\brst S = 0$, where the BRST differential is defined by
\begin{equations}
\brst \h_{\mu\nu} &= \mathcal{L}_{\cc} \left( \g_{\mu\nu} + \h_{\mu\nu} \right) = \nabla^{\g}_\mu \cc_\nu + \nabla^{\g}_\nu \cc_\mu + \cc^\rho \nabla^{\g}_\rho \h_{\mu\nu} + \h_{\mu\rho} \nabla^{\g}_\nu \cc^\rho + \h_{\nu\rho} \nabla^{\g}_\mu \cc^\rho \eqend{,} \\
\brst \cc^\mu &= \cc^\rho \nabla^{\g}_\rho \cc^\mu \eqend{,} \\
\brst \ac^\mu &= F^\mu \eqend{.}
\end{equations}

To construct the quantum theory, we split the action in a quadratic and an interacting part, $S = S_0 + V$. Thanks to the gauge fixing, the free Euler--Lagrange equations derived from $S_0$ admit unique retarded and advanced propagators, and choosing a state we can construct the corresponding Feynman propagator $\Delta_\text{F}$. Using this propagator, we define the time-ordering operator $\mathcal{T} \coloneqq \exp\left[ \mathi \int_{\mathcal M^{\otimes 2}} \langle \frac{\delta}{\delta \varphi}, \Delta_F \frac{\delta}{\delta \varphi} \rangle \right]$, which is a priori well-defined on regular functionals only, since on local functionals the loops appearing as the coincidence limit of products of Feynman propagators are UV divergent. Time-ordering can be extended to local functionals using the Epstein--Glaser renormalization procedure, and we refer to Refs.~\cite{AAQFT15,Brunetti2009,Brunetti1999,Brunetti2001,HollandsWald2001,HollandsWald2002,Rejzner2016} for further details.

To define the generating functionals we supplement the action with sources for both the fields ($J$) and their BRST transformations ($\Sigma$)
\begin{equation}
\label{eq:sources}
J + \Sigma \coloneqq \int_\mathcal{M} j_A(x) \varphi^A(x) + \int_\mathcal{M} \sigma_A(x) \brst \varphi^A(x) \eqend{,}
\end{equation}
and with the regulator action for fields ($Q_k$) and their BRST transformations ($H$)
\begin{splitequation}
\label{eq:regulator}
Q_k + H &\coloneqq - \int_\mathcal{M} \varphi_A^*(x) q^{AB}_k(x) \varphi_B(x) + \frac{1}{2} \int_\mathcal{M} \eta^{AB}(x) \brst \left[ \varphi_A^*(x) \varphi_B(x) \right] \\
&= - \frac{1}{2} \int_\mathcal{M} \h_{\rho\sigma} q_k^{\rho\sigma\mu\nu}(\g) \h_{\mu \nu} - \int_\mathcal{M} \ac_\mu \tilde{q}_k^{\mu\nu}(\g) c_\nu + \frac{1}{2} \int_\mathcal{M} \eta^{AB}(x) \brst \left[ \varphi_A^*(x) \varphi_B(x) \right] \eqend{.}
\end{splitequation}
Here we take $j$, $q_k$, $\sigma$ and $\eta$ to be smooth functions with compact support such that the integrals are IR finite. For the same reason, a cutoff function is introduced in the interaction $V$, which we will keep implicit. Of course, at the end the adiabatic limit where all these functions tend to a (spacetime) constant needs to be taken.

The addition of sources for the BRST variations of the fields and their squares allows to extend the BRST differential into a $k$-dependent BRST differential~\cite{DAngelo:2023tis}
\begin{equation}
\brst_k = \brst + \int_\mathcal{M} q^A_k(x) \frac{\delta}{\delta \varphi^A(x)} - \int_\mathcal{M} j^A(x) \frac{\delta}{\delta \sigma^A(x)} \eqend{,}
\end{equation}
encoding the extended symmetry of the extended action $S_\text{ext} = S + J + \Sigma + Q_k + H$:
\begin{equation}
\label{eq:extended-CME}
\brst_k S_\text{ext} = 0 \eqend{.}
\end{equation}
Furthermore, since $(q_k, \eta)$ and $(j, \sigma)$ form contractible pairs, the cohomology of the scale-dependent differential $\brst_k$ coincides with the cohomology of the BRST differential $\brst$. In particular, observables (which are elements of the cohomology at ghost number 0) are the same for all $k$.

We can now define the regularized generating functional of time-ordered correlation functions as
\begin{equation}
\label{eq:generating-functional}
Z_k(\g; j, \sigma, \eta) \coloneqq \omega\left( \mathcal{T}\left( \mathe^{-\mathi \, \mathcal{T}^{-1}(V)} \right) \mathcal{T}\left( \mathe^{\mathi \, \mathcal{T}^{-1}(V + \Sigma + J + Q_k + H) } \right) \right) \eqend{,}
\end{equation}
where $\omega$ is the expectation value in the state that we chose; in a quasifree (Gaussian) state it simply corresponds to the evaluation on the vanishing field configuration $\varphi = 0$. As usual, the regularized generating functional for connected, time-ordered correlation functions is $W_k = \mathi \ln Z_k$, and the \emph{effective average action} $\Gamma_k$ is defined as the modified Legendre transform
\begin{equation}
\Gamma_k(\g, \phi) \coloneqq W_k(j^\phi) - \int_\mathcal{M} j^\phi_A(x) \phi^A(x) - Q_k(\phi) \quad\text{with}\quad \phi^A(x) \coloneqq \left. \frac{\delta W_k(j)}{\delta j_A(x)} \right\rvert_{j = j^\phi} \eqend{.}
\end{equation}
The second relation defines the source $j^\phi$ as the source associated to the classical field $\phi$, which is the independent variable on which $\Gamma_k$ depends.

The extended symmetry~\eqref{eq:extended-CME} gets translated into an extended Slavnov--Taylor identity for the Legendre effective action $\tilde{\Gamma}_k = \Gamma_k + Q_k$, which takes the form of a standard, unregularized Zinn--Justin equation with an additional gauge-fixing term:
\begin{equation}
\int_\mathcal{M} \left[ \frac{\delta \tilde \Gamma_k(\phi)}{\delta \sigma_A(x)} \frac{\delta \tilde \Gamma_k(\phi)}{\delta \phi^A(x)} + q_k^{AB}(x) \frac{\delta \tilde \Gamma_k(\phi)}{\delta \eta^{AB}(x)} \right] = 0 \eqend{.}
\end{equation}

The FRGE can now be derived by standard methods~\cite{Wetterich:1992yh,Reuter:1996cp} and reads~\cite{DAngelo:2022vsh,DAngelo:2023wje}
\begin{equations}[eq:FRGE]
\partial_k \Gamma_k(\g; \phi) = \frac{\mathi}{2} \int_\mathcal{M} \partial_k q^{AB}_k(x) \, \normord{ G^{BA}_k(x,x) } \eqend{,} \\
\int_\mathcal{M} \frac{\delta^2 (\Gamma_k + Q_k)}{\delta \phi^A(x) \delta \phi^B(z)} G^{BC}_k(z,y) = \delta(x,y) \delta_A^C \eqend{,}
\end{equations}
where $G^{AB}_k$ is the scale-dependent full propagator defined by the second line, and $\normord{  }$ denotes Hadamard normal ordering, i.e., subtraction of the Hadamard parametrix which contains the universal singular terms of the propagator.

\subsection{Differences from the Euclidean case}
\label{sec:differences}

The FRGE in Lorentzian spacetimes takes the same form as the well-known Wetterich equation with local regulator, also known as the functional Callan-Symanzik equation~\cite{Alexandre:2000eg}. In Lorentzian spacetimes, however, there are two additional subtleties arising from the non-positivity of the wave operator.

The first issue is that, since the wave operator $\frac{\delta^2 (\Gamma_k + Q_k)}{\delta \phi^A(x) \delta \phi^B(z)}$ is, in general, a hyperbolic operator, it admits an infinite family of fundamental solutions whose difference is a smooth function. This corresponds directly to the non-uniqueness of the quantum state, and means that the FRGE must be supplemented by some additional condition to fix the choice of the interacting propagator $G_k$. In Ref.~\cite{DAngelo:2022vsh}, this ambiguity is resolved by choosing a reference Hadamard state for the free theory and imposing that $G_k$ reduces to the free Feynman propagator in the regions of spacetime where the interaction is turned off, which uniquely fixes $G_k$. The Hadamard condition~\cite[and references therein]{DeWitt:1960fc,Decanini:2005gt,Hack:2012qf,Hollands:2014eia} on the state ensures that the UV behaviour is the same as in Minkowski spacetime, and guarantees the existence of a well-defined renormalized stress-energy tensor~\cite{Hollands:2014eia}. Since the UV behavior is the same for all Hadamard states, the UV divergences in the FRGE can be removed by subtracting the universal singular terms (the Hadamard parametrix), and employing the Hadamard normal ordering in the FRGE~\eqref{eq:FRGE}. Once a Hadamard state for the free theory is chosen, it is possible to construct the interacting propagator $G_k$. In particular, in the local potential approximation (LPA), the interacting propagator coincides with the free Feynman propagator, with a mass shifted by the regulator function $q_k$ and a field dependent contribution coming from the effective potential.

We emphasize again that we deviate from the Wilsonian spirit of integrating out degrees of freedom, since all degrees of freedom contribute to the FRGE~\eqref{eq:FRGE}. Instead, we consider how the theory reacts to a change in an auxiliary mass term, which can be chosen to preserve local covariance (and thus Lorentz invariance). This represents a complementary analysis to Ref.~\cite{Ferrero:2022hor}.

The second issue is the choice of the regulator, which must satisfy the requirements of UV and IR finiteness of the FRGE, causality, unitarity, and local covariance (Lorentz invariance in flat space). In our formalism, since the time-ordered products are renormalized through the Epstein--Glaser procedure, the regulator function $q_k$ does not need to regulate UV divergences~\cite{DAngelo:2022vsh}. We are thus free to choose a mass-like (also known as Callan-Symanzik~\cite{Braun:2022mgx,Fehre:2021eob}) regulator in the form $q^{AB}_k(x) = \chi^{AB}(\xi,\zeta) k^2 f(x)$ for a smooth compactly supported function $f$, preserving causality and Lorentz invariance in the adiabatic limit $f \to \text{const}$. The matrix $\chi^{AB}(\xi, \zeta)$ will be fixed in the next section and is adapted to the choice of gauge parameters $\xi$ and $\zeta$. Since the integrand on the right-hand side of the FRGE~\eqref{eq:FRGE} is proportional to $\partial_k q_k$, the compact support of $f$ guarantees that the spacetime integral is finite for any field configuration; in this sense $q_k$ acts as an IR regulator. UV finiteness on the other hand is guaranteed by the Hadamard normal ordering of the interacting propagator $G_k$, which subtracts the universal short-distance divergences from the interacting propagator and results in the finite coincidence limit $\normord{ G_k(x,x) }$.

The subtraction of the Hadamard parametrix, containing the universal singular terms common to any Hadamard state, introduces a new arbitrary constant, the Hadamard scale $\ell$. This scale results from logarithmic terms in the parametrix that appear for even spacetime dimensions (and are ultimately responsible, for example, for the trace anomaly of the stress tensor), and parametrizes the freedom that one has in the normal-ordered two-point function $\normord{ G_k }$. In other applications of the Callan-Symanzik regulator to the FRG, it is often the case that the necessary additional UV regularization also introduces a further scale related to the UV cutoff, which can be dimensionful or dimensionless depending on the UV regularization scheme~\cite{Braun:2022mgx}. This additional parameter can be fixed to some convenient value, to be chosen at the end of the RG flow; for example, this is the choice made in the investigation of the graviton spectral function~\cite{Fehre:2021eob}. Here, in order to investigate the parametric dependence of the flow on the Hadamard scale $\ell$, we keep it arbitrary.

In practice, in the Einstein--Hilbert truncation, the FRGE in Lorentzian spacetimes requires two steps. First, one needs to compute the Feynman propagator (in Hadamard form) of the chosen state for massive tensor and vector fields, corresponding to gravitons and ghosts. The Hadamard condition guarantees that the short-distance singularity of the propagator has a universal structure. The interacting propagator in the LPA then simply corresponds to the free Feynman propagator, with a field- and regulator-dependent mass. Since the LPA modifies the mass term only, the interacting propagator still has the Hadamard form. Subtracting the Hadamard parametrix then allows to compute the finite coincidence limit of the normal-ordered interacting propagator.

In the following sections, we specialize this general framework to the case of the Einstein--Hilbert truncation on a de~Sitter background. The choice of a background allows to completely evaluate the interacting propagator, providing also terms that were missing in the background independent investigation in Ref.~\cite{DAngelo:2023wje}. The de~Sitter background corresponds to the widely used spherical backgrounds used in previous investigations of asymptotic safety in Euclidean spaces (see for example~\cite{Codello:2007bd,Codello:2008vh,Falls:2014tra,Falls:2017lst}, especially for applications to higher order truncations). Imposing invariance under the de~Sitter symmetries together with the Hadamard condition selects a unique vacuum state known as \emph{Bunch-Davies} vacuum~\cite{Schomblond:1976xc,Allen:1985ux}. The presence of a regulator- and field-dependent mass does not spoil unicity. Moreover, we can now study the gauge dependence and the Hadamard scale dependence of the RG flow. Therefore, we have a 3-parameter space, determined by the gauge parameters $\xi$ and $\zeta$ and the Hadamard scale $\ell$. We will study different flows for specific values of the parameters, choosing in particular the subspaces determined by the gauges with $\zeta = \frac{1}{2}$ and the harmonic gauges with $\zeta = 1$ to investigate the corresponding two-dimensional parameter subspace.

\section{Massive propagators in de Sitter: scalars, vectors and tensors}
\label{sec:propagators}

The first task that has been thrust upon us is thus to compute the propagators for massive fields in de~Sitter spacetime. For massive scalar and vector fields (including a linear gauge-fixing term), the propagators are well-known and have been constructed using a variety of methods, and we refer the reader to Ref.~\cite{Frob:2016hkx} and references therein. A recent work~\cite{Farnsworth:2024iwc} presents solutions of the equations of motion for massive vector and tensor fields, which, however, fail to satisfy the Hadamard condition~\cite[and references therein]{DeWitt:1960fc,Decanini:2005gt,Hack:2012qf,Hollands:2014eia}. The Hadamard condition ensures that the high-energy behaviour of the state is the same as in Minkowski spacetime. Since curvature is negligible at very high energies (compared to a typical curvature scale of the background spacetime), this is a physically reasonable condition, and in fact it is necessary to have finite correlation functions of composite operators such as the stress tensor~\cite{Fewster:2013lqa}. Canonical quantization of fields in the unique de~Sitter-invariant (Bunch--Davies) vacuum state~\cite{Chernikov:1968zm,Frob:2016hkx,Higuchi:1986py, Gerard:2024eef} shows that this state is Hadamard, and the solutions presented in~\cite{Farnsworth:2024iwc} disagree with the canonical quantization of massive vector fields~\cite{Frob:2016hkx}. For a full resolution of the issue and the determination of the correct massive propagators we refer the reader to our work~\cite{DFF2024b}; we give here only an overview, to fix ideas and notations and to provide the expressions for the propagators relevant for the FRG.

\subsection{Scalar propagator}

The scalar propagator $G^\text{F}_{m^2}$ of mass $m$ solves the inhomogeneous Klein--Gordon equation
\begin{equation}
\label{eq:kleingordon}
\left( \nabla^2 - m^2 \right) G^\text{F}_{m^2}(x,x') = \delta(x,x') \eqend{,}
\end{equation}
where $\delta(x,x') \equiv \delta^4(x-x')/\sqrt{-g}$ is the covariant Dirac $\delta$ distribution in $4$ dimensions.

Since de Sitter spacetime is maximally symmetric, the propagators in the unique de Sitter-invariant Hadamard state (the Bunch--Davies vacuum) only depend on the geodesic distance between the two points, with the tensor structure given by invariant bitensors~\cite{Allen:1985wd}. For practical computations it is convenient to introduce $Z = \cos(H \mu)$, where $H$ is the Hubble constant, the inverse de Sitter radius, and $\mu$ is the geodesic distance.

Making the ansatz $\mathi G^\text{F}_{m^2}(x,x') = f(Z(x,x'))$, and using the identities
\begin{equation}
\label{eq:z_identities}
\nabla_\mu \nabla_\nu Z = - Z H^2 g_{\mu\nu} \eqend{,} \quad \nabla_\mu Z \nabla^\mu Z = H^2 (1-Z^2) \eqend{,}
\end{equation}
outside of the coincidence limit Eq.~\eqref{eq:kleingordon} reduces to
\begin{equation}
\label{eq:kleingordon_hypergeom_eq}
H^2 (1-Z^2) f''(Z) - 4 Z H^2 f'(Z) - m^2 f(Z) = 0 \eqend{.}
\end{equation}
This is a hypergeometric equation, whose general solution reads
\begin{splitequation}
\label{eq:kleingordon_hypergeom_z}
f(Z) &= c \hypergeom{2}{1}\left( \frac{3}{2} + \nu, \frac{3}{2} - \nu; 2; \frac{1+Z}{2} \right) \\
&\quad+ d \hypergeom{2}{1}\left( \frac{3}{2} + \nu, \frac{3}{2} - \nu; 2; \frac{1-Z}{2} \right)
\end{splitequation}
with two constants $c$ and $d$ and the parameter
\begin{equation}
\label{eq:param_nu_def}
\nu \equiv \sqrt{ \frac{9}{4} - \frac{m^2}{H^2} } \eqend{.}
\end{equation}

We now need to impose the Hadamard condition. In $4$ dimensions, whenever $x$ and $x'$ lie in a convex geodesic neighborhood, such that they can be connected by a geodesic, the Hadamard condition requires that the propagator has the short-distance expression~\cite{DeWitt:1960fc,Decanini:2005gt,Hack:2012qf,Hollands:2014eia}
\begin{equation}
\label{eq:scalar_hadamard}
\mathi G^\text{F}_{m^2}(x,x') = \frac{1}{8 \pi^2} \left[ \frac{U(x,x')}{\sigma(x,x') + \mathi \epsilon} + V(x,x') \ln\left[ \ell^{-2} \sigma(x,x') + \mathi \epsilon \right] + W(x,x') \right] \eqend{,}
\end{equation}
In this expression, $\sigma = \frac{1}{2} \mu^2$ is the Synge world function, equal to one half of the geodesic distance squared; $U$, $V$ and $W$ are smooth symmetric biscalars~\cite{Moretti:1999fb,Hack:2012qf,Kaminski:2019adk}, of which $U$ and $V$ are state-independent and determined by the geometry, while $W$ encodes the state; $\ell$ is a scale needed to make the argument of the logarithm dimensionless, and the distributional limit $\epsilon \to 0$ is understood.

In particular, the propagator for a Hadamard state is divergent only when $\sigma = 0$, i.e., when $x$ and $x'$ are light-like related.\footnote{In a mathematically precise formulation, we say that the singular support of the distribution $G^\text{F}_{m^2}(x,x')$ is the set $\{ (x,x') \in \mathcal{M}^2 \colon \sigma(x,x') = 0 \}$. The Hadamard form~\eqref{eq:scalar_hadamard} can then be shown to be equivalent to a microlocal spectrum condition~\cite{Radzikowski:1996pa,Moretti:2021pzz}, a refinement of the singular support taking into account also in which directions in momentum space the distribution is singular. For our computations, the Hadamard form~\eqref{eq:scalar_hadamard} is sufficient.} 

The hypergeometric equation~\eqref{eq:kleingordon_hypergeom_eq} has the regular singular points $0$, $1$ and $\infty$, which translates into $Z = \pm 1$ and $Z = \infty$. From the relation $Z = \cos(H \mu)$, we see that $Z = 1$ corresponds to $\mu = 0$, which is the singularity of the Hadamard form~\eqref{eq:scalar_hadamard} that we want to keep. However, $Z = -1$ corresponds to $H \mu = \pi$, which is the geodesic distance between two antipodal points on the de Sitter hyperboloid (or more generally, whenever the antipodal point of $x'$ lies on the light-cone of $x$). Since a Hadamard state must be regular in this case, we must discard the solution in Eq.~\eqref{eq:kleingordon_hypergeom_z} which diverges for $Z = -1$. Finally, the limit $Z \to \pm \infty$ corresponds to infinite timelike or spacelike separation, for which the relation between $Z$ and the geodesic distance $\mu$ is more complicated~\cite{Allen:1985wd,Perez-Nadal:2009jcz}, and where the Hadamard form~\eqref{eq:scalar_hadamard} does not impose a restriction on $f$. Since the hypergeometric function becomes singular when its argument becomes $1$, we see that we have to set $d = 0$ to exclude a singularity at $Z = -1$. Using the known expansion of the hypergeometric function near $1$, we then obtain
\begin{equation}
f(Z) \approx \frac{2 c}{\Gamma\left( \frac{3}{2} + \nu \right) \Gamma\left( \frac{3}{2} - \nu \right)} (1-Z)^{-1} \eqend{,}
\end{equation}
and since $1-Z = 1 - \cos(H \mu) \approx H^2 \sigma$ and $U(x,x') = 1$ for $\sigma = 0$, we can match the Hadamard form~\eqref{eq:scalar_hadamard} by choosing
\begin{equation}
c = \frac{\Gamma\left( \frac{3}{2} + \nu \right) \Gamma\left( \frac{3}{2} - \nu \right)}{(4 \pi)^2} H^2 = \frac{m^2 - 2 H^2}{16 \pi \cos(\pi \nu)} \eqend{.}
\end{equation}
The logarithmic singularity is matched by subleading terms in the expansion of the hypergeometric function near $Z = 1$. The logarithmic terms arise because the singularity at argument $1$ is confluent, since the difference $\frac{3}{2} + \nu + \frac{3}{2} - \nu - 2 = 1$ between the parameters is an integer. We have thus determined the unique de Sitter-invariant Hadamard state for massive scalar fields, whose propagator is given by
\begin{equation}
\label{eq:scalar_propagator}
\mathi G^\text{F}_{m^2}(x,x') = F_\nu(Z - \mathi \epsilon) \eqend{,}
\end{equation}
with the function
\begin{equation}
\label{eq:fnu_def}
F_\nu(Z) \equiv \frac{H^2 \Gamma\left( \frac{3}{2} + \nu \right) \Gamma\left( \frac{3}{2} - \nu \right)}{(4 \pi)^2} \hypergeom{2}{1}\left( \frac{3}{2} + \nu, \frac{3}{2} - \nu; 2; \frac{1+Z}{2} \right) \eqend{.}
\end{equation}
The $\mathi \epsilon$ prescription imposes that we need the value of the hypergeometric function from below the branch cut at $Z \in [1,\infty)$, and coincides with the prescription of the Hadamard form~\eqref{eq:scalar_hadamard} of the Feynman propagator. As $m \to 0$ such that $\nu \to \frac{3}{2}$, $F_\nu$ is divergent due to the $\Gamma$ functions in front, which is the well-known IR divergence of the massless and minimally coupled scalar in de~Sitter spacetime~\cite{Ford:1977in}. However, the divergent part is only a constant independent of $Z$, and for later use we define
\begin{equation}
\label{eq:hatfnu_def}
\hat{F}_\nu(Z) \equiv F_\nu(Z) - \frac{9 H^4}{24 \pi^2 m^2} + \frac{51 H^2}{160 \pi^2} \to \frac{H^2}{8 \pi^2} \left( \frac{1}{1-Z} + \ln \frac{2}{1-Z} + \frac{13}{60} \right) \quad (m \to 0) \eqend{.}
\end{equation}
The Bunch-Davies scalar propagator is of course well-known, but serves as a simple example of the procedure that needs to be followed. In the next subsections, we use the same method to determine the propagators of massive vector and tensor fields.

\subsection{Vector propagator}

The equation for the vector propagator $G^\text{F}_{m^2,\alpha,\mu\rho'}$ with mass parameter $m$ and gauge-fixing parameter $\alpha$ reads
\begin{equation}
\label{eq:eom_vector}
\left( \nabla^2 - m^2 - 3 H^2 \right) G^\text{F}_{m^2,\alpha,\mu\rho'}(x,x') - \left( 1 - \frac{1}{\alpha} \right) \nabla_\mu \nabla^\nu G^\text{F}_{m^2,\alpha,\nu\rho'}(x,x') = g_{\mu\rho'} \delta(x,x') \eqend{,}
\end{equation}
where $g_{\mu\rho'}$ is the parallel propagator~\cite{Allen:1985wd,Poisson:2011nh}. Taking a derivative of this equation and commuting covariant derivatives, we obtain
\begin{equation}
\label{eq:eom_vector_div}
\left( \frac{1}{\alpha} \nabla^2 - m^2 \right) \nabla^\mu G^\text{F}_{m^2,\alpha,\mu\rho'}(x,x') = - \nabla_{\rho'} \delta(x,x') \eqend{,}
\end{equation}
where we used that~\cite{DeWitt:1960fc}
\begin{equation}
\label{eq:div_delta_relation}
\nabla^\mu \left[ g_{\mu\rho'} \delta(x,x') \right] = - \nabla_{\rho'} \delta(x,x') \eqend{.}
\end{equation}
The solution of Eq. \eqref{eq:eom_vector_div} is given by
\begin{equation}
\label{eq:vector_div}
\nabla^\mu G^\text{F}_{m^2,\alpha,\mu\rho'}(x,x') = - \alpha \nabla_{\rho'} G^\text{F}_{\alpha m^2}(x,x') \eqend{,}
\end{equation}
where we used the equation of motion~\eqref{eq:kleingordon} of the scalar propagator. We thus have to solve equation~\eqref{eq:eom_vector} together with the constraint~\eqref{eq:vector_div}.

The vector propagator is a bitensor, transforming like the vector $A_\mu$ at $x$ and like $A_{\rho'}$ at $x'$, and in a de Sitter-invariant state the tensor structure of the propagator must be given by de Sitter-invariant bitensors. A basis of such bitensors is given by the unit tangent vectors to the geodesic at $x$ and $x'$, which are given by the covariant derivatives of the geodesic distance $\mu$, and the parallel propagator~\cite{Allen:1985wd}. For our purposes, it is more practical to work with covariant derivatives of $Z$ instead, and so we make the ansatz\footnote{This form of the ansatz results from decoupling the resulting equations, see~\cite{DFF2024b} for more details.}
\begin{splitequation}
\label{eq:vectorprop_ansatz}
\mathi G^\text{F}_{m^2,\alpha,\mu\rho'}(x,x') &= G_\text{A}(Z(x,x')) \nabla_\mu \nabla_{\rho'} Z(x,x') \\
&\quad+ \nabla_\mu Z(x,x') \nabla_{\rho'} \left[ G_\text{A}(Z(x,x')) + G_\text{B}(Z(x,x')) \right] \eqend{,}
\end{splitequation}
where $G_\text{A}$ and $G_\text{B}$ are scalar functions that need to be determined. Let us first consider the case $\alpha = 1$, which corresponds to Feynman gauge. We insert the ansatz~\eqref{eq:vectorprop_ansatz} into the equation~\eqref{eq:eom_vector} for $x \neq x'$, commute covariant derivatives and use the identities~\eqref{eq:z_identities}. Since the two tensor structures in~\eqref{eq:vectorprop_ansatz} are independent, the coefficient of each must vanish, and taking suitable linear combinations we obtain
\begin{equations}[eq:eom_vector_ga_gb]
H^2 (1-Z^2) G_\text{A}''(Z) - 6 H^2 Z G_\text{A}'(Z) - (m^2 + 4 H^2) G_\text{A}(Z) &= 2 H^2 Z G_\text{B}'(Z) \eqend{,} \\
\partial_Z \left[ H^2 (1-Z^2) G_\text{B}''(Z) - 4 H^2 Z G_\text{B}'(Z) - (m^2 + 2 H^2) G_\text{B}(Z) \right] &= 0 \eqend{.}
\end{equations}
Moreover, the divergence constraint~\eqref{eq:vector_div} results for Feynman gauge $\alpha = 1$ in
\begin{equation}
\label{eq:vector_div_constraint}
- (1-Z^2) G_\text{A}''(Z) - (1-Z^2) G_\text{B}''(Z) + 6 Z G_\text{A}'(Z) + 5 Z G_\text{B}'(Z) + 4 G_\text{A}(Z) = H^{-2} F_\nu'(Z - \mathi \epsilon) \eqend{,}
\end{equation}
where we used again the identities~\eqref{eq:z_identities} and the solution~\eqref{eq:scalar_propagator} for the scalar propagator. Note that the second equation of motion~\eqref{eq:eom_vector_ga_gb} for $G_\text{B}$ is solved by solving the second-order equation in brackets and adding an arbitrary constant, which is consistent with the fact that both in the ansatz~\eqref{eq:vectorprop_ansatz} and the constraint~\eqref{eq:vector_div_constraint} only the derivative of $G_\text{B}$ enters. Taking into account the equations~\eqref{eq:eom_vector_ga_gb} for $G_\text{A}$ and $G_\text{B}$, the constraint results in a unique solution for $G_\text{A}$:
\begin{equation}
G_\text{A}(Z) = - \frac{1}{m^2} \left[ H^2 Z G_\text{B}'(Z) + (m^2 + 2 H^2) G_\text{B}(Z) + F_\nu'(Z - \mathi \epsilon) \right] \eqend{.}
\end{equation}
As in the case of the scalar propagator, to obtain a Hadamard state we have to impose that the solution remains regular at $Z = -1$ and that the branch cut starting from the singularity at $Z = 1$ has the correct $\mathi \epsilon$ prescription for a Feynman propagator. This gives the solution for $G_\text{B}$, which reads
\begin{equation}
G_\text{B}(Z) = c F_\rho(Z - \mathi \epsilon) + c'
\end{equation}
with constants $c$ and $c'$ and the parameter
\begin{equation}
\label{eq:param_rho_def}
\rho \equiv \sqrt{ \frac{1}{4} - \frac{m^2}{H^2} }
\end{equation}
in addition to $\nu$~\eqref{eq:param_nu_def}. The constant $c$ is determined by requiring that the propagator has the Hadamard form. For $\alpha = 1$, where the vector propagator is in Feynman gauge, this means that~\cite{DeWitt:1960fc,Decanini:2005gt}
\begin{equation}
\label{eq:vector_hadamard}
\mathi G^\text{F}_{m^2,1,\mu\rho'}(x,x') = \frac{1}{8 \pi^2} \left[ \frac{U_{\mu\rho'}(x,x')}{\sigma(x,x') + \mathi \epsilon} + V_{\mu\rho'}(x,x') \ln\left[ \ell^{-2} \sigma(x,x') + \mathi \epsilon \right] + W_{\mu\rho'}(x,x') \right] \eqend{,}
\end{equation}
which as in the scalar case is valid for $x$ and $x'$ lying in a convex geodesic neighbourhood. In particular, since $U_{\mu\rho'}(x,x) = g_{\mu\rho'}$, the propagator does not diverge faster than $1/(1-Z)$ near coincidence. Since $\lim_{x' \to x} \nabla_\mu Z = 0$ and $\lim_{x' \to x} \nabla_\mu \nabla_{\rho'} Z = H^2 g_{\mu\rho'}$, with the ansatz~\eqref{eq:vectorprop_ansatz} we need to impose that
\begin{equation}
G_\text{A}(Z) \approx \frac{1}{8 \pi^2} \frac{1}{1-Z}
\end{equation}
where we used again that $1 - Z = 1 - \cos(H \mu) \approx H^2 \sigma$ close to coincidence. Using the known expansion of the hypergeometric function near 1, this holds for $c = - H^{-2}$.

Putting all these considerations together, we obtain the unique de Sitter-invariant, Hadamard vector propagator in Feynman gauge:
\begin{splitequation}
\label{eq:vector_propagator_feynman_gauge}
\mathi G^\text{F}_{m^2,1,\mu\rho'}(x,x') &= \left[ H_{m^2}(Z - \mathi \epsilon) + \frac{1}{H^2} F_\rho(Z - \mathi \epsilon) \right] \nabla_\mu \nabla_{\rho'} Z + H'_{m^2}(Z - \mathi \epsilon) \nabla_\mu Z \nabla_{\rho'} Z \\
&= \frac{1}{H^2} F_\rho(Z - \mathi \epsilon) \nabla_\mu \nabla_{\rho'} Z + \nabla_\mu \left[ H_{m^2}(Z - \mathi \epsilon) \nabla_{\rho'} Z \right] \eqend{.}
\end{splitequation}
Here we defined the linear combination
\begin{equation}
\label{eq:hm2_def}
H_{m^2}(Z) \equiv \frac{1}{m^2} \left[ Z F'_\rho(Z) - F'_\nu(Z) + 2 F_\rho(Z) \right] \eqend{,}
\end{equation}
which is logarithmically divergent in the coincidence limit, and has a finite limit as $m \to 0$. Furthermore, we chose $c' = 0$ for simplicity.

To derive the propagator in a general gauge, we add a longitudinal term to the Feynman gauge propagator, such that
\begin{equation}
\label{eq:vector_propagator_general_gauge_ansatz}
G^\text{F}_{m^2,\alpha,\mu\rho'}(x,x') = G^\text{F}_{m^2,1,\mu\rho'}(x,x') + \nabla_\mu \nabla_{\rho'} G_\text{C}(x,x') \eqend{.}
\end{equation}
Taking a divergence of this equation and using the result~\eqref{eq:vector_div}, we obtain
\begin{equation}
\nabla^2 G_\text{C}(x,x') = G^\text{F}_{m^2}(x,x') - \alpha G^\text{F}_{\alpha m^2}(x,x') \eqend{.}
\end{equation}
The solution to this is easily seen to be
\begin{equation}
\label{eq:vector_propagator_general_gauge_ansatz_sol}
G_\text{C}(x,x') = \frac{1}{m^2} \left[ G^\text{F}_{m^2}(x,x') - G^\text{F}_{\alpha m^2}(x,x') \right] \eqend{,}
\end{equation}
such that the unique de Sitter-invariant vector propagator in a general linear gauge reads
\begin{splitequation}
\label{eq:vector_propagator_general_gauge}
\mathi G^\text{F}_{m^2,\alpha,\mu\rho'}(x,x') &= \left[ H_{m^2}(Z - \mathi \epsilon) + \frac{1}{H^2} F_\rho(Z - \mathi \epsilon) \right] \nabla_\mu \nabla_{\rho'} Z + H'_{m^2}(Z - \mathi \epsilon) \nabla_\mu Z \nabla_{\rho'} Z \\
&\quad+ \frac{1}{m^2} \nabla_\mu \nabla_{\rho'} \left[ F_\nu(Z - \mathi \epsilon) - F_{\nu^\alpha}(Z - \mathi \epsilon) \right]
\end{splitequation}
with the new parameter
\begin{equation}
\nu^\alpha \equiv \sqrt{ \frac{9}{4} - \alpha m^2 } \eqend{.}
\end{equation}
As a check on the computation, one verifies that this propagator agrees with the one derived in Ref.~\cite{Frob:2016hkx} via canonical quantization.

\subsection{Tensor propagator}

For the tensor propagator, an additional difficulty as compared to the vector propagator is that we have to separate the spin-0 (scalar) sector from the spin-2 (tensor) one. Moreover, the spin-2 sector depends on two gauge-fixing parameters $\xi$ and $\zeta$, and needs to be rescaled to have a Hadamard expansion. The equation that we need to solve reads
\begin{equation}
\label{eq:eom_tensor_xizeta}
P_{\xi,\zeta}^{\mu\nu\alpha\beta} G^\text{F}_{m^2,M^2,\zeta,\xi,\alpha\beta\rho'\sigma'}(x,x') = g^\mu{}_{(\rho'} g^\nu{}_{\sigma')} \delta(x,x')
\end{equation}
with the differential operator
\begin{splitequation}
\label{eq:p_xizeta_def}
P_{\xi,\zeta}^{\rho\sigma\mu\nu} &\equiv \frac{1}{2} \left[ g^{\rho(\mu} g^{\nu)\sigma} - \frac{1}{2} \left( 2 - \frac{1}{\xi \zeta^2} \right) g^{\rho\sigma} g^{\mu\nu} \right] \nabla^2 \\
&\quad- \left( 1 - \frac{1}{\xi} \right) \nabla^{(\rho} g^{\sigma)(\mu} \nabla^{\nu)} + \frac{1}{2} \left( 1 - \frac{1}{\xi \zeta} \right) \left( g^{\mu\nu} \nabla^\rho \nabla^\sigma + g^{\rho\sigma} \nabla^\mu \nabla^\nu \right) \\
&\quad- \frac{m^2 + 2 H^2}{2} g^{\rho(\mu} g^{\nu)\sigma} + \frac{M^2 + m^2 - 4 H^2}{8} g^{\rho\sigma} g^{\mu\nu} \eqend{,}
\end{splitequation}
which also depends on two masses $m$ and $M$. As in the vector case, we obtain a constraint by taking two derivatives of this equation and commuting covariant derivatives, and another by taking its trace. This gives
\begin{splitequation}
\label{eq:tensor_trddiv1_constraint}
&\left[ \frac{1 - 2 \zeta}{4 \xi \zeta^2} \nabla^2 + \frac{\xi \zeta (m^2 + M^2) - 12 H^2}{8 \xi \zeta} \right] \nabla^2 G^\text{F}_{m^2,M^2,\zeta,\xi}{}^\alpha{}_{\alpha\rho'\sigma'}(x,x') \\
&= - \nabla_{\rho'} \nabla_{\sigma'} \delta(x,x') - \left[ \frac{1 - 2 \zeta}{2 \xi \zeta} \nabla^2 - \frac{6 H^2 - \xi m^2}{2 \xi} \right] \nabla^\alpha \nabla^\beta G^\text{F}_{m^2,M^2,\zeta,\xi,\alpha\beta\rho'\sigma'}(x,x')
\end{splitequation}
and
\begin{splitequation}
\label{eq:tensor_trddiv2_constraint}
&\left[ \left( 2 + \frac{1}{\xi \zeta} - \frac{2}{\xi \zeta^2} \right) \nabla^2 - (M^2 - 6 H^2) \right] G^\text{F}_{m^2,M^2,\zeta,\xi}{}^\alpha{}_{\alpha\rho'\sigma'}(x,x') \\
&= - 2 g_{\rho'\sigma'} \delta(x,x') + 2 \left( 1 + \frac{1}{\xi} - \frac{2}{\xi \zeta} \right) \nabla^\alpha \nabla^\beta G^\text{F}_{m^2,M^2,\zeta,\xi,\alpha\beta\rho'\sigma'}(x,x') \eqend{,}
\end{splitequation}
and taking a single derivative of the equation of motion~\eqref{eq:eom_tensor_xizeta} and using the above results we also obtain the constraint
\begin{splitequation}
\label{eq:tensor_div_constraint}
&\left( \nabla^2 - \xi m^2 + 3 H^2 \right) \nabla^\alpha G^\text{F}_{m^2,M^2,\zeta,\xi,\alpha\nu\rho'\sigma'}(x,x') \\
&= - 2 \xi g_{\nu(\rho'} \nabla_{\sigma')} \delta(x,x') + \frac{\xi (1-\zeta)}{\zeta - 2 + \xi \zeta} g_{\rho'\sigma'} \nabla_\nu \delta(x,x') \\
&\quad- \frac{2 + \xi - \zeta + 2 \xi \zeta (\zeta-2)}{2 \zeta (\zeta - 2 + \xi \zeta)} \nabla_\nu \nabla^2 G^\text{F}_{m^2,M^2,\zeta,\xi}{}^\alpha{}_{\alpha\rho'\sigma'}(x,x') \\
&\quad- \frac{12 H^2 (\zeta - 2) (\xi \zeta - 1) + M^2 (\xi - 1) \xi \zeta^2 + m^2 \xi \zeta (\zeta - 2 + \xi \zeta)}{4 \zeta (\zeta - 2 + \xi \zeta)} \nabla_\nu G^\text{F}_{m^2,M^2,\zeta,\xi}{}^\alpha{}_{\alpha\rho'\sigma'}(x,x') \eqend{.}
\end{splitequation}

As for the vector propagator we first consider the case $\xi = \zeta = 1$, which corresponds to the Feynman gauge (also known as de Donder gauge in the classical theory). In this case, the constraint~\eqref{eq:tensor_trddiv2_constraint} reduces to
\begin{equation}
\left( \nabla^2 - M^2 + 6 H^2 \right) G^\text{F}_{m^2,M^2,1,1}{}^\alpha{}_{\alpha\rho'\sigma'}(x,x') = - 2 g_{\rho'\sigma'} \delta(x,x') \ ,
\end{equation}
with solution
\begin{equation}
\label{eq:tensor_feynman_trace}
G^\text{F}_{m^2,M^2,1,1}{}^\alpha{}_{\alpha\rho'\sigma'}(x,x') = - 2 g_{\rho'\sigma'} G^\text{F}_{M^2 - 6 H^2}(x,x') \eqend{.}
\end{equation}
Thus, we have separated the spin-0 propagator, which is proportional to a scalar propagator. The other constraint~\eqref{eq:tensor_div_constraint} is singular in this gauge, but the singular part actually vanishes using the solution~\eqref{eq:tensor_feynman_trace}. After some rearrangements and the use of the vector equation of motion~\eqref{eq:eom_vector}, we obtain
\begin{splitequation}
\label{eq:tensor_feynman_div}
\left( \nabla^2 - m^2 + 3 H^2 \right) &\biggl[ \nabla^\alpha \left[ G^\text{F}_{m^2,M^2,1,1,\alpha\nu\rho'\sigma'}(x,x') - \frac{1}{4} g_{\alpha\nu} G^\text{F}_{m^2,M^2,1,1}{}^\beta{}_{\beta\rho'\sigma'}(x,x') \right] \\
&\quad+ 2 \nabla_{(\rho'} G^\text{F}_{m^2-6H^2,1,|\mu|\sigma')}(x,x') + \frac{1}{2} g_{\rho'\sigma'} \nabla_\nu G^\text{F}_{m^2 - 6 H^2}(x,x') \biggr] = 0 \eqend{.}
\end{splitequation}
We therefore make the decomposition
\begin{equation}
G^\text{F}_{m^2,M^2,1,1,\mu\nu\rho'\sigma'}(x,x') = 2 G^\text{F}_{m^2,\mu\nu\rho'\sigma'}(x,x') - \frac{1}{2} g_{\mu\nu} g_{\rho'\sigma'} G^\text{F}_{M^2 - 6 H^2}(x,x') \eqend{,}
\end{equation}
where $G^\text{F}_{m^2,\mu\nu\rho'\sigma'}(x,x')$ is the traceless spin-2 propagator, satisfying the equation of motion
\begin{equation}
\label{eq:eom_tensor}
\left( \nabla^2 - m^2 - 2 H^2 \right) G^\text{F}_{m^2,\mu\nu\rho'\sigma'}(x,x') = \left( g_{\mu(\rho'} g_{\sigma')\nu} - \frac{1}{4} g_{\mu\nu} g_{\rho'\sigma'} \right) \delta(x,x') \ ,
\end{equation}
obtained by inserting this decomposition into Eq.~\eqref{eq:eom_tensor_xizeta} for $\xi = \zeta = 1$, and the constraint
\begin{equation}
\label{eq:tensor_div}
\nabla^\mu G^\text{F}_{m^2,\mu\nu\rho'\sigma'}(x,x') = - \nabla_{(\rho'} G^\text{F}_{m^2-6H^2,1,|\mu|\sigma')}(x,x') - \frac{1}{4} g_{\rho'\sigma'} \nabla_\nu G^\text{F}_{m^2 - 6 H^2}(x,x') \ ,
\end{equation}
obtained from Eq.~\eqref{eq:tensor_feynman_div}.

The most general de Sitter-invariant ansatz is given by\footnote{This form of the ansatz results again from decoupling the resulting equations, see~\cite{DFF2024b} for more details.}
\begin{splitequation}
\label{eq:tensorprop_ansatz}
\mathi G^\text{F}_{m^2,\mu\nu\rho'\sigma'}(x,x') &= G_\text{A}(Z(x,x')) T^{(1)}_{\mu\nu\rho'\sigma'} + \Bigl[ G_\text{B}'(Z(x,x')) + 2 G_\text{A}'(Z(x,x')) \Bigr] T^{(2)}_{\mu\nu\rho'\sigma'} \\
&\quad+ \Bigl[ G_\text{C}''(Z(x,x')) + \frac{1}{2} G_\text{B}''(Z(x,x')) + \frac{1}{2} G_\text{A}''(Z(x,x')) \Bigr] T^{(3)}_{\mu\nu\rho'\sigma'}
\end{splitequation}
with
\begin{equations}[eq:tensor_factors]
\begin{split}
T^{(1)}_{\mu\nu\rho'\sigma'} &= \nabla_\mu \nabla_{(\rho'} Z \nabla_{\sigma')} \nabla_\nu Z + \frac{H^2}{4} g_{\mu\nu} \nabla_{\rho'} Z \nabla_{\sigma'} Z \\
&\quad+ \frac{H^2}{4} g_{\rho'\sigma'} \nabla_\mu Z \nabla_\nu Z - \frac{(5-Z^2) H^4}{16} g_{\mu\nu} g_{\rho'\sigma'} \eqend{,}
\end{split} \\
\begin{split}
T^{(2)}_{\mu\nu\rho'\sigma'} &= \nabla_{(\mu} Z \nabla_{\nu)} \nabla_{(\rho'} Z \nabla_{\sigma')} Z + \frac{H^2 Z}{4} g_{\mu\nu} \nabla_{\rho'} Z \nabla_{\sigma'} Z \\
&\quad+ \frac{H^2 Z}{4} g_{\rho'\sigma'} \nabla_\mu Z \nabla_\nu Z - \frac{H^4 Z}{16} (1-Z^2) g_{\mu\nu} g_{\rho'\sigma'} \eqend{,}
\end{split} \\
T^{(3)}_{\mu\nu\rho'\sigma'} &= \left[ \nabla_\mu Z \nabla_\nu Z - \frac{1}{4} g_{\mu\nu} H^2 (1-Z^2) \right] \left[ \nabla_{\rho'} Z \nabla_{\sigma'} Z - \frac{1}{4} g_{\rho'\sigma'} H^2 (1-Z^2) \right] \eqend{,}
\end{equations}
where the condition of tracelessness has reduced the a priori five independent tensor structures to three.

Plugging the ansatz~\eqref{eq:tensorprop_ansatz} into the equation of motion~\eqref{eq:eom_tensor}, we obtain
\begin{equations}[eq:eom_tensor_ga_gb_gc]
H^2 (1-Z^2) G_\text{A}''(Z) - 8 H^2 Z G_\text{A}'(Z) - \left( m^2 + 4 H^2 \right) G_\text{A}(Z) &= 2 H^2 Z G_\text{B}'(Z) \eqend{,} \\
\partial_Z \left[ H^2 (1-Z^2) G_\text{B}''(Z) - 6 H^2 Z G_\text{B}'(Z) - m^2 G_\text{B}(Z) \right] &= 8 H^2 Z G_\text{C}''(Z) \eqend{,} \\
\partial_Z^2 \left[ H^2 (1-Z^2) G_\text{C}''(Z) - 4 H^2 Z G_\text{C}'(Z) - m^2 G_\text{C}(Z) \right] &= 0 \eqend{,}
\end{equations}
and as in the vector case the latter two equations are solved by solving the second-order equations in brackets, and adding an arbitrary constant to $G_\text{B}$ or an arbitrary linear function of $Z$ to $G_\text{C}$. Contrary to the vector case, they must be non-vanishing to obtain a finite limit $m \to 0$. Again, we also have further constraints, given by Eq.~\eqref{eq:tensor_div}, which as in the vector case give us a unique solution for $G_\text{A}$ and $G_\text{B}$ after employing the equations of motion~\eqref{eq:eom_tensor_ga_gb_gc}. This solution reads
\begin{splitequation}
G_\text{A}(Z) &= - \frac{2}{3 (m^2 - 2 H^2)} \left[ 4 K'_{m^2}(Z - \mathi \epsilon) + \frac{2}{m^2} Z F''_\sigma(Z - \mathi \epsilon) + \frac{1}{H^2} F'_\sigma(Z - \mathi \epsilon) \right] \\
&\quad+ \frac{4 H^2}{m^2 - 2 H^2} \left[ \frac{2 H^2}{3 m^2} G_\text{C}''(Z) + Z G_\text{C}'(Z) + \frac{m^2 + 2 H^2}{2 H^2} G_\text{C}(Z) \right] \eqend{,}
\end{splitequation}
\begin{splitequation}
G_\text{B}(Z) &= \frac{2}{H^2 m^2} \left[ F'_\sigma(Z - \mathi \epsilon) - 2 H^4 Z G_\text{C}'(Z) - 2 H^2 (m^2 + 3 H^2) G_\text{C}(Z) \right] + c
\end{splitequation}
with a constant $c$, and where we have defined the new parameters
\begin{equation}
\label{eq:param_sigmatau_def}
\sigma \equiv \sqrt{ \frac{25}{4} - \frac{m^2}{H^2} } \eqend{,} \quad \tau \equiv \sqrt{ \frac{33}{4} - \frac{m^2}{H^2} }
\end{equation}
and the linear combination
\begin{equation}
K_{m^2}(Z) = \frac{1}{m^2 - 6 H^2} \left[ Z F'_\sigma(Z) - F'_\tau(Z) - \left( 1 - \frac{m^2}{2 H^2} \right) F_\sigma(Z) \right] \eqend{.}
\end{equation}
$K_{m^2}(Z)$ is finite as $m \to 0$, and it diverges like $1/(1-Z)$ in the coincidence limit.

The remaining equation of motion~\eqref{eq:eom_tensor_ga_gb_gc} for $G_\text{C}$ is easily solved, namely
\begin{equation}
G_\text{C}(Z) = c' F_\nu(Z - \mathi \epsilon) + c'' Z + c''' \ ,
\end{equation}
with constants $c'$, $c''$, $c'''$, and the appropriate $\mathi \epsilon$ prescription for a Feynman propagator. The constant $c'$ is again determined by requiring that the propagator has the Hadamard form. In Feynman harmonic gauge $\xi = \zeta = 1$, this is a simple generalization of the vector one~\eqref{eq:vector_hadamard}
\begin{equation}
\label{eq:tensor_hadamard}
\mathi G^\text{F}_{m^2,\mu\nu\rho'\sigma'}(x,x') = \frac{1}{8 \pi^2} \left[ \frac{U_{\mu\nu\rho'\sigma'}(x,x')}{\sigma(x,x') + \mathi \epsilon} + V_{\mu\nu\rho'\sigma'}(x,x') \ln\left[ \ell^{-2} \sigma(x,x') + \mathi \epsilon \right] + W_{\mu\nu\rho'\sigma'}(x,x') \right] \eqend{,}
\end{equation}
which again is valid for $x$ and $x'$ lying in a convex geodesic neighborhood. Because we consider the traceless propagator, we have $U_{\mu\nu\rho'\sigma'}(x,x) = g_{\mu(\rho'} g_{\sigma')\nu} - \frac{1}{4} g_{\mu\nu} g_{\rho'\sigma'}$ and the propagator does not diverge faster than $1/(1-Z)$ near coincidence. Using as in the vector case that $\lim_{x' \to x} \nabla_\mu Z = 0$ and $\lim_{x' \to x} \nabla_\mu \nabla_{\rho'} Z = H^2 g_{\mu\rho'}$ and that $1 - Z = 1 - \cos(H \mu) \approx H^2 \sigma$ close to coincidence, with the ansatz~\eqref{eq:tensorprop_ansatz} we need to impose that
\begin{equation}
G_\text{A}(Z) \approx \frac{1}{8 \pi^2 H^2} \frac{1}{1-Z} \eqend{.}
\end{equation}
Using the known expansion of the hypergeometric function near 1~\cite[Eq.~15.4.23]{dlmf}, this holds for $c' = 1/(2 H^4)$.

Putting all together, we obtain the unique de Sitter-invariant, Hadamard, traceless spin-2 tensor propagator in Feynman gauge:
\begin{splitequation}
\label{eq:tensor_propagator_feynman_gauge}
&\mathi G^\text{F}_{m^2,\mu\nu\rho'\sigma'}(x,x') \\
&= - \frac{2}{3 (m^2 - 2 H^2)} \Biggl[ 4 K'_{m^2}(Z - \mathi \epsilon) + 2 H^2 Z L'_{m^2}(Z - \mathi \epsilon) \\
&\qquad\qquad+ m^2 L_{m^2}(Z - \mathi \epsilon) - \left( 2 + \frac{m^2}{2 H^2} \right) \hat{F}_\nu(Z - \mathi \epsilon) \Biggr] T^{(1)}_{\mu\nu\rho'\sigma'} \\
&\quad- \frac{4}{3 (m^2 - 2 H^2)} \Biggl[ 4 K''_{m^2}(Z - \mathi \epsilon) + 2 H^2 Z L''_{m^2}(Z - \mathi \epsilon) \\
&\qquad\qquad+ \left( 5 H^2 - m^2 \right) L'_{m^2}(Z - \mathi \epsilon) - \left( 2 + \frac{m^2}{2 H^2} \right) F'_\nu(Z - \mathi \epsilon) \Biggr] T^{(2)}_{\mu\nu\rho'\sigma'} \\
&\quad- \frac{1}{3 (m^2 - 2 H^2)} \Biggl[ 4 K'''_{m^2}(Z - \mathi \epsilon) + 2 H^2 Z L'''_{m^2}(Z - \mathi \epsilon) \\
&\qquad\qquad+ 2 \left( 5 H^2 - m^2 \right) L''_{m^2}(Z - \mathi \epsilon) + \left( 1 - \frac{2 m^2}{H^2} \right) F''_\nu(Z - \mathi \epsilon) \Biggr] T^{(3)}_{\mu\nu\rho'\sigma'}
\end{splitequation}
where the tensor factors $T^{(k)}_{\mu\nu\rho'\sigma'}$ are given in Eqs.~\eqref{eq:tensor_factors}, $\hat{F}_\nu(Z)$ is defined by~\eqref{eq:hatfnu_def}, we made the choice $c''' \neq 0$ in such a way that the limit $m \to 0$ is finite, and we defined the linear combination
\begin{equation}
L_{m^2}(Z) = \frac{1}{m^2 H^2} \left[ F'_\sigma(Z) - Z F'_\nu(Z) - \frac{m^2 + 3 H^2}{H^2} F_\nu(Z) - \frac{3 H^4}{4 \pi^2 m^2} + \frac{9 H^2}{20 \pi^2} + \frac{51 m^2}{320 \pi^2} \right] \eqend{,}
\end{equation}
which is finite as $m \to 0$ and diverges like $1/(1-Z)$ in the coincidence limit.

In a generic gauge, analogously to the vector case we add longitudinal and trace terms to the traceless spin-2 propagator, and make the ansatz
\begin{splitequation}
\label{eq:tensor_propagator_ansatz}
G^\text{F}_{m^2,M^2,\zeta,\xi,\mu\nu\rho'\sigma'}(x,x') &= 2 G^\text{F}_{m^2,\mu\nu\rho'\sigma'}(x,x') + \nabla_{(\rho'} \nabla_{|(\mu} G^\text{A}_{\nu)|\sigma')}(x,x') + g_{\mu\nu} g_{\rho'\sigma'} G^\text{B}(x,x') \\
&\quad+ \left( g_{\mu\nu} \nabla_{\rho'} \nabla_{\sigma'} + g_{\rho'\sigma'} \nabla_\mu \nabla_\nu \right) G^\text{C}(x,x') + g_{\mu(\rho'} g_{\sigma')\nu} G^\text{D}(x,x') \eqend{,}
\end{splitequation}
which is the most general form compatible with the symmetries. Here, $G^\text{A}_{\mu\rho'}(x,x')$ is a spin-1 contribution and $G^\text{B}$, $G^\text{C}$ and $G^\text{D}$ are spin-0 contributions, all of which can depend on the masses $m$ and $M$ and the gauge parameters $\zeta$ and $\xi$. Substituting this ansatz into the equation of motion~\eqref{eq:eom_tensor_xizeta} and the constraints~\eqref{eq:tensor_trddiv1_constraint}--\eqref{eq:tensor_div_constraint}, there exists a unique solution for all these contributions, but in a generic gauge they are very involved. We obtain a relatively simple result for $\zeta = \frac{1}{2}$, which reads
\begin{splitequation}
\label{eq:tensor_propagator_zeta12}
&G^\text{F}_{m^2,M^2,\frac{1}{2},\xi,\mu\nu\rho'\sigma'}(x,x') \\
&= 2 G^\text{F}_{m^2,\mu\nu\rho'\sigma'}(x,x') + \frac{4}{m^2} \nabla_{(\rho'} \nabla_{|(\mu} \left[ G^\text{F}_{m^2 - 6 H^2,1,\nu)|\sigma')}(x,x') - G^\text{F}_{\xi m^2 - 6 H^2,1,\nu)|\sigma')}(x,x') \right] \\
&\quad+ \frac{2}{3 (m^2 - 2 H^2)} \left( g_{\mu\nu} \nabla_{\rho'} \nabla_{\sigma'} + g_{\rho'\sigma'} \nabla_\mu \nabla_\nu \right) \left( G^\text{F}_{m^2 - 6 H^2}(x,x') - G^\text{F}_{\hat{m}^2}(x,x') \right) \\
&\quad- \frac{1}{6 (m^2 - 2 H^2)} g_{\mu\nu} g_{\rho'\sigma'} \left[ (m^2 - 6 H^2) G^\text{F}_{m^2 - 6 H^2}(x,x') - 2 \frac{\xi m^2 - 6 H^2}{3-\xi} G^\text{F}_{\hat{m}^2}(x,x') \right] \\
&\quad+ \frac{4}{3 m^2 (m^2 - 2 H^2)} \nabla_\mu \nabla_\nu \nabla_{\rho'} \nabla_{\sigma'} \biggl[ G^\text{F}_{m^2 - 6 H^2}(x,x') - \frac{3 (m^2 - 2 H^2)}{(\xi m^2 - 6 H^2)} G^\text{F}_{\xi m^2 - 6 H^2}(x,x') \\
&\hspace{16em}+ \frac{m^2 (3-\xi)}{(\xi m^2 - 6 H^2)} G^\text{F}_{\hat{m}^2}(x,x') \biggr]
\end{splitequation}
with the Feynman gauge vector propagator~\eqref{eq:vector_propagator_feynman_gauge}, the scalar propagator~\eqref{eq:scalar_propagator} and the mass
\begin{equation}
\hat{m}^2 = \frac{2 (M^2 - 6 H^2) (\xi m^2 - 6 H^2)}{(M^2 - 3 m^2) (3-\xi)} \eqend{.}
\end{equation}
The full expressions for the propagator in a general gauge can be found in Ref.~\cite{DFF2024b}.

\subsection{Hadamard expansion}

The quantity that enters on the right-hand side of the Lorentzian flow equation~\eqref{eq:FRGE} is the coincidence value of the Hadamard-subtracted propagator. In our case, this concerns both the tensor propagator for the metric perturbations and the corresponding vector ghost propagator, and since in a general gauge both of these also depend on derivatives of lower-spin propagator, we need to compute the coincidence value for the scalar propagator as well. Concretely, we need the limit $x' \to x$ of the normal-ordered scalar propagator
\begin{equation}
\label{eq:scalar_hadamard_normalordered}
\normord{ G^\text{F}_{m^2}(x,x') } = G^\text{F}_{m^2}(x,x') + \frac{\mathi}{8 \pi^2} \left[ \frac{U(x,x')}{\sigma(x,x') + \mathi \epsilon} + V(x,x') \ln\left[ \ell^{-2} \sigma(x,x') + \mathi \epsilon \right] \right]
\end{equation}
following from the scalar Hadamard expansion~\eqref{eq:scalar_hadamard}, as well as the limit after taking two or four covariant derivatives. For the vector, we need the limit $x' \to x$ of
\begin{splitequation}
\label{eq:vector_hadamard_normalordered}
&\normord{ G^\text{F}_{m^2,1,\mu\rho'}(x,x') } \\
&= G^\text{F}_{m^2,1,\mu\rho'}(x,x') + \frac{\mathi}{8 \pi^2} \left[ \frac{U_{\mu\rho'}(x,x')}{\sigma(x,x') + \mathi \epsilon} + V_{\mu\rho'}(x,x') \ln\left[ \ell^{-2} \sigma(x,x') + \mathi \epsilon \right] \right]
\end{splitequation}
following from the vector Hadamard expansion~\eqref{eq:vector_hadamard} as well as the limit after taking two covariant derivatives, while for the tensor the limit $x' \to x$ of
\begin{splitequation}
\label{eq:tensor_hadamard_normalordered}
&\normord{ G^\text{F}_{m^2,\mu\nu\rho'\sigma'}(x,x') } \\
&= G^\text{F}_{m^2,\mu\nu\rho'\sigma'}(x,x') + \frac{\mathi}{8 \pi^2} \left[ \frac{U_{\mu\nu\rho'\sigma'}(x,x')}{\sigma(x,x') + \mathi \epsilon} + V_{\mu\nu\rho'\sigma'}(x,x') \ln\left[ \ell^{-2} \sigma(x,x') + \mathi \epsilon \right] \right]
\end{splitequation}
following from the tensor Hadamard expansion~\eqref{eq:tensor_hadamard} is sufficient. For the vector propagator in a general gauge, we then use the formulae~\eqref{eq:vector_propagator_general_gauge_ansatz} and~\eqref{eq:vector_propagator_general_gauge_ansatz_sol} and perform the Hadamard subtraction on each propagator separately, such that
\begin{equation}
\label{eq:vector_propagator_hadamard}
\normord{ G^\text{F}_{m^2,\alpha,\mu\rho'}(x,x) } = \normord{ G^\text{F}_{m^2,1,\mu\rho'}(x,x) } + \frac{1}{m^2} \nabla_\mu \nabla_{\rho'} \left[ \normord{ G^\text{F}_{m^2}(x,x) } - \normord{ G^\text{F}_{\alpha m^2}(x,x) } \right] \eqend{.}
\end{equation}
We follow the same procedure for the tensor propagator in a general gauge.

Let us start with the scalar case. To actually compute the limit~\eqref{eq:scalar_hadamard_normalordered}, we need to know the bitensors $U(x,x')$ and $V(x,x')$, at least up to some finite order in the separation of the points. For this, we use the known result~\cite{DeWitt:1960fc,Poisson:2011nh}
\begin{equation}
U(x,x') = \sqrt{\Delta(x,x')}
\end{equation}
with the van-Vleck--Morette determinant $\Delta$. In de~Sitter spacetime, we have the explicit expression~\cite{Allen:1985wd}
\begin{equation}
\Delta = \left( \frac{H \mu}{\sin(H \mu)} \right)^3 = \left( \frac{\arccos Z}{\sqrt{1-Z^2}} \right)^3 \eqend{,}
\end{equation}
where we used the relation $Z = \cos(H \mu)$. In a generic spacetime, the second biscalar $V(x,x')$ can be expanded near coincidence limit in a series whose coefficients are obtained by well-known recursion relations~\cite{Allen:1987bn,Decanini:2005gt}, which in principle would also be enough for the above limits. However, in de~Sitter spacetime, where we know the explicit expression for the propagator, we can even determine $V$ exactly. For this, we use a hypergeometric transformation~\cite[Eq.~15.8.10]{dlmf} to rewrite $F_\nu(Z)$~\eqref{eq:fnu_def} as
\begin{splitequation}
\label{eq:fnu_hadamard}
F_\nu(Z) &= \frac{H^2}{(4 \pi)^2} \frac{2}{1-Z} + \frac{m^2 - 2 H^2}{(4 \pi)^2} \hypergeom{2}{1}\left( \frac{3}{2} + \nu, \frac{3}{2} - \nu; 2; \frac{1-Z}{2} \right) \ln\left( \frac{1-Z}{2} \right) \\
&\quad- \frac{m^2 - 2 H^2}{(4 \pi)^2} \sum_{k=0}^\infty \frac{\Gamma\left( \frac{3}{2} + \nu + k \right) \Gamma\left( \frac{3}{2} - \nu + k \right)}{\Gamma\left( \frac{3}{2} + \nu \right) \Gamma\left( \frac{3}{2} - \nu \right) k! (k+1)!} \biggl[ \psi(k+1) + \psi(k+2) \\
&\hspace{10em}- \psi\left( \frac{3}{2} + \nu + k \right) - \psi\left( \frac{3}{2} - \nu + k \right) \biggr] \left( \frac{1-Z}{2} \right)^k \eqend{,}
\end{splitequation}
which exhibits explicitly the logarithmic singularity near coincidence $Z \approx 1$. Using that $1-Z = \frac{1}{4} H^2 \sigma + \bigo{\sigma^2}$, which follows from $Z = \cos(H \mu)$ and $\sigma = \frac{1}{2} \mu^2$ and comparing with the Hadamard expansion~\eqref{eq:scalar_hadamard}, we can thus identify $V$ (up to a constant factor) with the term in front of the logarithm, namely
\begin{equation}
\label{eq:hadamard_v_desitter}
V(x,x') = \frac{m^2 - 2 H^2}{2} \hypergeom{2}{1}\left( \frac{3}{2} + \nu, \frac{3}{2} - \nu; 2; \frac{1-Z}{2} \right) \eqend{.}
\end{equation}
Note that $V$ vanishes for a conformal scalar, which (in de~Sitter spacetime) is obtained by replacing $m^2 \to \xi R^2$ with $\xi = \frac{1}{6}$ and using that $R = 12 H^2$. This is in agreement with the fact that the two-point function is a pure power law for conformal fields.

Thus, it follows that the normal-ordered scalar propagator~\eqref{eq:scalar_hadamard_normalordered} is given by
\begin{splitequation}
\label{eq:scalar_hadamard_normalordered_inz}
\mathi \normord{ G^\text{F}_{m^2}(x,x') } &= \mathi G^\text{F}_{m^2}(x,x') - \frac{1}{8 \pi^2} \left[ \frac{U(x,x')}{\sigma(x,x') + \mathi \epsilon} + V(x,x') \ln\left[ \ell^{-2} \sigma(x,x') + \mathi \epsilon \right] \right] \\
&= F_\nu(Z - \mathi \epsilon) - \frac{1}{8 \pi^2} \left( \frac{\arccos Z}{\sqrt{1-Z^2}} \right)^\frac{3}{2} \frac{2 H^2}{\bigl[ \arccos Z \bigr]^2 + \mathi \epsilon} \\
&\quad- \frac{1}{8 \pi^2} \frac{m^2 - 2 H^2}{2} \hypergeom{2}{1}\left( \frac{3}{2} + \nu, \frac{3}{2} - \nu; 2; \frac{1-Z}{2} \right) \ln\left[ \frac{\bigl( \arccos Z \bigr)^2}{2 H^2 \ell^2} + \mathi \epsilon \right] \\
&= \frac{H^2}{8 \pi^2} \left[ \frac{1}{1-Z} - 2 (1-Z^2)^{-\frac{3}{4}} \bigl( \arccos Z \bigr)^{-\frac{1}{2}} \right] \\
&\quad+ \frac{m^2 - 2 H^2}{(4 \pi)^2} \hypergeom{2}{1}\left( \frac{3}{2} + \nu, \frac{3}{2} - \nu; 2; \frac{1-Z}{2} \right) \ln\left( \frac{H^2 \ell^2}{\bigl( \arccos Z \bigr)^2} (1-Z) \right) \\
&\quad- \frac{m^2 - 2 H^2}{(4 \pi)^2} \sum_{k=0}^\infty \frac{\Gamma\left( \frac{3}{2} + \nu + k \right) \Gamma\left( \frac{3}{2} - \nu + k \right)}{\Gamma\left( \frac{3}{2} + \nu \right) \Gamma\left( \frac{3}{2} - \nu \right) k! (k+1)!} \biggl[ \psi(k+1) + \psi(k+2) \\
&\hspace{10em}- \psi\left( \frac{3}{2} + \nu + k \right) - \psi\left( \frac{3}{2} - \nu + k \right) \biggr] \left( \frac{1-Z}{2} \right)^k \eqend{,}
\end{splitequation}
where we used the expression for the scalar propagator itself~\eqref{eq:scalar_propagator} in terms of $F_\nu$, the relation $\sigma = \frac{1}{2} H^{-2} \bigl( \arccos Z \bigr)^2$, and finally inserted the expression~\eqref{eq:fnu_hadamard} for $F_\nu$. Note that we could take the limit $\epsilon \to 0$ because we have subtracted all divergent terms. While this is not completely obvious from the expression~\eqref{eq:scalar_hadamard_normalordered_inz}, we can easily perform an expansion near coincidence $Z = 1$ and obtain
\begin{splitequation}
\label{eq:scalar_hadamard_normalordered_series}
\mathi \normord{ G^\text{F}_{m^2}(x,x') } &= \frac{4 H^2 - 3 m^2}{48 \pi^2} - \frac{174 H^4 - 220 H^2 m^2 + 75 m^4}{1920 \pi^2 H^2} (1-Z) \\
&\quad- \frac{4286 H^6 - 3969 H^4 m^2 + 105 H^2 m^4 + 525 m^6}{120960 \pi^2 H^4} (1-Z)^2 \\
&\quad+ \frac{m^2 - 2 H^2}{16 \pi^2} \left[ 1 + \frac{m^2}{4 H^2} (1-Z) + \frac{m^2 (m^2 + 4 H^2)}{48 H^4} (1-Z)^2 \right] L_\nu \\
&\quad+ \bigo{(1-Z)^3}
\end{splitequation}
with the combination
\begin{equation}
\label{eq:hadamard_lnu_def}
L_\nu \equiv \ln\left( \frac{H^2 \ell^2}{2} \right) + \psi\left( \frac{3}{2} - \nu \right) + \psi\left( \frac{3}{2} + \nu \right) + 2 \gamma \eqend{.}
\end{equation}
We remark that using the definition of $\nu$~\eqref{eq:param_nu_def}, it follows that for $m \to 0$ we have
\begin{equation}
\label{eq:hadamard_lnu_div}
L_\nu \sim - \frac{3 H^2}{m^2} + \bigo{m^0} \eqend{,}
\end{equation}
which reflects the well-known IR divergence of the massless and minimally coupled scalar field in de~Sitter spacetime. However, this only affects the leading term in~\eqref{eq:scalar_hadamard_normalordered_series}, and all derivatives have a finite coincidence limit even in the massless case.

The coincidence limit now follows by taking $Z=1$, and the normal-ordered scalar propagator \eqref{eq:scalar_hadamard_normalordered_series} remains finite, as it should. For the coincidence limit of covariant derivatives we need to use in addition that
\begin{equation}
\label{eq:hadamard_dz_limit}
\lim_{x' \to x} \nabla_\mu Z = 0 = \lim_{x' \to x} \nabla_{\rho'} Z \eqend{,} \quad \lim_{x' \to x} \nabla_\mu \nabla_{\rho'} Z = H^2 g_{\mu\rho'} \eqend{.}
\end{equation}
With these limits, it is easily seen that terms of order $(1-Z)^3$ do not make a contribution to the coincidence limit of up to four covariant derivatives acting on~\eqref{eq:scalar_hadamard_normalordered_series}, which is all that we need.

For the vector case, we follow the same steps, using in addition that
\begin{equation}
U_{\mu\rho'}(x,x') = \sqrt{\Delta(x,x')} \, g_{\mu\rho'}
\end{equation}
and that
\begin{equation}
g_{\mu\rho'} = H^{-2} \left( \nabla_\mu \nabla_{\rho'} Z - \frac{\nabla_\mu Z \nabla_{\rho'} Z}{1+Z} \right) \eqend{.}
\end{equation}
We refer to~\cite{DFF2024b} for the details of the computation, and only give the end result
\begin{splitequation}
\label{eq:vector_hadamard_normalordered_series}
\mathi \normord{ G^\text{F}_{m^2,1,\mu\rho'}(x,x') } &= \frac{(m^2 + 2 H^2)}{48 H^2 \pi^2} L_\rho \left[ 1 + \frac{5 m^2 + 12 H^2}{18 H^2} (1-Z) \right] \nabla_\mu \nabla_{\rho'} Z \\
&\quad+ \frac{(m^2 - 2 H^2)}{64 H^2 \pi^2} \left( L_\nu + \frac{3 H^2}{m^2} \right) \left[ 1 + \frac{m^2 + 4 H^2}{6 H^2} (1-Z) \right] \nabla_\mu \nabla_{\rho'} Z \\
&\quad- \frac{1}{32 \pi^2} \biggl[ \frac{11 H^2 + 12 m^2}{6 H^2} + \frac{242 H^4 + 615 H^2 m^2 + 225 m^4}{180 H^4} (1-Z) \\
&\qquad\qquad+ \bigo{ (1-Z)^2 } \biggr] \nabla_\mu \nabla_{\rho'} Z \\
&\quad- \frac{(m^2 + 2 H^2)}{32 \pi^2} L_\rho \left[ \frac{m^2 - 12 H^2}{12 H^4} + \bigo{ 1-Z } \right] \nabla_\mu Z \nabla_{\rho'} Z \\
&\quad- \frac{(m^2 - 2 H^2)}{32 \pi^2} \left( L_\nu + \frac{3 H^2}{m^2} \right) \left[ \frac{m^2 + 4 H^2}{12 H^4} + \bigo{ 1-Z } \right] \nabla_\mu \nabla_{\rho'} Z \\
&\quad+ \frac{1}{16 \pi^2} \left[ \frac{28 H^2 + 39 m^2}{72 H^2} + \bigo{ 1-Z } \right] \nabla_\mu Z \nabla_{\rho'} Z \eqend{,}
\end{splitequation}
where $L_\nu$ was defined in~\eqref{eq:hadamard_lnu_def}. Again, using the limits~\eqref{eq:hadamard_dz_limit} it is easily seen that this expression is all we need for the coincidence limit of up to two covariant derivatives acting on it. Moreover, using the small-mass expansion of $L_\nu$~\eqref{eq:hadamard_lnu_div}, we see that the vector propagator has a finite coincidence limit even in the massless limit (using also that $L_\rho$ remains finite).\footnote{To avoid confusion, we note that the massless limit is discontinuous, and Ward identities that connect the vector and scalar propagators force the strictly massless vector field to also have an IR divergence in most covariant gauges~\cite{Glavan:2022dwb}. Whether this influences physical observables or can be circumvented by a construction analogous to~\cite{Faizal:2008ns,Gibbons:2014zya} remains to be seen.}

Lastly, for the tensor case in the Feynman gauge, the same procedure is straightforward but lengthy, using that for the traceless spin-2 propagator we have
\begin{equation}
U_{\mu\nu\rho'\sigma'}(x,x') = \sqrt{\Delta(x,x')} \left( g_{\mu(\rho'} g_{\sigma')\nu} - \frac{1}{4} g_{\mu\nu} g_{\rho'\sigma'} \right) \eqend{.}
\end{equation}
Since in this case we do not need to consider any derivatives acting on the Hadamard-subtracted propagator, we can take directly the coincidence limit and obtain
\begin{splitequation}
\label{eq:tensor_hadamard_normalordered_series}
\mathi \lim_{x' \to x} \normord{ G^{m^2}_{\mu\nu\rho'\sigma'}(x,x') } &= \Bigg[ \frac{5 (4 H^2 + m^2)}{144 \pi^2} L_\nu - \frac{4 H^2 - m^2}{48 \pi^2} L_\sigma - \frac{8 H^2 - m^2}{144 \pi^2} L_\tau \\
&\quad+ \frac{184 H^6 - 128 H^4 m^2 + 34 H^2 m^4 - 3 m^6}{48 \pi^2 (2 H^2 - m^2) (6 H^2 - m^2)} \Bigg] \left[ g_{\mu(\rho'} g_{\sigma')\nu} - \frac{1}{4} g_{\mu\nu} g_{\rho'\sigma'} \right] \eqend{.}
\end{splitequation}
For details of the computation, we refer again to~\cite{DFF2024b}. Since also $L_\sigma$ is divergent in the massless limit, namely we have
\begin{equation}
L_\sigma \sim - \frac{5 H^2}{m^2} + \bigo{m^0} \eqend{,}
\end{equation}
we see that in the limit $m \to 0$ also the coincidence value of the normal-ordered tensor propagator~\eqref{eq:tensor_hadamard_normalordered_series} remains finite due to cancellation with the divergent part of $L_\nu$.\footnote{As for the vector, this limit may differ from the strictly massless case depending on the gauge. However, a comprehensive study including all relevant Ward identities has not been done yet to our knowledge.}

\subsection{Coincidence limit to order $H^2$}
\label{sec:coincidence_h2}

Since we consider the flow in the Einstein--Hilbert truncation, we only need the coincidence values to order $R = 12 H^2$. This is a regular limit because the coincidence values of the biscalars $U(x,x')$ and $V(x,x')$ and their covariant derivatives (as well as the ones of $U_{\mu\rho'}$ and $V_{\mu\rho'}$ for vectors and $U_{\mu\nu\rho'\sigma'}$ and $V_{\mu\nu\rho'\sigma'}$ for tensors) that we subtracted from the propagator are polynomials in the curvature tensors and their derivatives.

For the coincidence limit of the scalar propagator~\eqref{eq:scalar_hadamard_normalordered_series} and its covariant derivatives, a straightforward computation yields
\begin{equations}
&\mathi \normord{ G^\text{F}_{m^2}(x,x) } = \frac{m^2 - 2 H^2}{16 \pi^2} \left[ 2 \gamma + \ln\left( \frac{m^2 \ell^2}{2} \right) - 1 \right] - \frac{H^2}{8 \pi^2} + \bigo{H^4} \eqend{,} \\
\begin{split}
&\lim_{x' \to x} \left[ \mathi \nabla_\mu \nabla_\nu \normord{ G^\text{F}_{m^2}(x,x') } \right] \\
&\quad= \frac{m^2}{64 \pi^2} \left\{ (m^2 - 2 H^2) \left[ 2 \gamma + \ln\left( \frac{m^2 \ell^2}{2} \right) - \frac{5}{2} \right] + H^2 \right\} g_{\mu\nu} + \bigo{H^4} \eqend{,}
\end{split} \\
\begin{split}
&\lim_{x' \to x} \left[ \mathi \nabla_{\rho'} \nabla_{\sigma'} \normord{ G^\text{F}_{m^2}(x,x') } \right] \\
&\quad= \frac{m^2}{64 \pi^2} \left\{ (m^2 - 2 H^2) \left[ 2 \gamma + \ln\left( \frac{m^2 \ell^2}{2} \right) - \frac{5}{2} \right] + H^2 \right\} g_{\rho'\sigma'} + \bigo{H^4} \eqend{,}
\end{split} \\
\begin{split}
&\lim_{x' \to x} \left[ \mathi \nabla_\mu \nabla_\nu \nabla_{\rho'} \nabla_{\sigma'} \normord{ G^\text{F}_{m^2}(x,x') } \right] \\
&\quad= \frac{m^4}{192 \pi^2} \left\{ (m^2 + 2 H^2) \left[ 2 \gamma + \ln\left( \frac{m^2 \ell^2}{2} \right) - \frac{10}{3} \right] + \frac{14 H^2}{3} \right\} g_{\mu(\rho'} g_{\sigma')\nu} \\
&\qquad- \frac{m^4}{384 \pi^2} \left\{ (m^2 - 4 H^2) \left[ 2 \gamma + \ln\left( \frac{m^2 \ell^2}{2} \right) - \frac{10}{3} \right] - \frac{H^2}{3} \right\} g_{\mu\nu} g_{\rho'\sigma'} + \bigo{H^4} \eqend{,}
\end{split}
\end{equations}
where we used that for $L_\nu$~\eqref{eq:hadamard_lnu_def} we have the expansion
\begin{equation}
L_\nu = 2 \gamma + \ln\left( \frac{m^2 \ell^2}{2} \right) - \frac{4 H^2}{3 m^2} + \bigo{H^4} \eqend{.}
\end{equation}
For the coincidence limit of the Feynman-gauge vector propagator~\eqref{eq:vector_hadamard_normalordered_series} and its covariant derivatives, we obtain analogously
\begin{equation}
\mathi \normord{ G^\text{F}_{m^2,1,\mu\rho'}(x,x) } = \frac{1}{16 \pi^2} \left\{ (m^2 + H^2) \left[ 2 \gamma + \ln\left( \frac{m^2 \ell^2}{2} \right) - 1 \right] + H^2 \right\} g_{\mu\rho'} + \bigo{H^4}
\end{equation}
and
\begin{splitequation}
&\lim_{x' \to x} \left[ \mathi \nabla_{(\rho'} \nabla_{|(\mu} \normord{ G^\text{F}_{m^2,1,\nu)|\sigma')}(x,x') } \right] \\
&\quad= - \frac{m^2}{64 \pi^2} \left\{ (m^2 + 6 H^2) \left[ 2 \gamma + \ln\left( \frac{m^2 \ell^2}{2} \right) - \frac{5}{2} \right] + 7 H^2 \right\} g_{\mu(\rho'} g_{\sigma')\nu} \\
&\qquad+ \frac{m^2 H^2}{32 \pi^2} \left[ 2 \gamma + \ln\left( \frac{m^2 \ell^2}{2} \right) - 1 \right] g_{\mu\nu} g_{\rho'\sigma'} + \bigo{H^4} \eqend{,}
\end{splitequation}
where we also used that
\begin{equation}
L_\rho = 2 \gamma + \ln\left( \frac{m^2 \ell^2}{2} \right) + \frac{2 H^2}{3 m^2} + \bigo{H^4} \eqend{.}
\end{equation}
Lastly, for the coincidence limit of the tensor propagator in the Feynman gauge~\eqref{eq:tensor_hadamard_normalordered_series}, we obtain
\begin{equation}
\mathi \normord{ G^{m^2}_{\mu\nu\rho'\sigma'}(x,x) } = \frac{m^2}{16 \pi^2} \left[ 2 \gamma + \ln\left( \frac{m^2 \ell^2}{2} \right) - 1 \right] \left[ g_{\mu(\rho'} g_{\sigma')\nu} - \frac{1}{4} g_{\mu\nu} g_{\rho'\sigma'} \right] + \bigo{H^4} \eqend{,}
\end{equation}
using also that
\begin{equations}
L_\sigma &= 2 \gamma + \ln\left( \frac{m^2 \ell^2}{2} \right) - \frac{16 H^2}{3 m^2} + \bigo{H^4} \eqend{,} \\
L_\tau &= 2 \gamma + \ln\left( \frac{m^2 \ell^2}{2} \right) - \frac{22 H^2}{3 m^2} + \bigo{H^4} \eqend{.}
\end{equations}

We can now combine these results to obtain the coincidence limit of the graviton and ghost propagators in a general gauge. In general, the expressions are very complicated, and only for special choices of gauge we obtain a presentable expression, which we display in the later sections.

\section{The flow in de Sitter space}
\label{sec:FRGE-dS}

\subsection{Gauge and parameter dependence}
\label{sec:gauge dependence}

It is well known that the off-shell effective action in quantum gravity typically depends on both the gauge choice and the field parametrization~\cite{Benedetti:2011ct,Gies:2015tca,Falls:2017cze,Martini:2023qkp,Falls:2024noj}. However, these unphysical dependencies vanish when going on-shell, which has been explicitly shown in scalar-tensor theories formulated in both the Jordan and Einstein frames~\cite{Falls:2017cze,Falls:2018olk}. Instead, when working off-shell, as in the formalism we are using here, the cancellations of the unphysical degrees of freedom are not guaranteed. In particular for what concerns the FRGE, in a general parametrization and for a generic gauge fixing, terms proportional to the equation of motion appear in the propagators, and the cancellations will not take place~\cite{Baldazzi:2021ydj,Baldazzi:2021orb,Falls:2024noj}. We note that many computations in quantum gravity have been (and continue to be) performed within the class of harmonic gauges with $\zeta = 1$, often specialized to the Feynman gauge $\zeta = \xi = 1$. This choice is favored because it offers several technical simplifications, such as the straightforward diagonalization of scalar modes. Another popular gauge is the Landau limit $\xi \to 0$, which enforces the gauge-fixing condition exactly, even inside time-ordered correlation functions. However, this limit only exists for the correlation functions and not (for example) for the differential operator $P_{\xi,\zeta}^{\rho\sigma\mu\nu}$ 
\eqref{eq:p_xizeta_def}, so one has to be careful when taking it.

In our work, we will deal with the dependence on the gauge fixing in a pragmatic manner, analyzing the sensitivity and stability of the UV behavior of quantum gravity with respect to variations of the gauge fixing parameters. We will focus on the question of the existence and the properties of a non-Gaussian UV fixed point, which implies that quantum gravity can be asymptotically safe. Furthermore, we will analyze the gauge-dependence of the critical exponents at the UV fixed points, which furnish information on the universality class of the theory. In particular, we study in detail the dependence of the flow on the gauge parameter $\xi$ for $\zeta = 1$ (the class of harmonic gauges) and $\zeta = \frac{1}{2}$. We also pay special attention to avoid the \emph{exceptional gauges}, first introduced in Refs.~\cite{Allen:1985ux, Allen:1985wd}, which harbor unphysical IR divergences. We refer to these references for details, and only note that for any given $\xi$ there exists a series of values for $\zeta$ for which the two-point-function turns out to be divergent in the coincidence limit.

Furthermore, as explained in Sec.~\ref{sec:differences}, and as we have seen explicitly in Sec.~\ref{sec:propagators}, the removal of the UV divergences using Hadamard subtraction introduces a new arbitrary length scale $\ell$ in the coincidence limit of the interacting propagator. This scale encodes the ambiguity of finite renormalizations, and can be fixed at some convenient value. Here, we study the dependence of the RG flow on this scale parameter as well. More precisely, we study the dependence of the UV fixed point and of its critical exponent on the dimensionless \emph{Hadamard parameter} $\alpha \coloneqq \ell k$.

Finally, we briefly remark that the right-hand side of the flow equation~\eqref{eq:FRGE} diverges in the Landau gauge limit $\xi \to 0$ due to our choice~\eqref{eq:regulator_choice} of the regulator $q^{AB}$. The choice we made is the same as the one usually taken in the Euclidean setting, which amounts to the replacement $\nabla^2 \to \nabla^2 - k^2$. However, in Landau-type gauges, this is not a sensible choice since the differential operator~\eqref{eq:p_xizeta_def} is divergent in this limit. We leave the analysis of the RG flow in the Landau gauge for future works.

\subsection{Effective average action, regulator, and physical masses}
\label{sec:physical masses}

We can now specialize the general framework of Sec.~\ref{sec:FRGE} to de~Sitter space, choose the specific form of regulator function alluded to above, and write the FRGE using the coincidence limit of propagators computed in Sec.~\ref{sec:propagators}.

As gravitational action $S_\text{grav}$, we take the Einstein--Hilbert action
\begin{equation}
\label{eq:action_eh}
S_\text{grav}(\g + h) = \frac{1}{16 \pi G_\text{N}} \int_\mathcal{M} (R - 2 \Lambda) \eqend{,}
\end{equation}
where $G_\text{N}$ is Newton's constant, $\Lambda$ is the cosmological constant, and all quantities depend on the full metric $\gamma = \g + \h$. The effective average action in the Einstein--Hilbert truncation then takes the form
\begin{equation}
\Gamma_k(\g, h, c, \bar c, \sigma, \eta) = \Gamma_k^\text{EH}(\g + h) + \Gamma_k^\text{gf} + \Gamma_k^\text{gh} + Q_k + \Sigma + H \eqend{,}
\end{equation}
where $Q_k$, $\Sigma$ and $H$ are defined in Eqs.~\eqref{eq:sources} and~\eqref{eq:regulator}, the Einstein--Hilbert term is
\begin{equation}
\label{eq:gamma_k_einsteinhilbert}
\Gamma_k^\text{EH}(\g + h) = \frac{1}{16 \pi G_{\text{N},k}} \int_\mathcal{M} \Bigl[ R(\g + h) - 2 \Lambda_k \Bigr] \eqend{,}
\end{equation}
$G_{\text{N},k}$ is the running Newton constant and $\Lambda_k$ is the running cosmological constant. That is, it is identical to the gravitational action~\eqref{eq:action_eh} with Newton's constant and the cosmological constant replaced by their running equivalents and the quantum fluctuation $\h$ replaced by the average fluctuation $h$. In the same way, the gauge-fixing and ghost terms are given by~\eqref{eq:action_gaugefixing} and~\eqref{eq:action_ghost} with $F^\nu = \bar{\nabla}_\mu h^{\mu\nu} - \frac{1}{2 \zeta} \bar{\nabla}^\nu h_\mu^\mu$, Newton's constant replaced by $G_{\text{N},k}$, and the quantum fields replaced by the average fields. Concretely, the ghost term takes the form
\begin{equation}
\label{eq:gammak_ghost}
\Gamma_k^\text{gh} = \frac{1}{32 \pi G_{\text{N},k}} \int_\mathcal{M} \bar{c}_\mu \left( \g^{\mu\nu} \nabla^2 + \bar{R}^{\mu\nu} + \frac{\zeta-1}{\zeta} \nabla^\mu \nabla^\nu \right) c_\nu \eqend{.}
\end{equation}
Since we only need second functional derivatives to determine the scale-dependent propagator in the flow equation~\eqref{eq:FRGE}, it is enough to expand $\Gamma_k^\text{EH}$ to second order in $h_{\mu\nu}$. Recalling the regulator term in Eq.~\eqref{eq:regulator}, the parts quadratic in $h_{\mu\nu}$ read
\begin{equation}
\label{eq:gammak_second_derivative}
\Gamma_k^{(2)} = \frac{1}{32 \pi G_{\text{N},k}} \int_\mathcal{M} h_{\rho\sigma} \left( \bar{P}_{\xi,\zeta}^{\rho\sigma\mu\nu} - 16 \pi G_{\text{N},k} q_k^{\rho\sigma\mu\nu} \right) h_{\mu\nu} \eqend{,}
\end{equation}
where the differential operator $\bar{P}$ is given by~\eqref{eq:p_xizeta_def} evaluated on the de~Sitter background $g = \g$ (such that $\bar{R}_{\mu\nu} = 3 H^2 \g_{\mu\nu}$) and with the choice $m^2 = M^2 = 2 ( 3 H^2 - \Lambda_k )$.

We now choose the regulator functions in such a way that they act as artificial masses for the fields $h$, $\bar{c}$ and $c$, dressing the d'Alembertian operator $\nabla^2 \to \nabla^2 - k^2$. We see from the expression~\eqref{eq:p_xizeta_def} for $P_{\xi,\zeta}$ that this entails the choice
\begin{equation}
\label{eq:regulator_choice}
q_k^{\rho\sigma\mu\nu} = \frac{1}{32 \pi G_{\text{N},k}} \left[ \g^{\rho(\mu} \g^{\nu)\sigma} - \frac{1}{2} \left( 2 - \frac{1}{\xi \zeta^2} \right) \g^{\rho\sigma} \g^{\mu\nu} \right] k^2 \eqend{.}
\end{equation}
It follows that the combination $\bar{P}_{\xi,\zeta}^{\rho\sigma\mu\nu} - 16 \pi G_{\text{N},k} q_k^{\rho\sigma\mu\nu}$ appearing in the quadratic part~\eqref{eq:gammak_second_derivative} of $\Gamma_k$ is again given by~\eqref{eq:p_xizeta_def} evaluated on the background $g = \g$, but now with the choice
\begin{equations}[eq:mass_k]
m^2 &= k^2 + 2 ( 3 H^2 - \Lambda_k ) \eqend{,} \\
M^2 &= k^2 \left( 3 - \frac{2}{\xi \zeta^2} \right) + 2 ( 3 H^2 - \Lambda_k ) \eqend{.}
\end{equations}
The scale-dependent graviton propagator is thus given by the tensor propagator $G^\text{F}_{m^2,M^2,\zeta,\xi}$ determined in Sec.~\ref{sec:propagators} with the choice~\eqref{eq:mass_k} of masses.

For the ghost sector, we see that the coefficient of the minimal kinetic operator $\nabla^2$ does not depend on the gauge parameter, such that we can choose
\begin{equation}
q_k^{\mu\nu} = g^{\mu\nu} k^2 \eqend{.}
\end{equation}
Comparing the equation of motion for the ghost derived from $\Gamma_k^\text{gh}$~\eqref{eq:gammak_ghost} with the vector one~\eqref{eq:eom_vector}, we see that the scale-dependent ghost propagator is (minus) the vector propagator $G^\text{F}_{m^2,\alpha}$~\eqref{eq:vector_propagator_general_gauge} with mass $m^2 = k^2$ and gauge parameter $\alpha = \zeta/(2\zeta-1)$.

Finally, we compute the coincidence limit of the Hadamard normal-ordered propagators $\normord{ G_k(x,x) }$ using the expressions from Sec.~\ref{sec:coincidence_h2}, and substitute them in the FRGE~\eqref{eq:FRGE}. Since we restrict to the Einstein--Hilbert truncation~\eqref{eq:gamma_k_einsteinhilbert} for the effective average action $\Gamma_k$, it is enough to work to order $\bar{R} = 12 H^2$. Comparing the terms of orders $H^0$ and $H^2$ on both sides of the FRGE (which corresponds to deriving it with respect to the volume form and the Ricci scalar), we obtain the $\beta$ functions of the dimensionless Newton and cosmological constants
\begin{equation}
\label{eq:dimless_constants}
g_k = G_{\text{N},k} k^2 \eqend{,} \quad \lambda_k = \frac{\Lambda_k}{k^2} \eqend{,}
\end{equation}
which then can be solved for the flow.

\subsection{The flow for \texorpdfstring{$\zeta = \frac{1}{2}$}{\textzeta = 1/2}}
\label{sec:zeta=1/2}

We first consider the flow fixing $\zeta = \frac{1}{2}$ and $\xi = 1$, but studying different values of the Hadamard scale $\alpha$. In this case, the coincidence limit of the graviton propagator~\eqref{eq:tensor_propagator_zeta12} that enters the right-hand side of the flow equation~\eqref{eq:FRGE} reads
\begin{splitequation}
&\mathi q_k^{\rho\sigma\mu\nu} \normord{ G^\text{F}_{6 H^2 + k^2 (1-2\lambda_k), 6 H^2 - k^2 (5+2\lambda_k), \frac{1}{2},1,\mu\nu\rho\sigma}(x,x) } \\
&= \frac{k^4 (1-2\lambda_k)}{256 \pi^2 (2-\lambda_k)^3} \Biggl[ (1-2\lambda_k) (212 + 31 \lambda_k + 26 \lambda_k^2) + 2 (5 + 2 \lambda_k) (5 - 5 \lambda_k + 8 \lambda_k^2) \ln\left( \frac{5+2\lambda_k}{4 (2-\lambda_k)} \right) \\
&\hspace{4em}+ 6 (179 - 261 \lambda_k + 138 \lambda_k^2 - 16 \lambda_k^3) \left[ \ln\left( \frac{1-2\lambda_k}{2 \alpha^2} \right) + 2 \gamma - 1 \right] \Biggr] \\
&\quad+ \frac{H^2 k^2}{128 \pi^2 (2-\lambda k)^4} \Biggl[ 5974 - 10838 \lambda_k + 8781 \lambda_k^2 - 4652 \lambda_k^3 + 292 \lambda_k^4 \\
&\hspace{4em}+ 3 (2311 - 4622 \lambda_k + 3336 \lambda_k^2 - 944 \lambda_k^3 + 208 \lambda k^4) \left[ \ln\left( \frac{1-2\lambda_k}{2 \alpha^2} \right) + 2 \gamma - 1 \right] \\
&\hspace{4em}+ (-235 + 470 \lambda_k - 744 \lambda_k^2 + 752 \lambda_k^3 + 176 \lambda_k^4) \ln\left( \frac{5+2\lambda_k}{4 (2-\lambda_k)} \right) \Biggr] + \bigo{H^4} \eqend{,}
\end{splitequation}
while the coincidence limit of the ghost propagator~\eqref{eq:vector_propagator_general_gauge} reads
\begin{splitequation}
&\mathi q_k^{\rho\mu} \normord{ - G^\text{F}_{k^2,\zeta/(2\zeta-1),\mu\rho}(x,x) } \\
&= \frac{3 k^4}{32 \pi^2} \left[ 2 \ln\left( 2 \alpha^2 \right) - 4 \gamma + 1 \right] + \frac{3 H^2 k^2}{8 \pi^2} \left[ \ln\left( 2 \alpha^2 \right) - 2 \gamma + 1 \right] \eqend{.}
\end{splitequation}
Substituting these expressions into the FRGE~\eqref{eq:FRGE}, we obtain the explicit expressions for the $\beta$ functions. We see that they are long and unwieldy, and become even more so in a general gauge, such that we do not report them explicitly. We note that while the $\beta$ functions in the Einstein--Hilbert truncation can be written down analytically in any gauge, as in the Euclidean case their solution for a general gauge unfortunately can only be found numerically.

\begin{figure}[h]
    \centering
    \includegraphics[width=0.33 \linewidth]{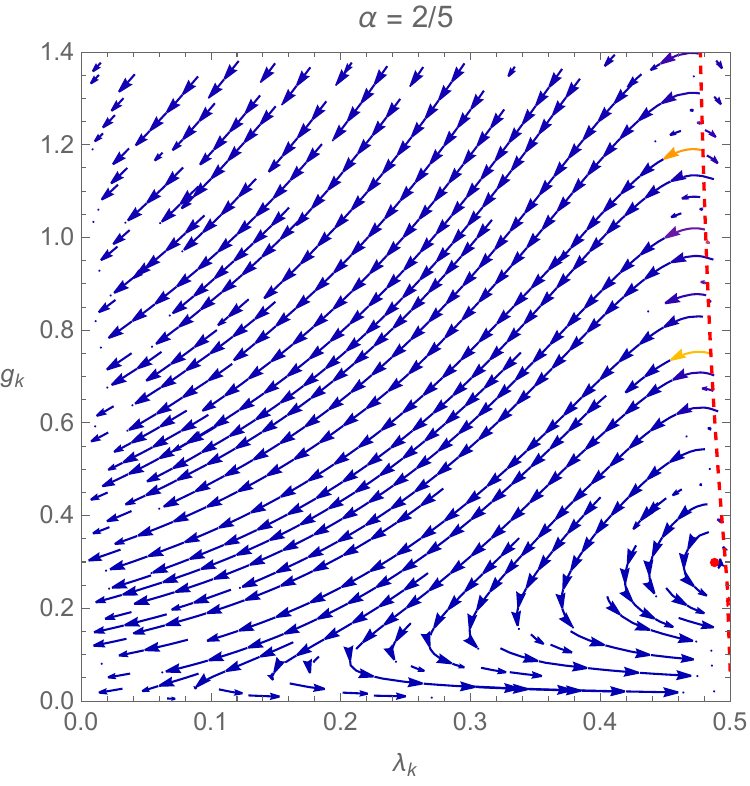}
    \includegraphics[width=0.32 \linewidth]{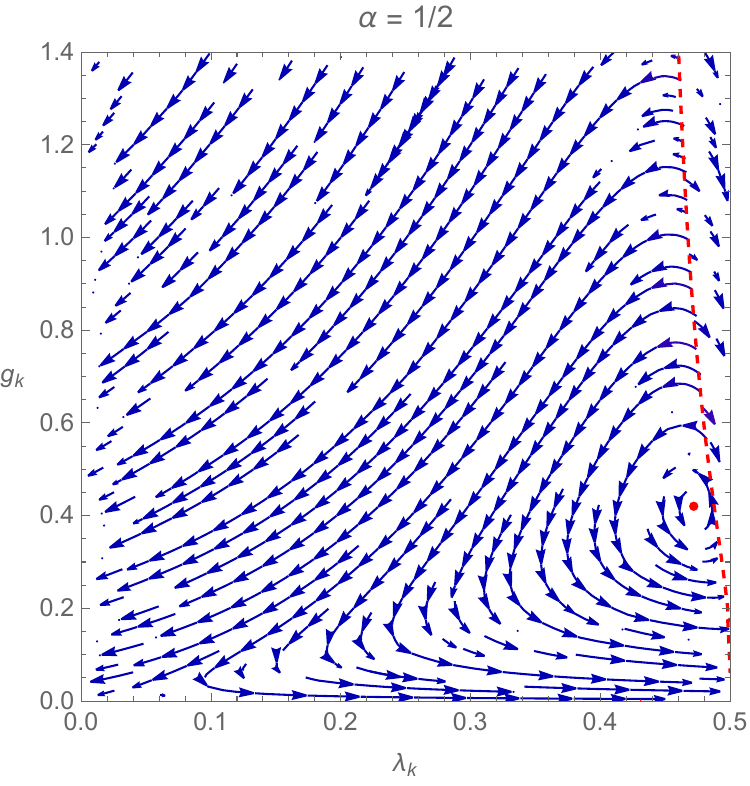}
    \includegraphics[width=0.32 \linewidth]{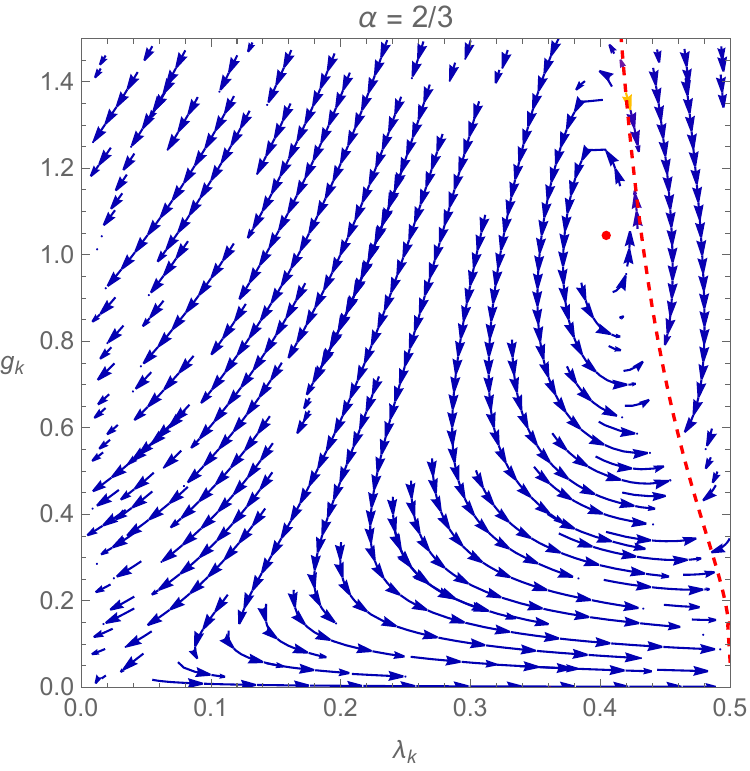}
    \caption{Flow diagram with $\zeta = \frac{1}{2}$ and $\xi = 1$ for different values of $\alpha$.} 
    \label{fig:zeta12-Feynman}
\end{figure}

\begin{figure}[h]
    \centering
    \includegraphics[scale=0.3]{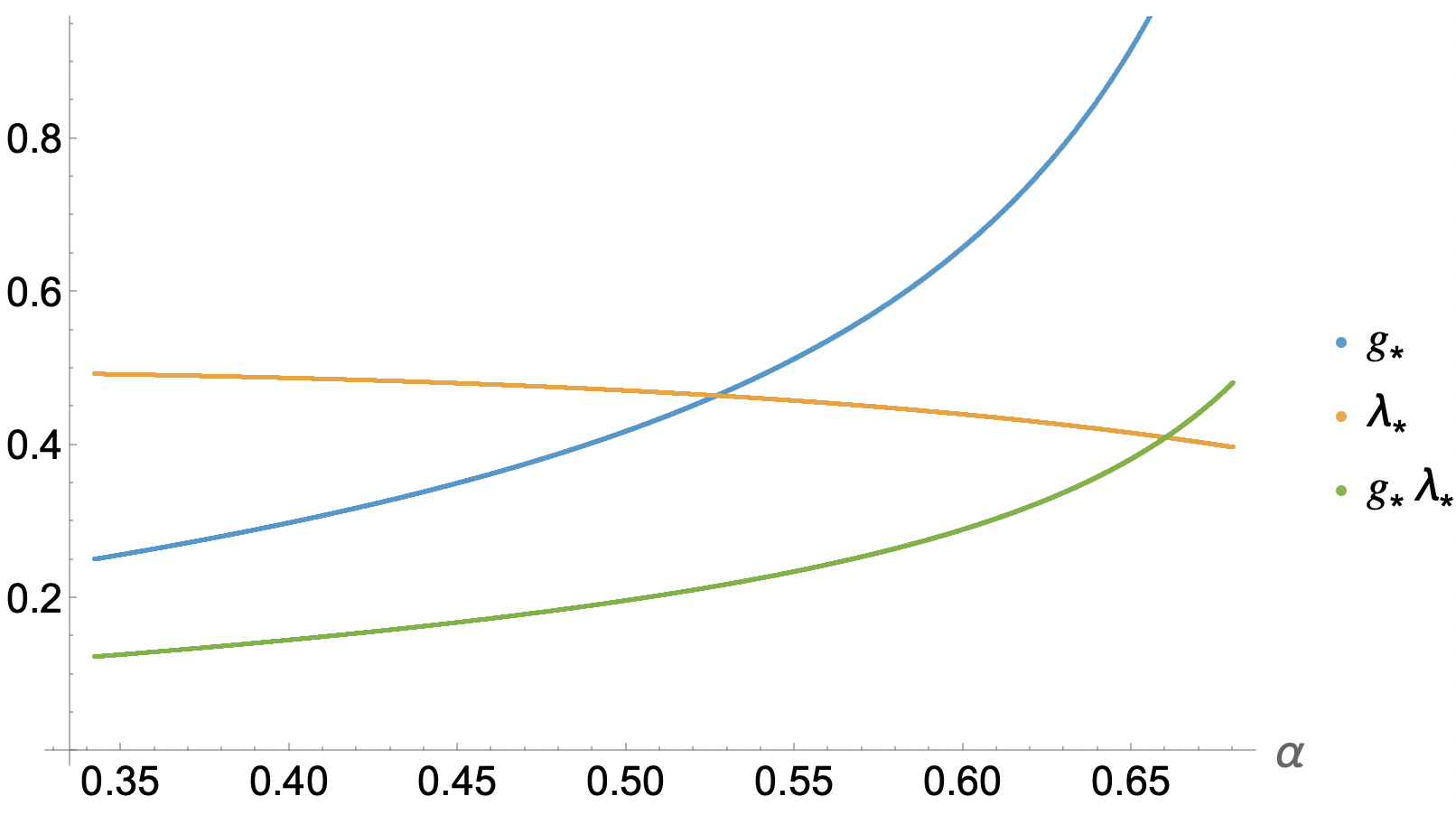}
    \caption{UV fixed points with $\zeta = \frac{1}{2}$ and $\xi = 1$ for different values of $\alpha$.} 
    \label{fig:fp-zeta12-Feynman-UV}
\end{figure}

The results are shown in Fig.~\ref{fig:zeta12-Feynman}, and we see that generically a UV fixed point $(g_k^*, \lambda_k*)$ exists.
However, the position of the UV fixed point depends quite strongly on $\alpha$. We can determine this dependence numerically by converting the fixed point equation $\beta_i(g_k^*,\lambda_k^*) = 0$ into a system of PDEs, taking the derivative with respect to $\alpha$, and then numerically integrating the PDEs with initial values given by a fixed point at a small value of $\alpha$. We consider in particular the dependence on $\alpha$ in the gauge $\xi = 1$, which is one of the most common in the literature. The results are shown in Fig.~\ref{fig:fp-zeta12-Feynman-UV}.
We see that the existence of a non-trivial UV fixed point is compatible only with a small range of values of the Hadamard parameter $\alpha$. As $\alpha$ tends to $0$, the UV fixed point approaches the vertical asymptote $\lambda_k = \frac{1}{2}$, which can also be seen in the flow diagram Fig.~\ref{fig:zeta12-Feynman}. This asymptote corresponds to a divergence of the flow, which arises from the vanishing of the argument of the logarithms present in the coincidence limit of the propagators. In the FRGE context, this is the manifestation of the Higuchi bound~\cite{Higuchi:1986py,Higuchi:1989gz}, which states that a massive spin-2 field with mass below a certain multiple of the cosmological constant must have negative-norm states, i.e., that the theory is not unitary. In our case, since we are working off-shell and with the scale-dependent mass~\eqref{eq:mass_k}, the Higuchi bound translates into the statement that 
\begin{equation}
\label{eq:higuchi-bound}
m^2 = k^2 (1-2\lambda_k) + 6 H^2 \geq 2 H^2 \eqend{,}
\end{equation}
which for large RG scale $k \gg H$ gives the condition $\lambda_k \leq \frac{1}{2}$.

The numerical evaluation of the flow breaks down at $\alpha \approx 0.77$, where the anomalous dimension $\eta = \partial_k \ln g_k - 2$ diverges. The critical exponents $\theta$ of the UV fixed point in the gauge $\zeta = \frac{1}{2}$, $\xi = 1$ are complex conjugates for the entire range of admissible $\alpha$, with the real part monotonically increasing in $\alpha$ from $\Re \theta \approx 1.5$ to $\Re \theta \approx 3$, see Fig. \ref{fig:fp-zeta12-Feynman-UV}.
\begin{figure}[h]
    \centering
    \includegraphics[scale=0.45]{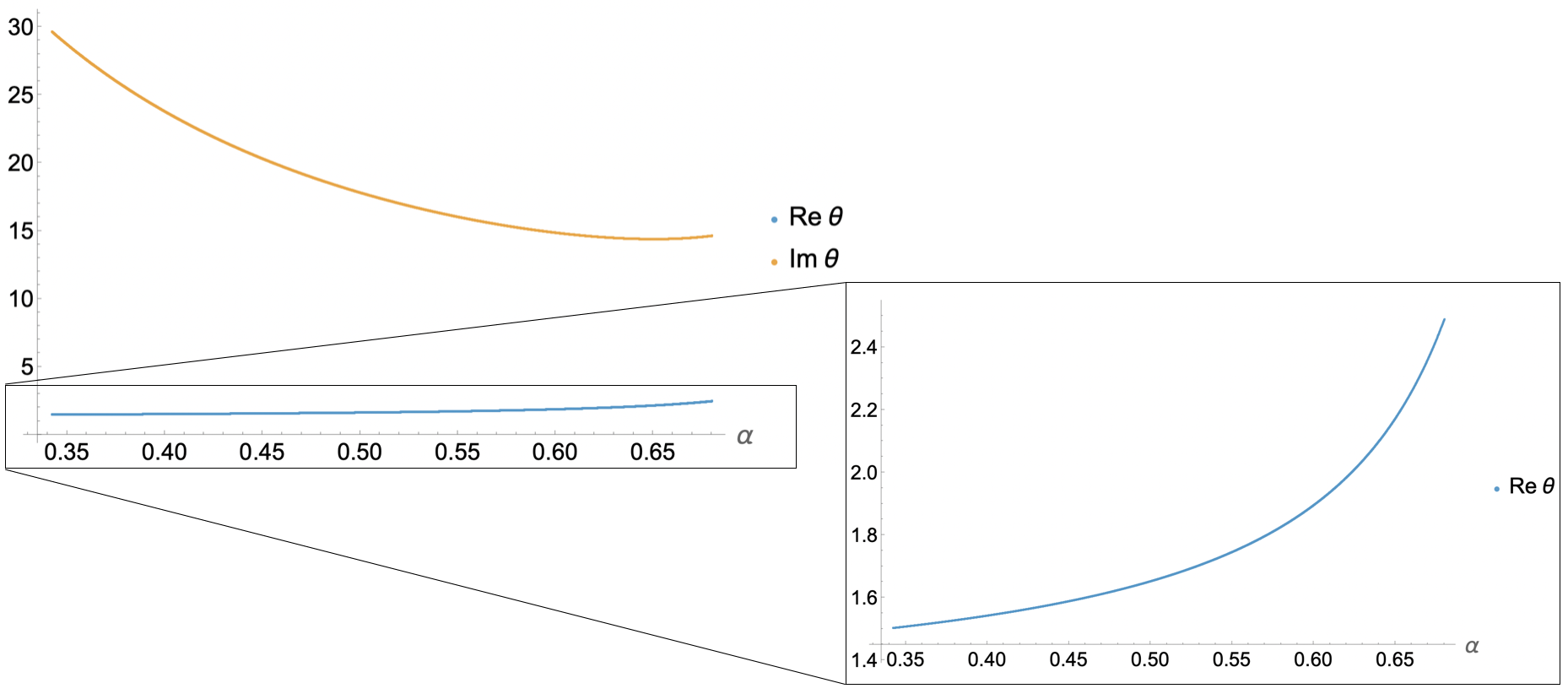}
    \caption{Critical exponents with $\zeta = \frac{1}{2}$ and $\xi = 1$ for different values of $\alpha$.}
    \label{fig:fp-zeta12-Feynman-UV-critical}
\end{figure}

\begin{figure}[h]
    \centering
    \includegraphics[width=0.32 \linewidth]{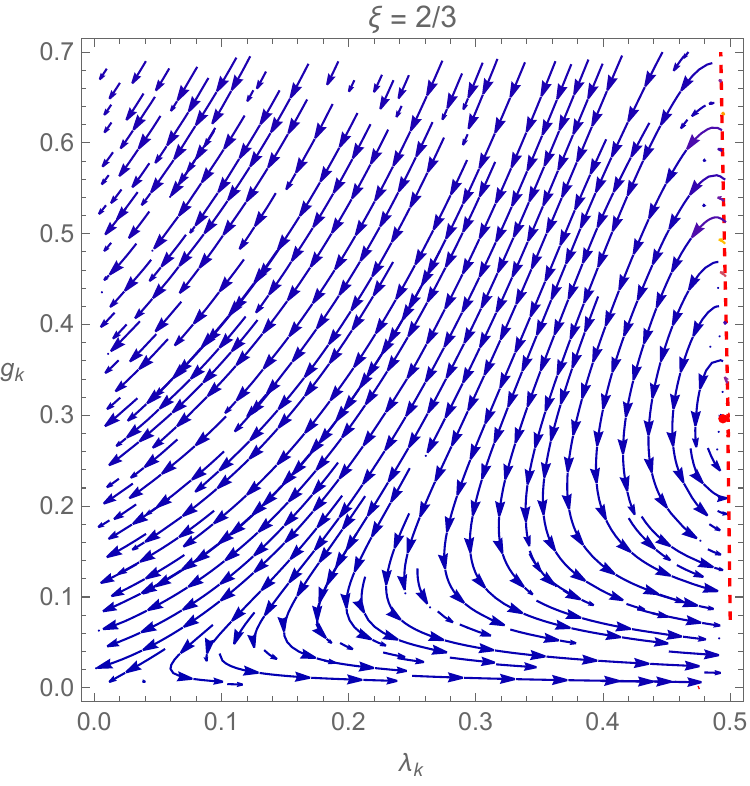}
    \includegraphics[width=0.32 \linewidth]{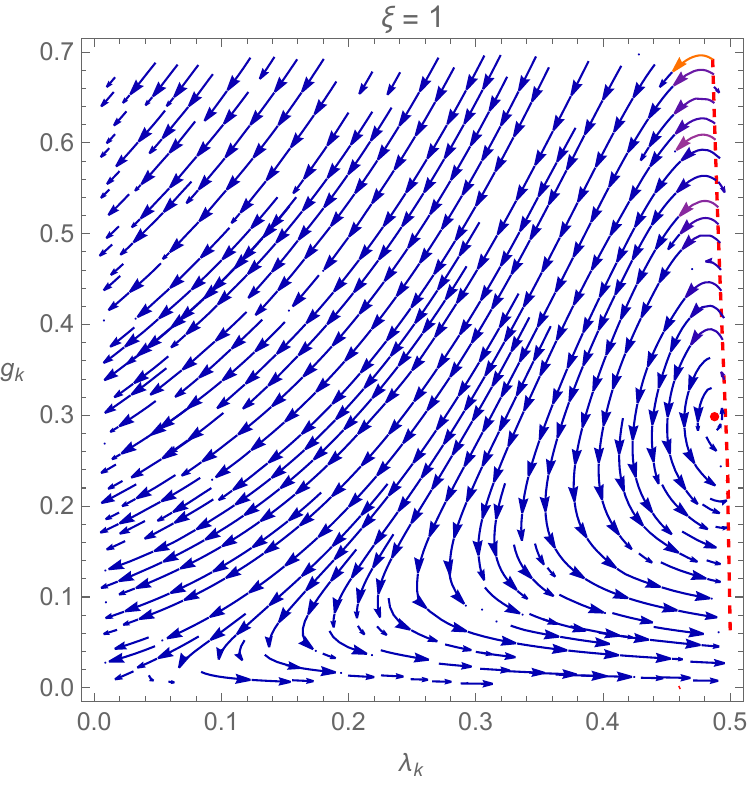}
    \includegraphics[width=0.32 \linewidth]{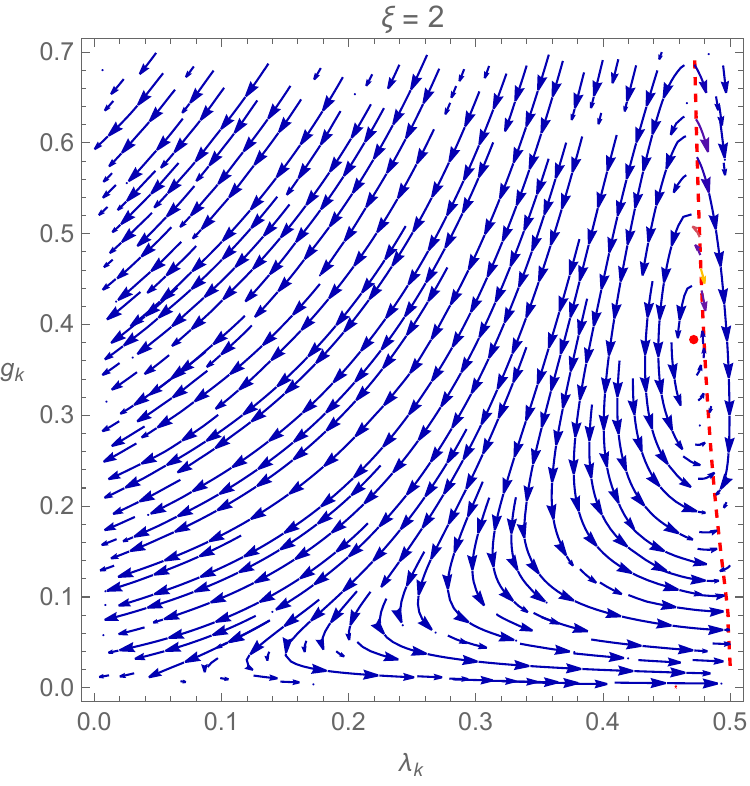}
    \caption{Flow diagram with $\zeta = \frac{1}{2}$ and $\alpha = \frac{2}{5}$, for different values of $\xi$.}
    \label{fig:zeta12-alpha25}
\end{figure}

\begin{figure}[h]
    \centering
    \includegraphics[scale=0.3]{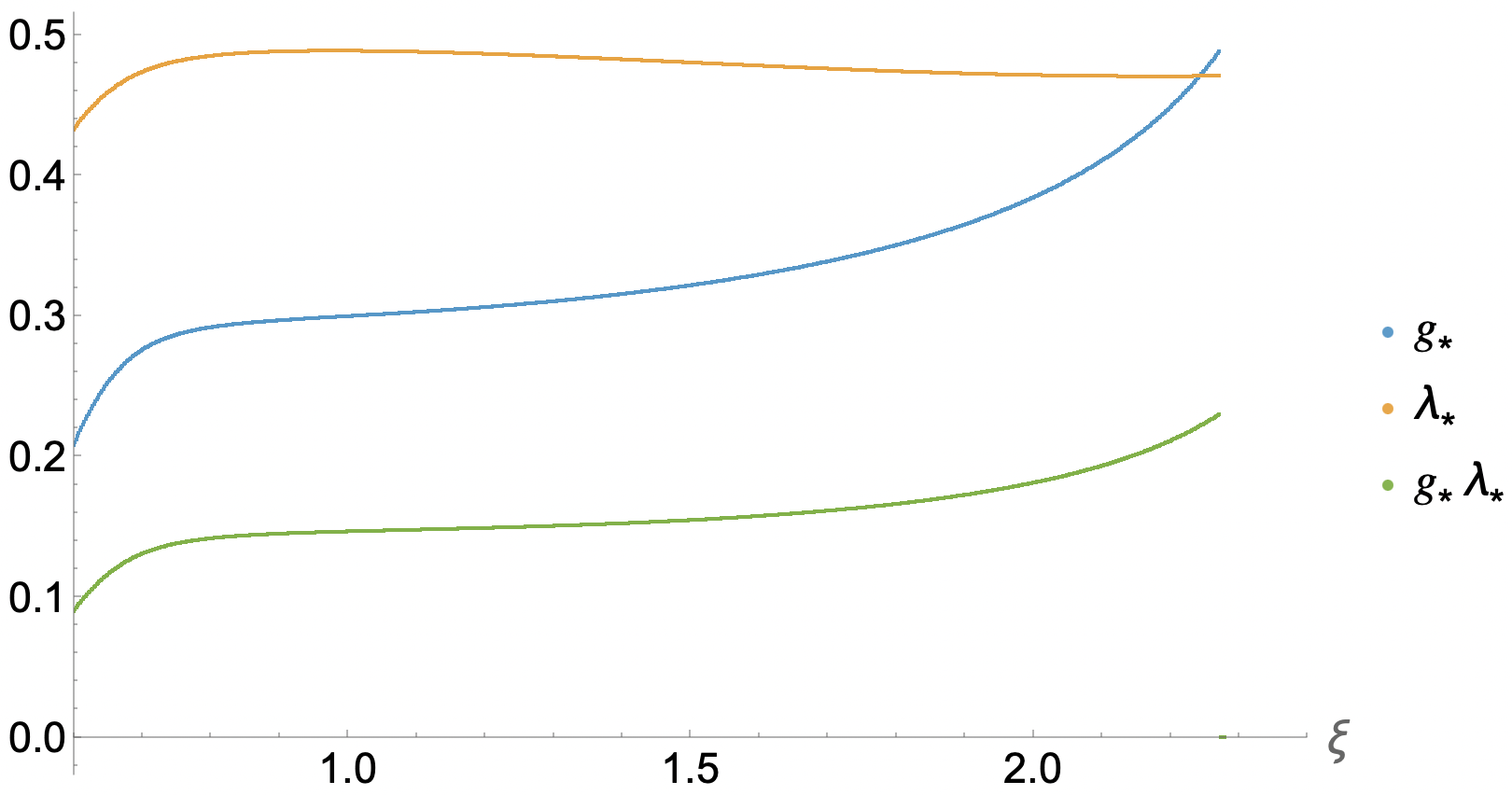}
    \caption{UV fixed point with $\zeta = \frac{1}{2}$ and $\alpha = \frac{2}{5}$ for different values of $\xi$.} 
    \label{fig:fp-zeta12-alpha-25}
\end{figure}

\begin{figure}[h]
    \centering
    \includegraphics[scale=0.45]{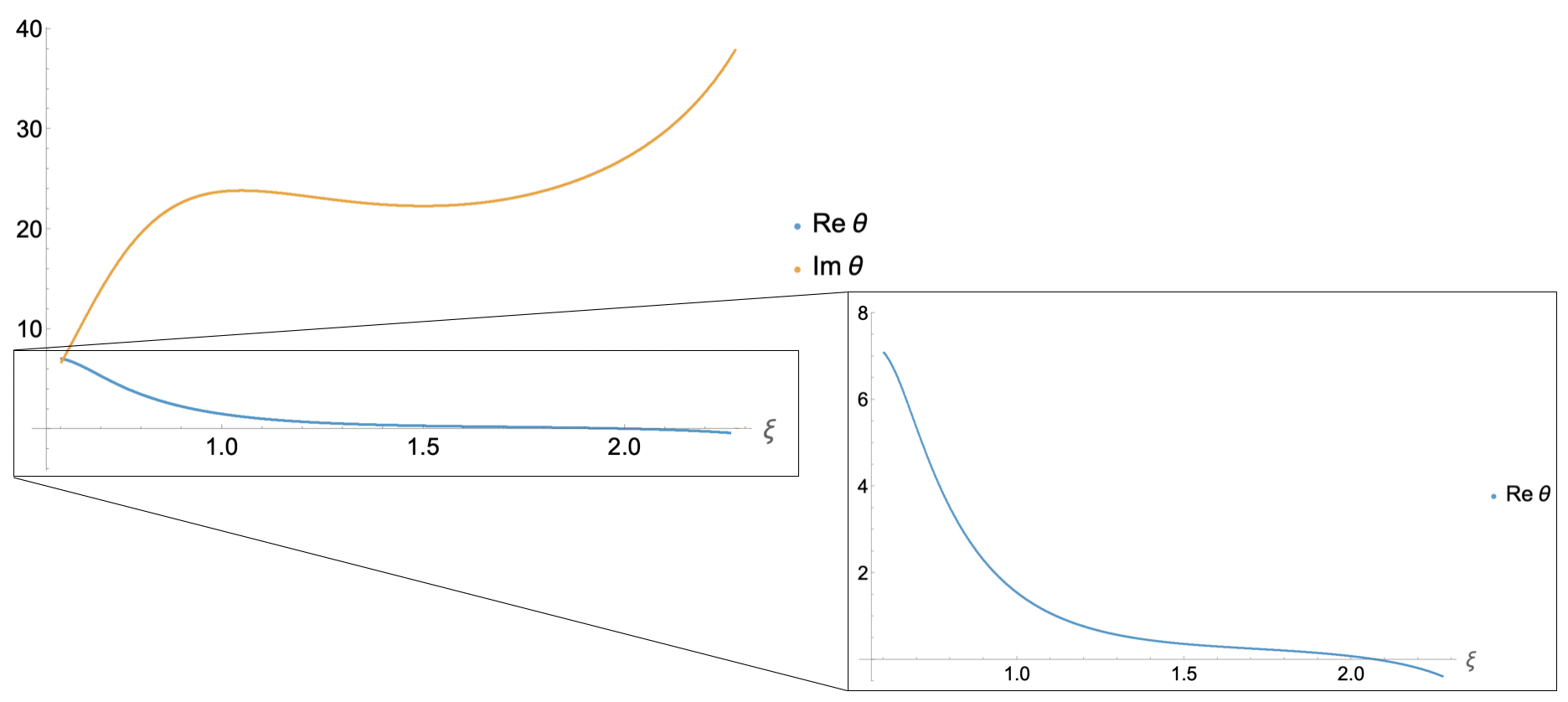}
    \caption{Critical exponents with $\zeta = \frac{1}{2}$ and $\alpha = \frac{2}{5}$ for different values of $\xi$.} 
    \label{fig:fp-zeta12-alpha-25-critical}
\end{figure}

We now turn our attention to the $\xi$ dependence of the phase diagram for a fixed value of the Hadamard parameter $\alpha = \frac{2}{5}$. The flow diagram is shown in Fig.~\ref{fig:zeta12-alpha25}, and the dependence of the fixed point $(g_k^*, \lambda_k*)$ on $\xi$ is shown in Fig.~\ref{fig:fp-zeta12-alpha-25}.
We note that the UV fixed point lies in close proximity of the vertical asymptote $\lambda_k = \frac{1}{2}$, which makes the numerical evaluation of the critical exponents difficult and restricts the range of admissible values for $\alpha$ and $\xi$. The critical exponents $\theta$ are again complex conjugates and are shown in Fig.~\ref{fig:fp-zeta12-alpha-25-critical}.
In the limit $\xi \to 3$\footnote{The corresponding gauge in QED is known as the Fried--Yennie gauge~\cite{Fried:1958zz}, which makes individual Feynman diagrams IR-finite (as opposed to only their sum), and is extremely useful in bound-state calculations~\cite{Eides:2000xc}.} the flow diverges, and the $\beta$ functions become complex for higher values of $\xi$. Again, this is due to the vanishing of the argument of the logarithms present in the coincidence limit of the propagators. From Fig.~\ref{fig:fp-zeta12-alpha-25-critical}, we see that the real part of the critical exponent $\Re \theta$ is positive for $0.5 \lesssim \xi \lesssim 2$, which includes the previous case $\xi = 1$.

\begin{figure}[h]
    \centering
    \includegraphics[scale=0.24]{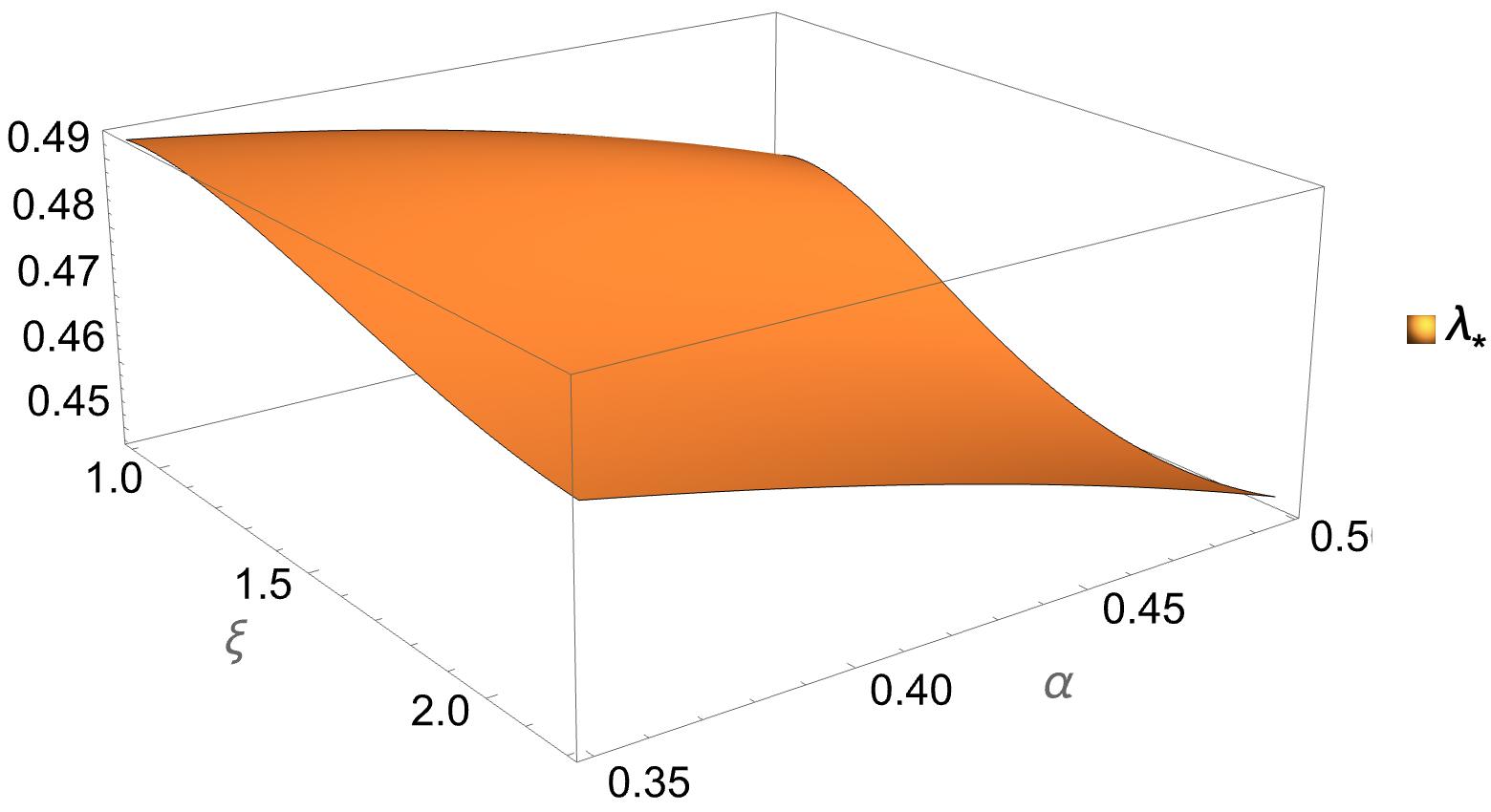}
    \includegraphics[scale=0.24]{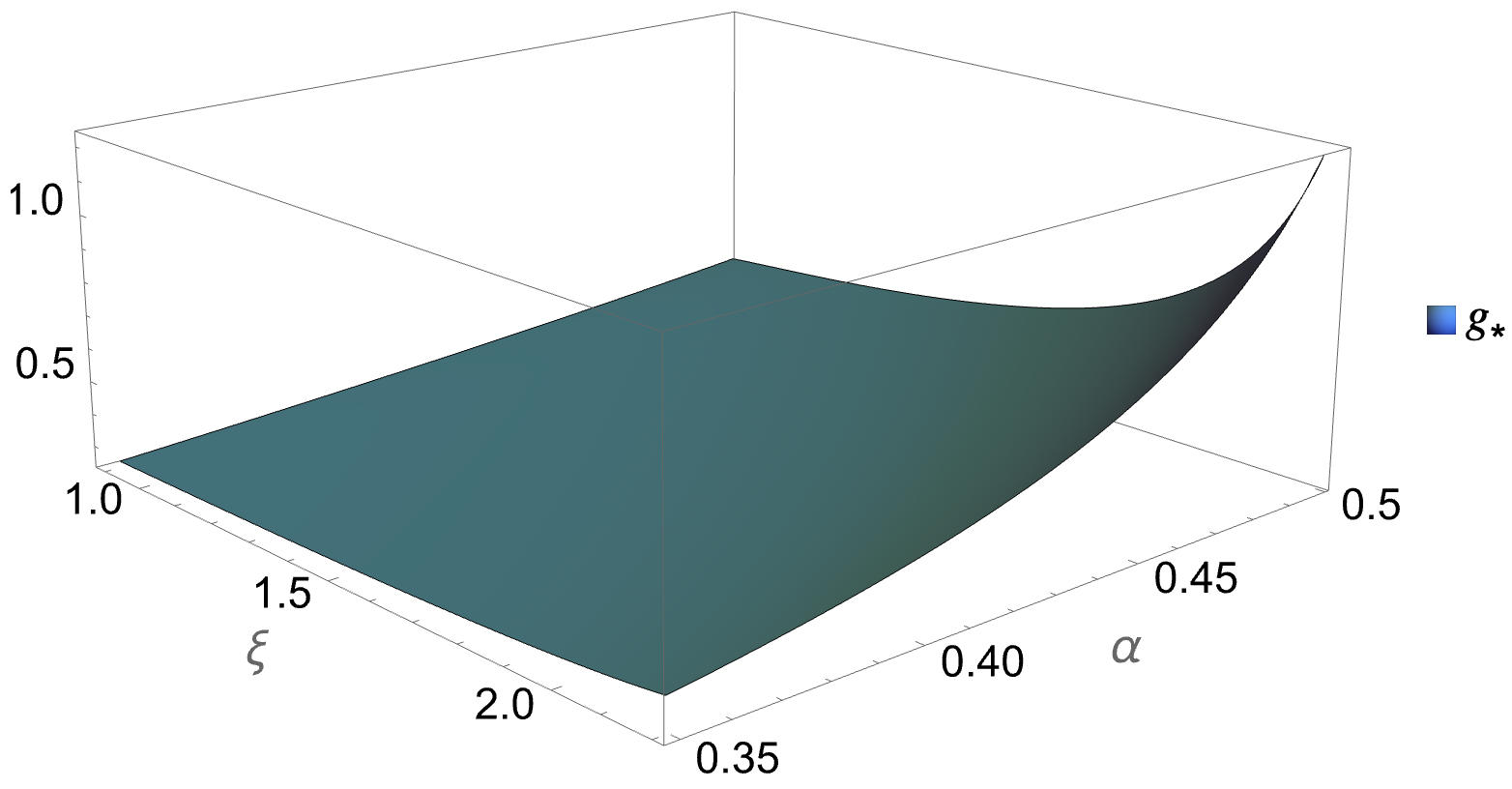}
    \includegraphics[scale=0.24]{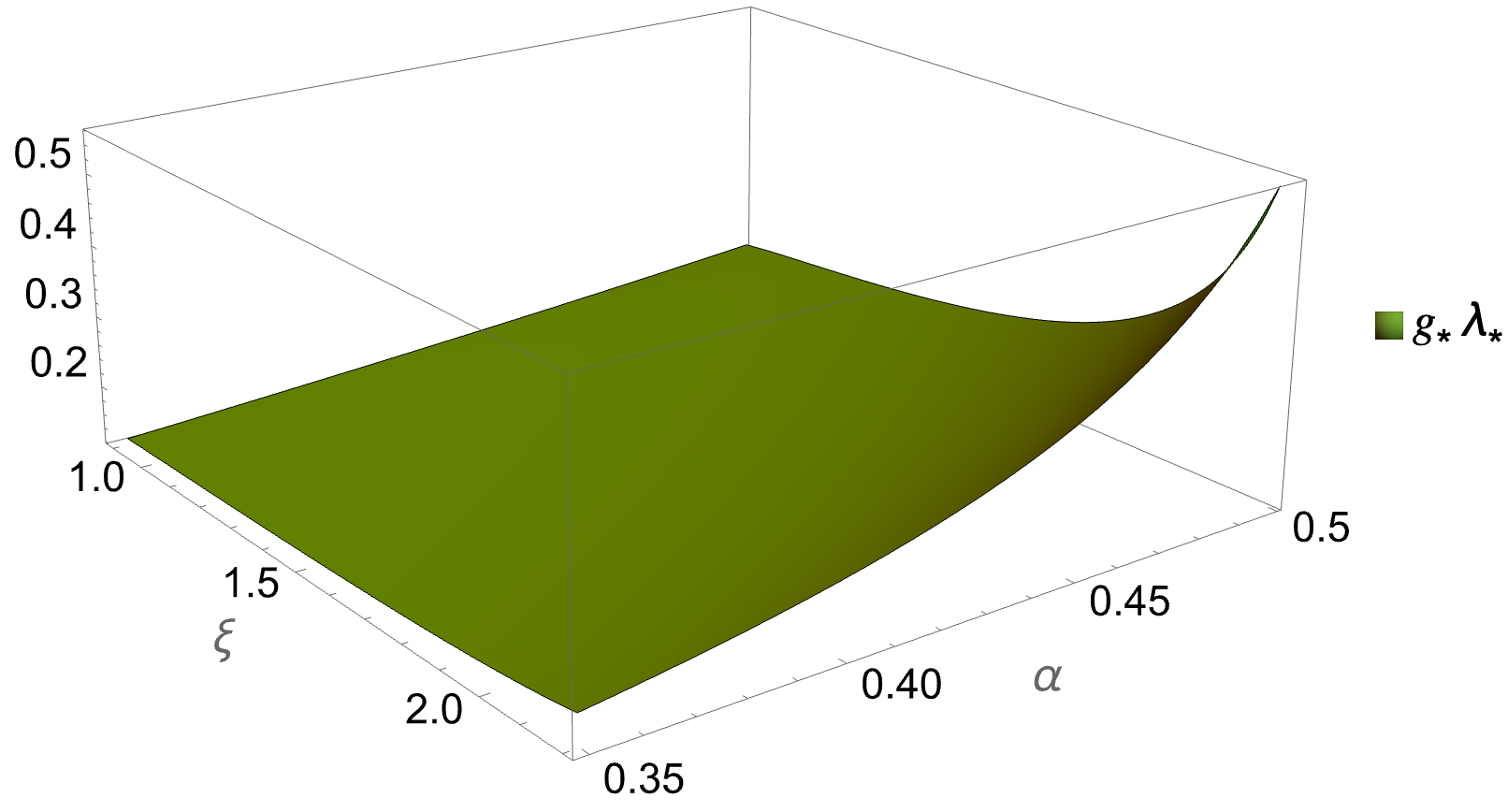}
    \caption{UV fixed points for $\zeta = \frac{1}{2}$ and with varying $\xi$ and $\alpha$.}
    \label{fig:FP-xi-alpha-1:2}
\end{figure}

Finally, we investigate the dependence of the RG flow in the full two-dimensional parameter space with coordinates $\alpha$ and $\xi$, for the fixed value of $\zeta = \frac{1}{2}$. We numerically evaluate the surfaces determined by the fixed-point equation and by the critical exponents as functions of $\alpha$ and $\xi$, and the results are given in Fig.~\ref{fig:FP-xi-alpha-1:2}.
\begin{figure}[h]
    \centering
    \includegraphics[scale=0.26]{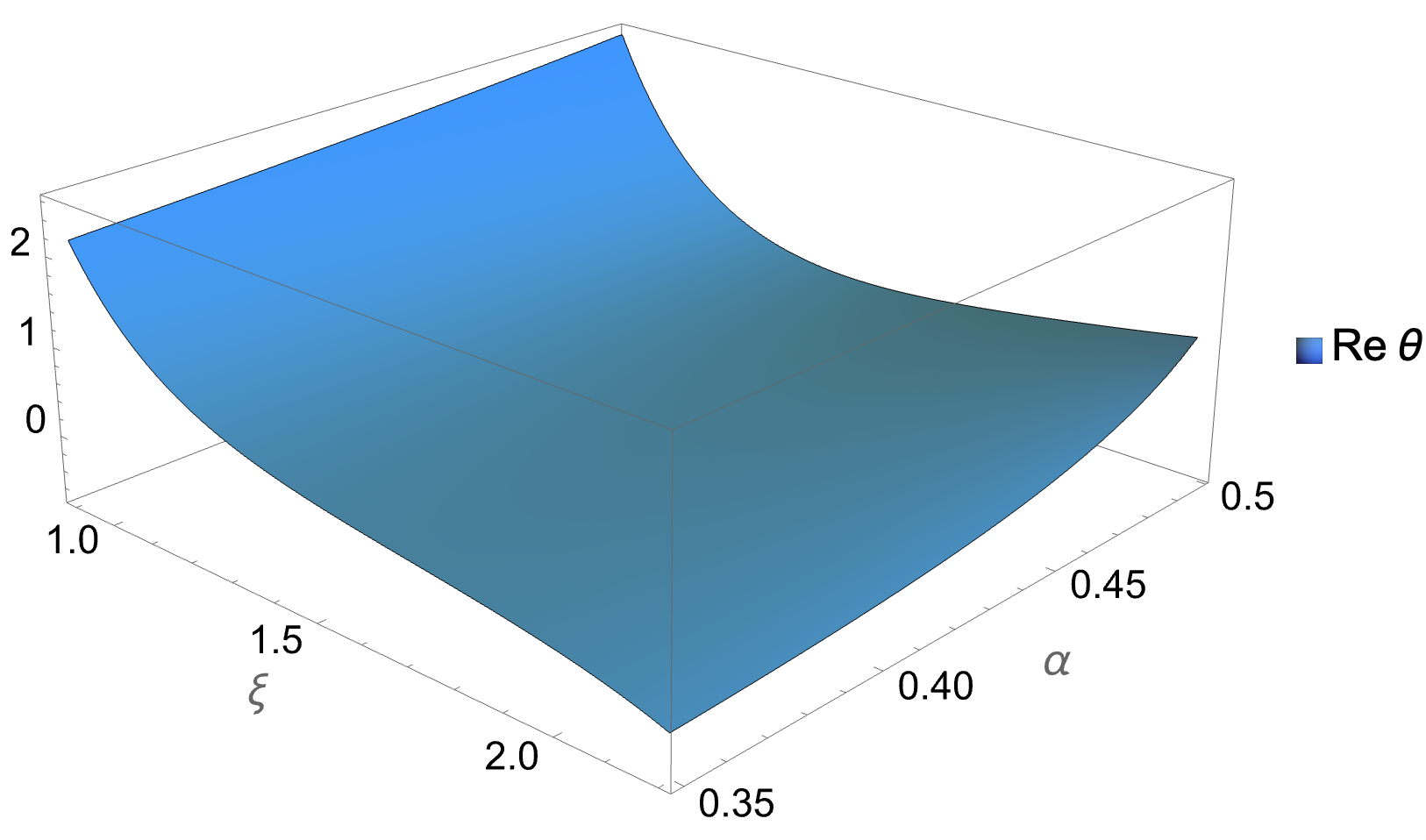}
    \includegraphics[scale=0.26]{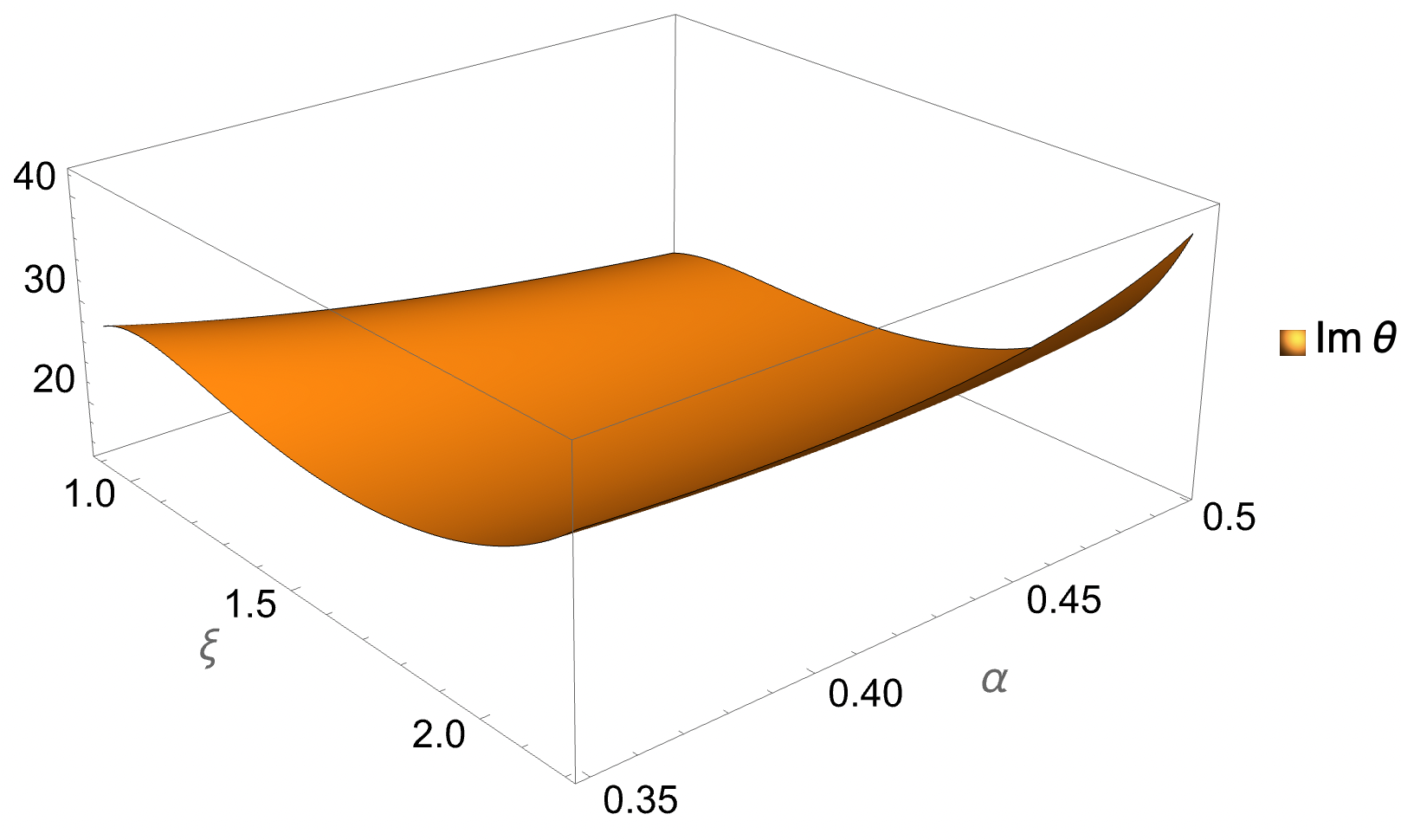}
    \caption{Critical exponents for $\zeta = \frac{1}{2}$ and with varying $\xi$ and $\alpha$.}
    \label{fig:cexp-xi-alpha-1:2}
\end{figure}
We see that the UV fixed points and critical exponents are smooth functions of $\alpha$ and $\xi$ (in the considered range of values), and in particular the critical exponents show only a mild dependence on the Hadamard parameter $\alpha$. Given this smooth dependence, it would be interesting to see if it is possible to choose the Hadamard parameter $\alpha$ as a function of the gauge parameter $\xi$ in such a way that the UV fixed points and the critical exponents become independent of it. While this seems possible for any single parameter (for example $g_k^*$, which is approximately constant if we decrease $\alpha$ with increasing $\xi$), eliminating (or minimizing) the parametrization dependence of the FRGE in this way would be possible only if \emph{all} fixed points and critical exponents become (approximately) $\xi$-independent for a given $\alpha = \alpha(\xi)$. Since the $\beta$ functions are very complicated, we leave a detailed study of this issue for future work.

\subsection{The flow for \texorpdfstring{$\zeta = 1$}{\textzeta = 1}}
\label{sec:harmonic gauge}

We perform a similar analysis for the harmonic gauge $\zeta = 1$, investigating the parameter dependence on $\alpha$ and $\xi$ of the $\beta-$functions, the fixed points and the critical exponents. Although the harmonic gauge is widely used in the literature for its simplicity, since the propagators are more complicated for $\zeta = 1$ than for $\zeta = \frac{1}{2}$, we find also more complicated expressions for the $\beta$ functions.

\begin{figure}[h]
    \centering
    \includegraphics[width=0.32\linewidth]{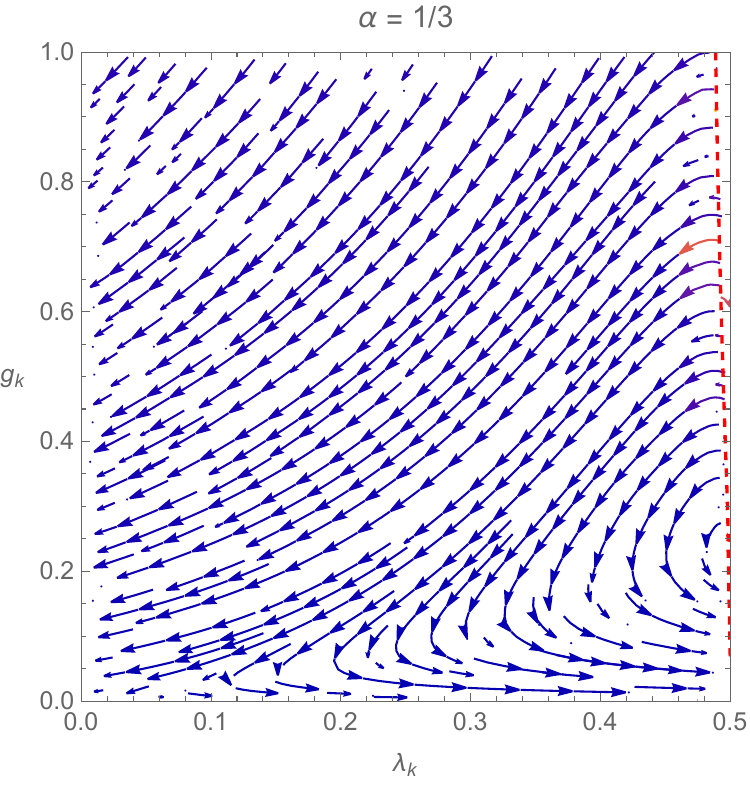}
    \includegraphics[width=0.32\linewidth]{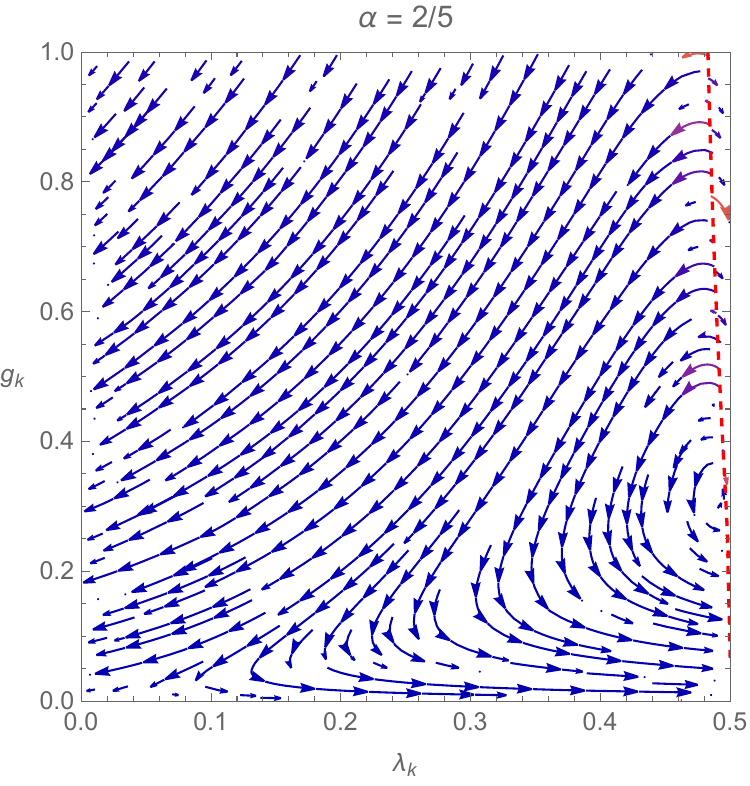}
    \includegraphics[width=0.32\linewidth]{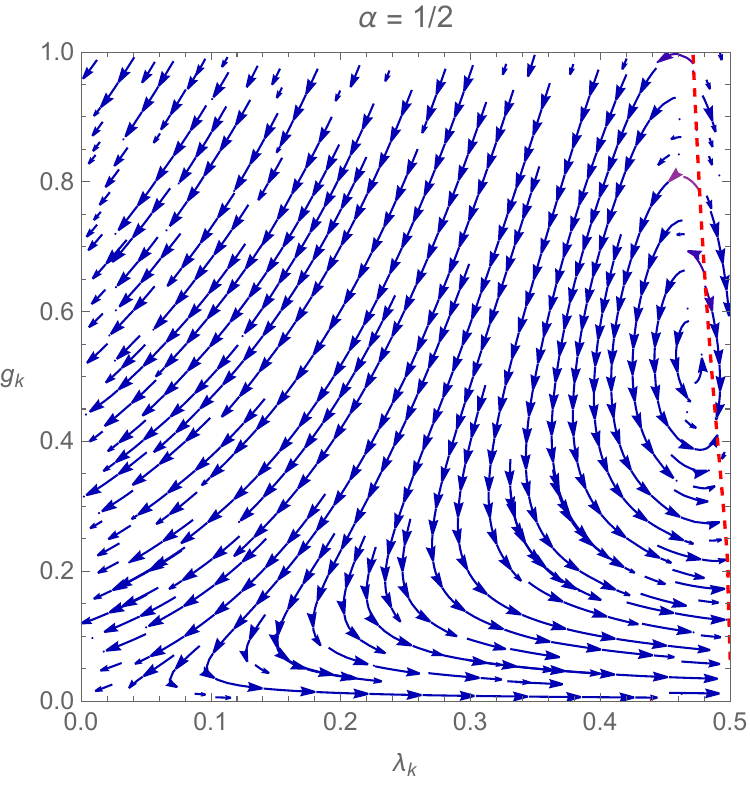}
    \caption{Flow diagram with $\zeta=1$ and $\xi =1$, for different values of $\alpha$.}
    \label{fig:phase-diagrams-H-F}
\end{figure}

\begin{figure}[h]
	\centering
	\includegraphics[scale=0.3]{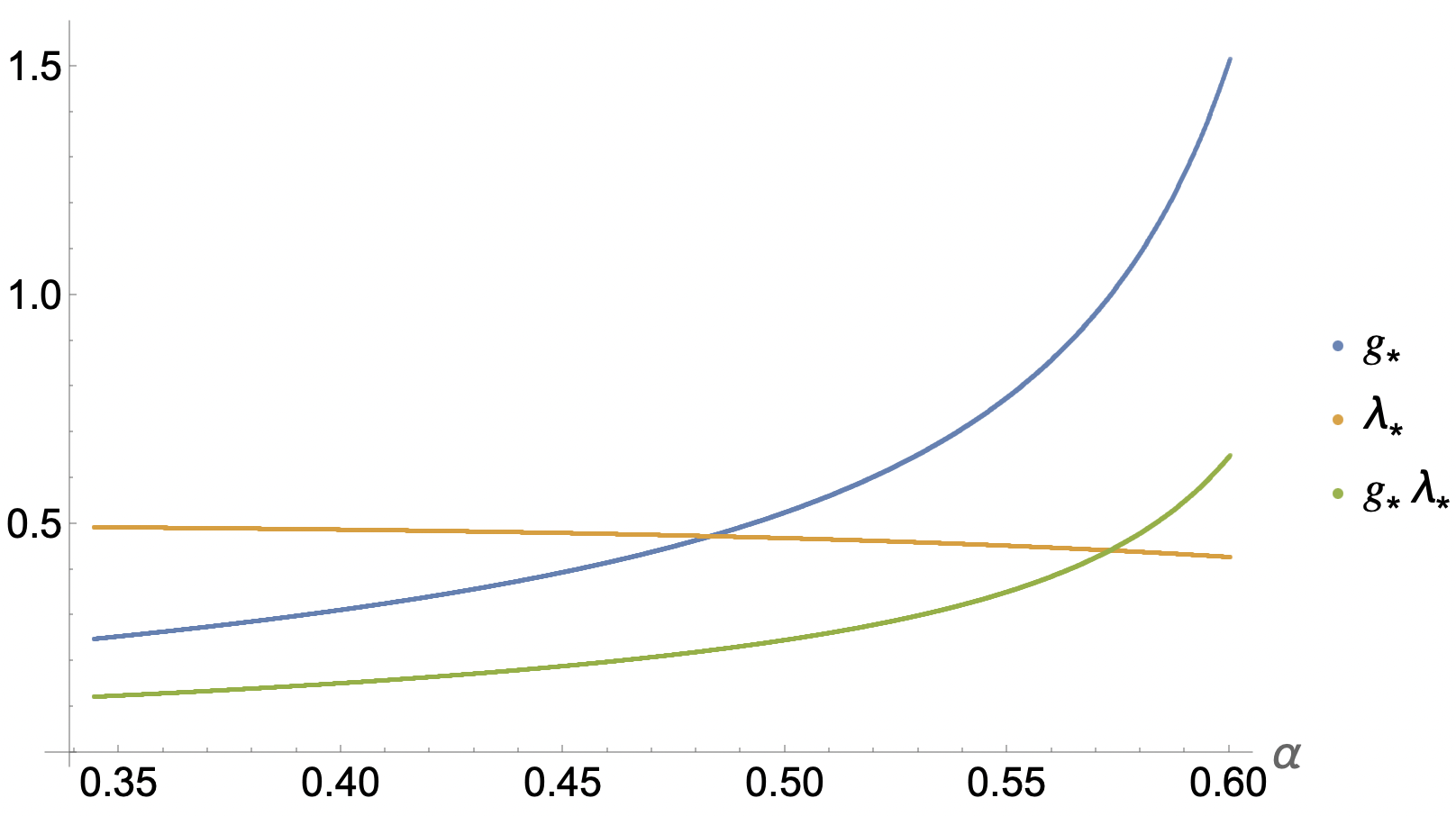}
	\caption{UV fixed point with $\zeta = 1$ and $\xi = 1$, for different values of $\alpha$.}
    \label{fig:fp-F-H}
\end{figure}

\begin{figure}[h]
	\centering
    \includegraphics[scale=0.45]{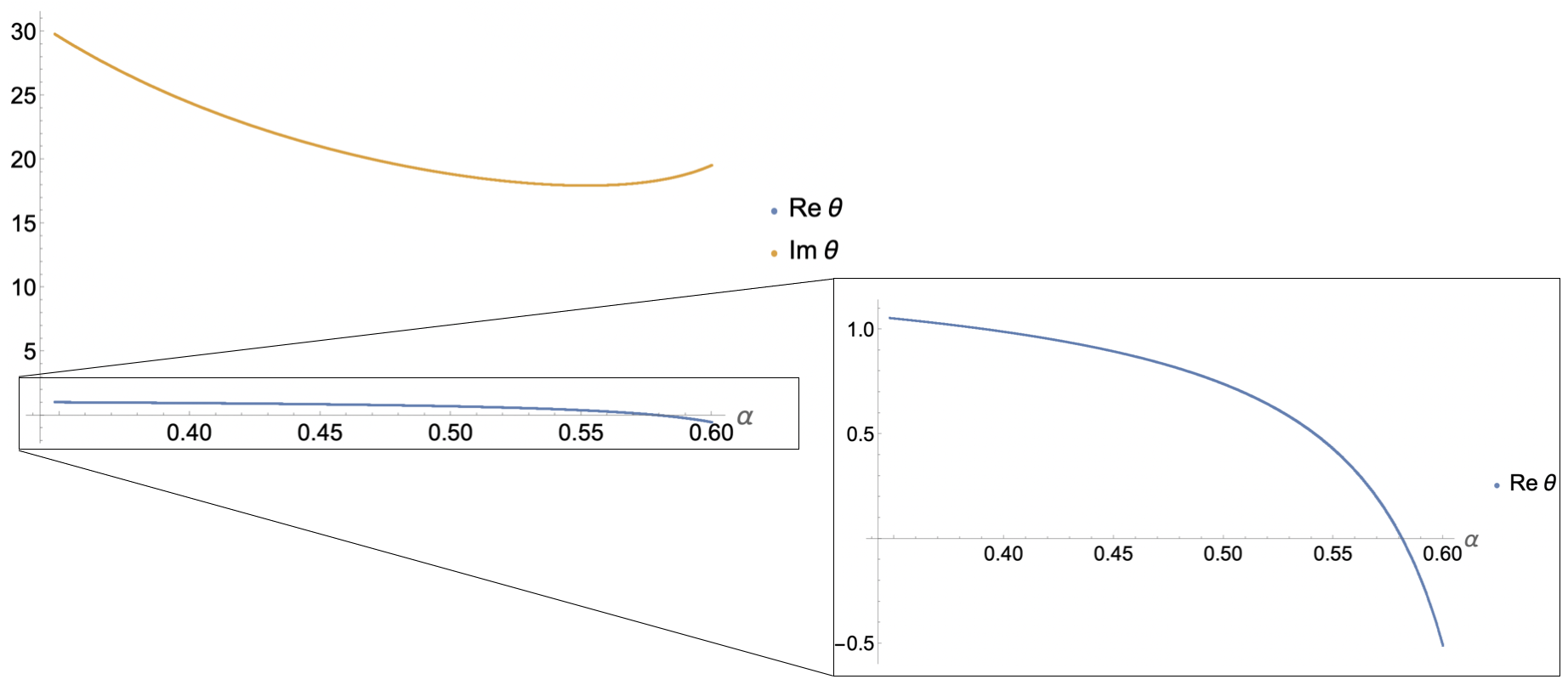}
	\caption{Critical exponents with $\zeta = 1$ and $\xi = 1$, for different values of $\alpha$.}
    \label{fig:fp-F-H-critical}
\end{figure}

As for the previous gauge $\zeta = \frac{1}{2}$, we find a non-trivial UV fixed point also in harmonic gauge whose exact location depends on the other parameters $\xi$ and $\alpha$, and which is shown in Fig.~\ref{fig:phase-diagrams-H-F}.
The $\alpha$ dependence of the UV fixed point is very similar to the case $\zeta = \frac{1}{2}$, and shown in Fig.~\ref{fig:fp-F-H}, fixing $\xi = 1$. In this case, the range of admissible values for the Hadamard parameter is $0.1 \lesssim \alpha \lesssim 0.64$; for smaller values the critical point crosses the asymptote $\lambda_k = \frac{1}{2}$ [the Higuchi bound~\eqref{eq:higuchi-bound}], while for greater values the numerical evaluation breaks down.
The critical exponents shown in Fig.~\ref{fig:fp-F-H-critical} are complex conjugates for all admissible values of $\alpha$. However, for $\alpha \gtrsim 0.56$, the real part of the critical exponents becomes negative. While the scaling solution appears reliable, the numerical values of the critical exponents are deformed by the asymptote.

\begin{figure}[h]
\centering
	\includegraphics[width=0.32\linewidth]{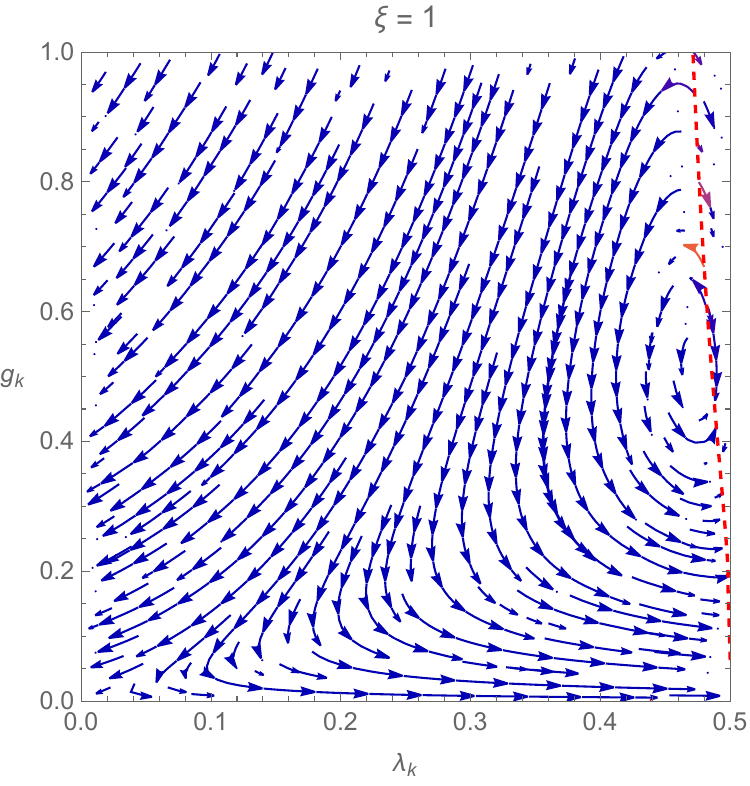}
    \includegraphics[width=0.32\linewidth]{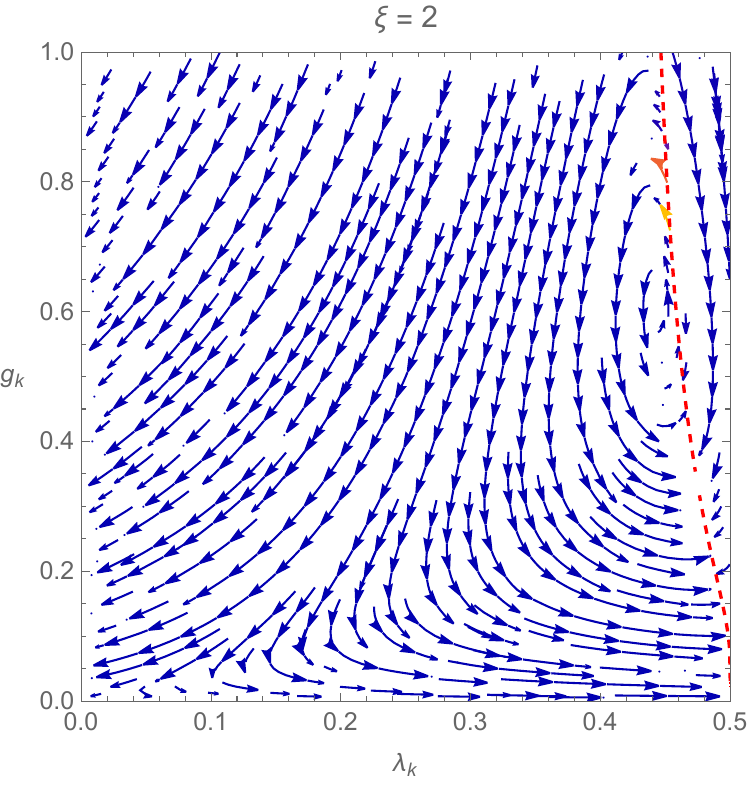}
	\includegraphics[width=0.32\linewidth]{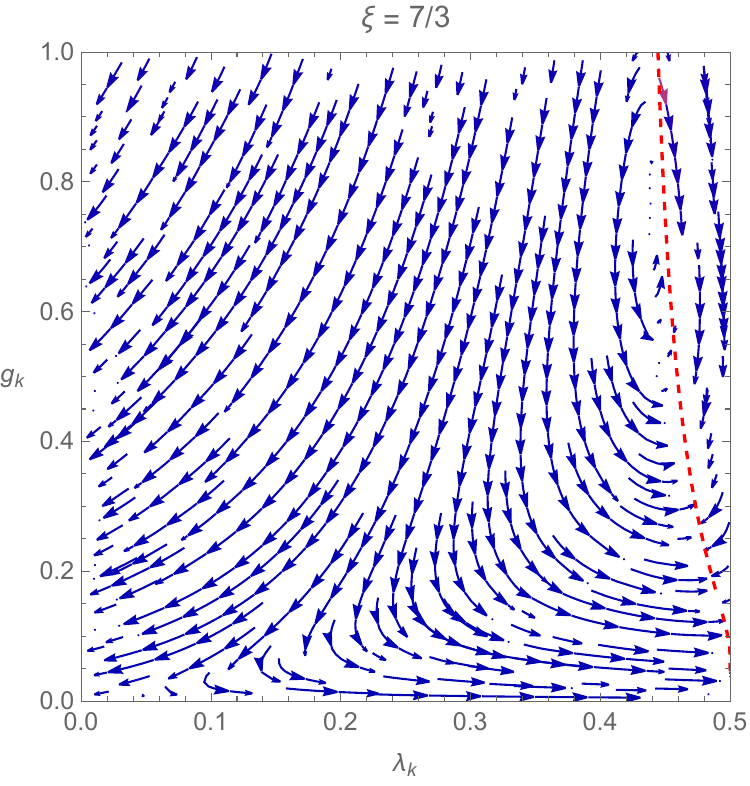}
    \caption{Flow diagram with $\zeta = 1$ and $\alpha = \frac{1}{2}$, for different values of $\xi$. The dashed red line indicates the locus of divergence of the anomalous dimension.}
    \label{fig:phase-diagrams-H-alpha}
\end{figure}
\begin{figure}[h]
	\centering
	\includegraphics[scale=0.3]{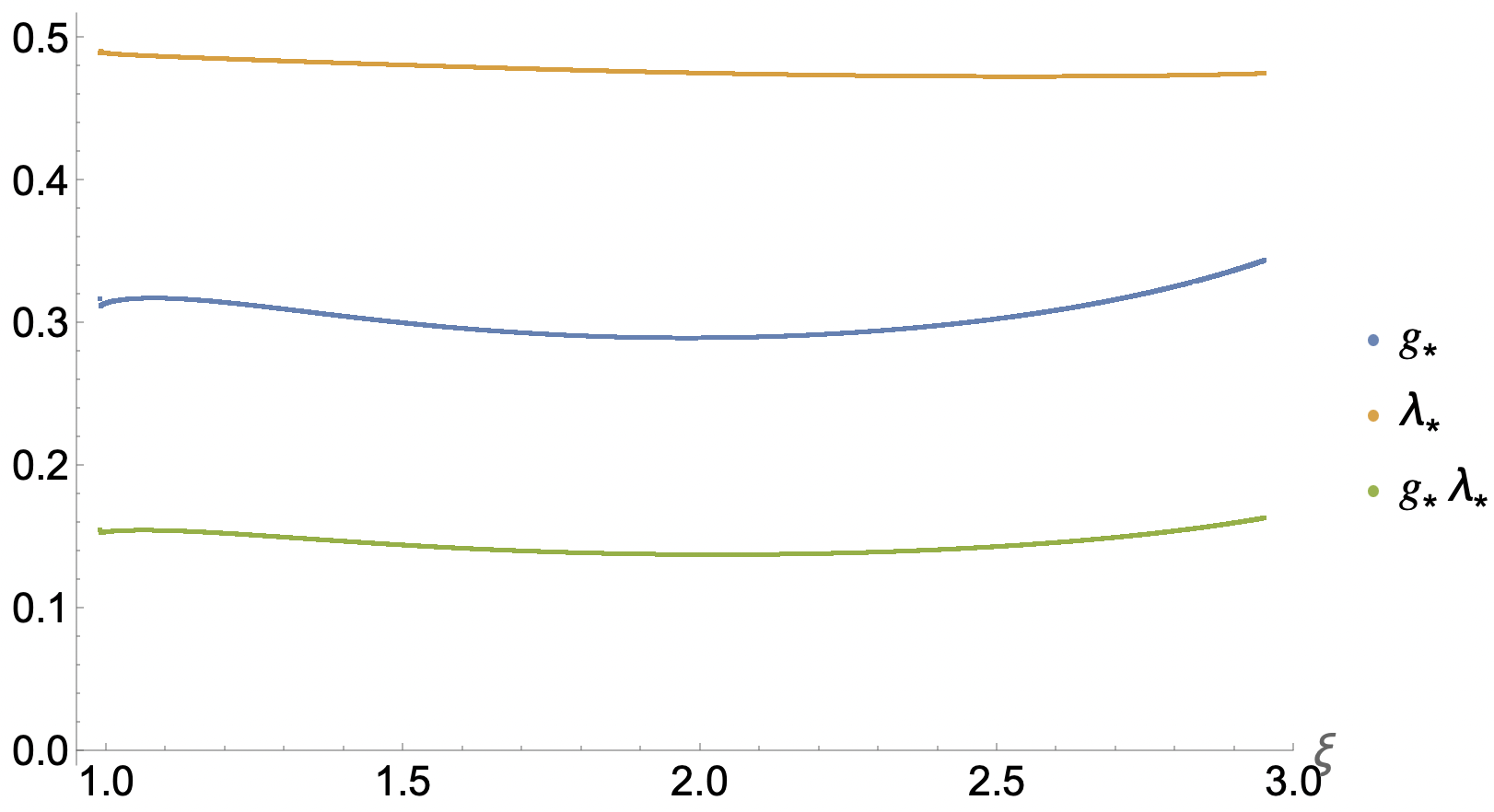}
	\caption{UV fixed point with $\zeta = 1$ and $\alpha = \frac{2}{5}$, for different values of $\xi$.} 
    \label{fig:fp-H-alpha-25}
\end{figure}
\begin{figure}[h]
	\centering
    \includegraphics[scale=0.45]{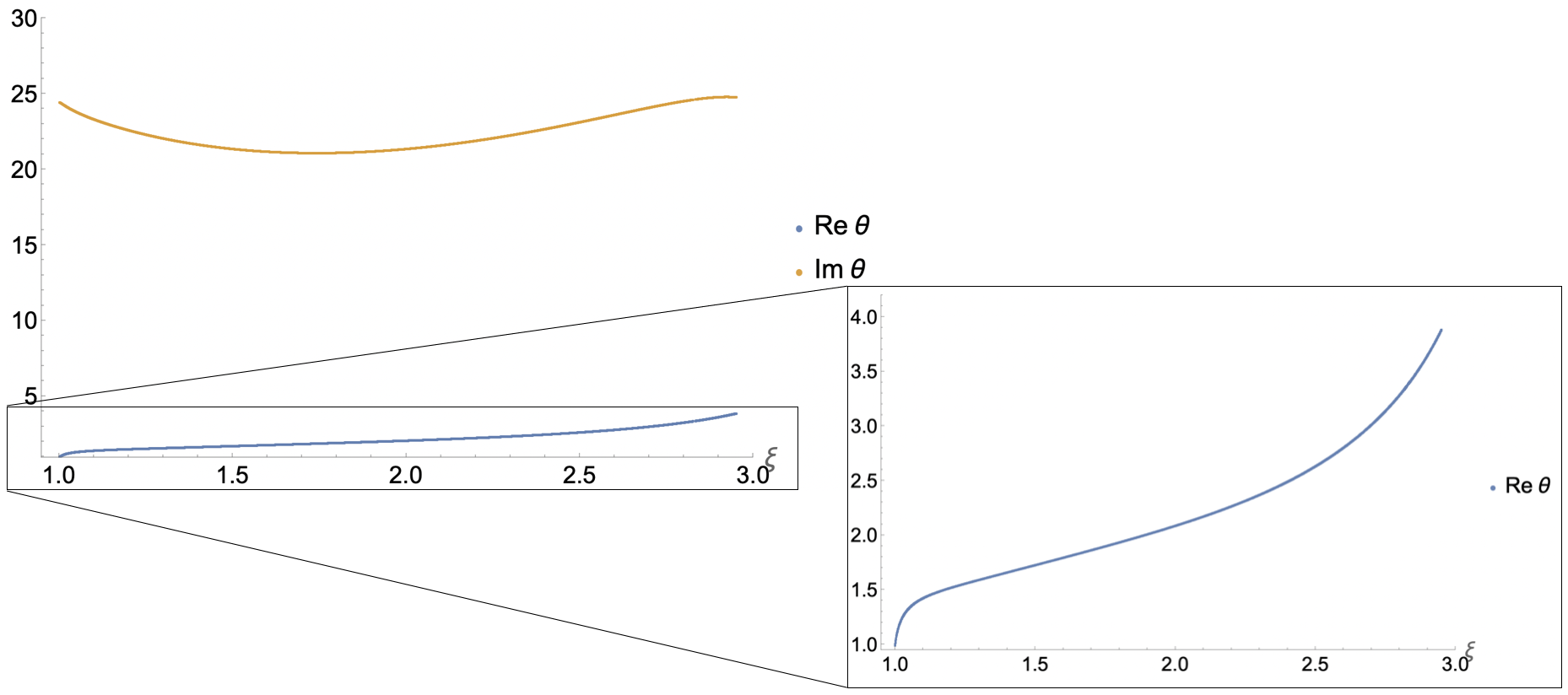}
	\caption{Critical exponents with $\zeta = 1$ and $\alpha = \frac{2}{5}$, for different values of $\xi$.} 
    \label{fig:fp-H-alpha-25-critical}
\end{figure}

The $\xi$ dependence of the flow (shown in Fig.~\ref{fig:phase-diagrams-H-alpha}) is complicated by the interplay with the asymptote as well. Indeed, for $\xi < 1$ the vertical asymptote at fixed value $\lambda_k = \lambda_A$ gets closer to the origin, effectively reducing the values of positive $\lambda_k$ (the ones compatible with a de~Sitter background) for which we have a real-valued flow. Similar to the Higuchi bound, this asymptote results from negative-norm gauge modes, which contribute to the off-shell flow. For this reason, we only study the range $1 \leq \xi \leq 3$. The UV fixed points shown in Fig.~\ref{fig:fp-H-alpha-25} are remarkably stable in this case, while the real part of the critical exponents increases monotonically in $\xi$ (Fig.~\ref{fig:fp-H-alpha-25-critical}).

\begin{figure}[h]
    \centering
	\includegraphics[scale=0.24]{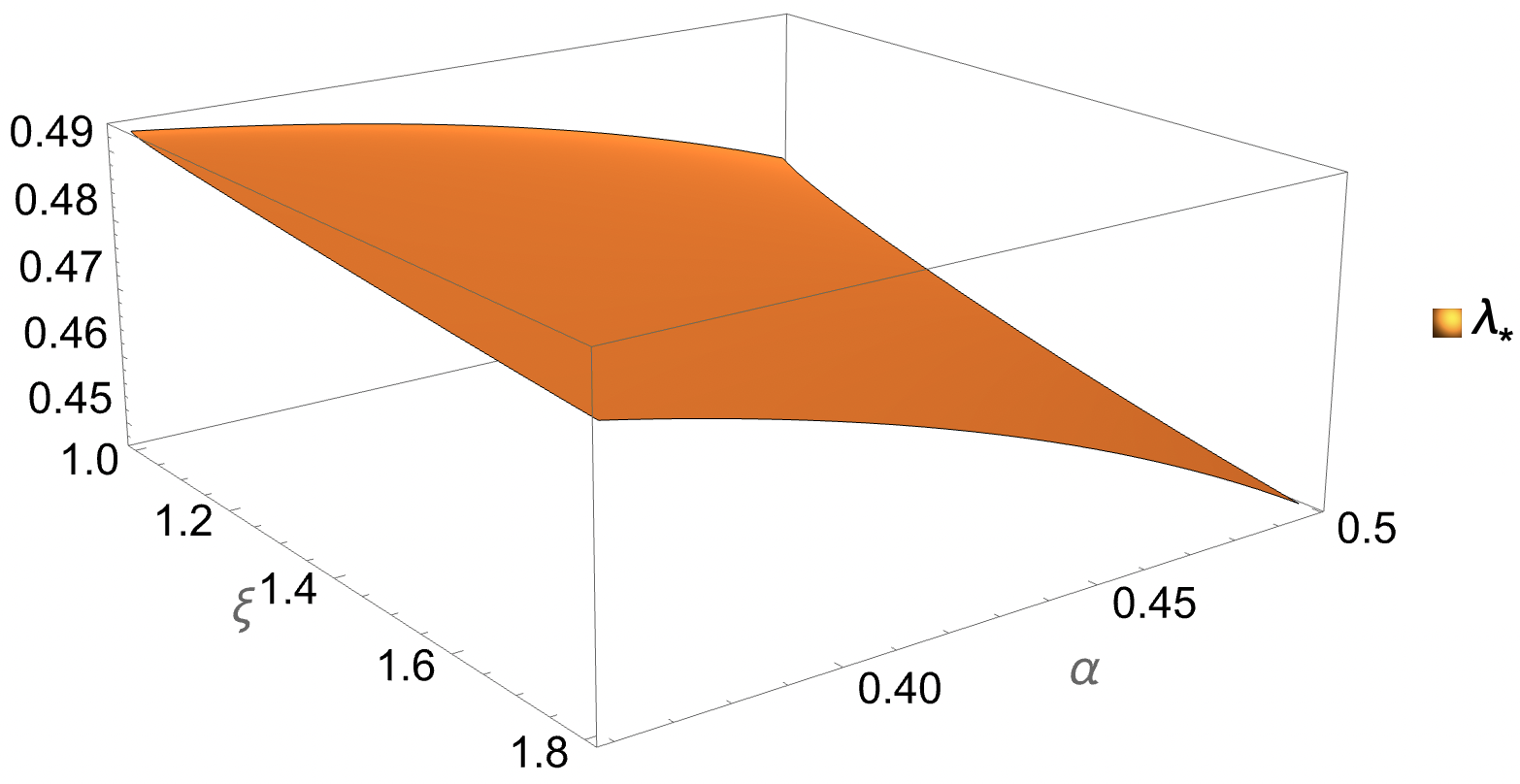}
    \includegraphics[scale=0.24]{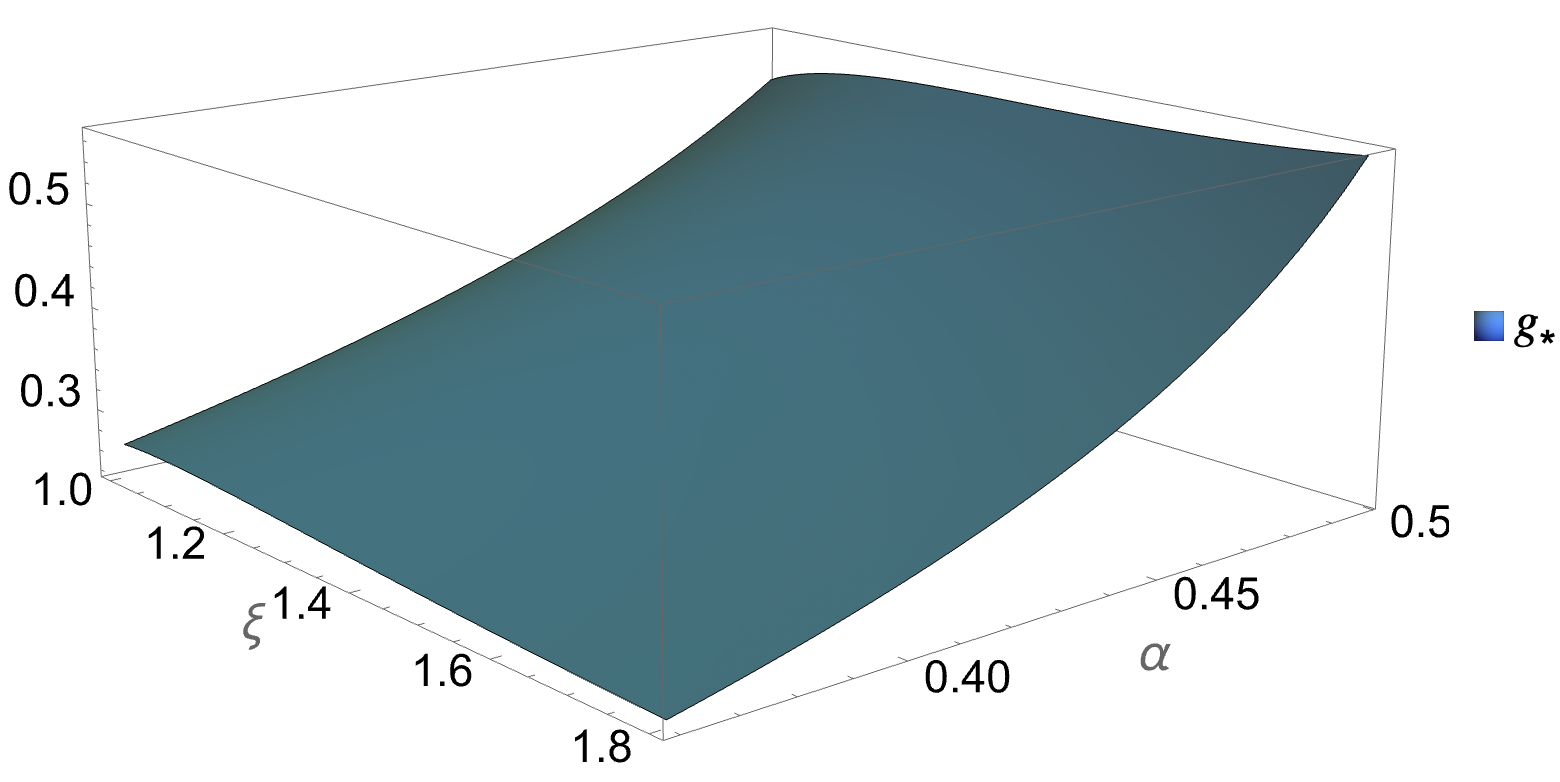}
    \includegraphics[scale=0.24]{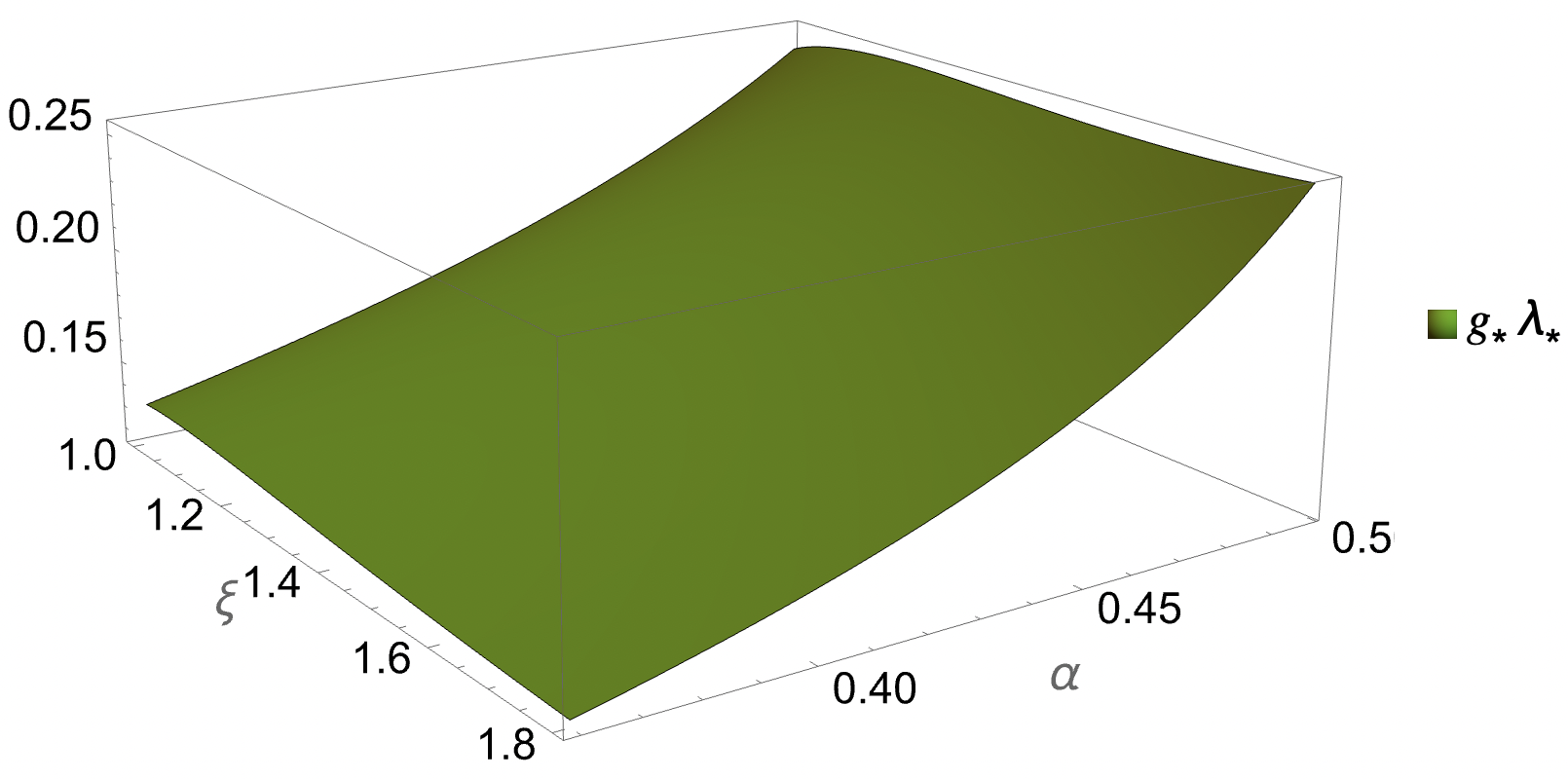}
    \caption{UV fixed points varying $\xi$ and $\alpha$ for $\zeta = 1$.}
    \label{fig:FP-xi-kappa}
\end{figure}

\begin{figure}[h]
    \centering
    \includegraphics[scale=0.26]{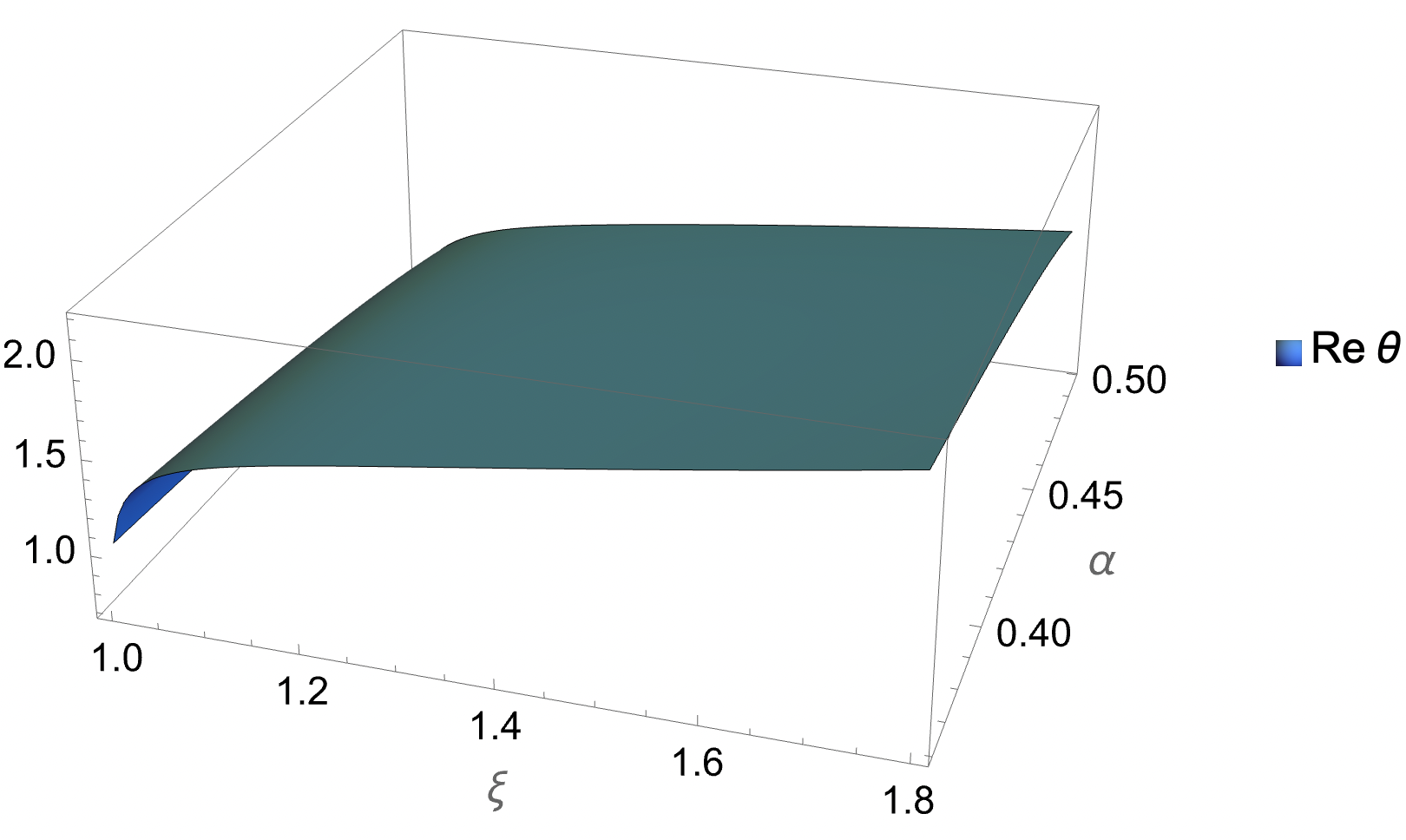}
    \includegraphics[scale=0.26]{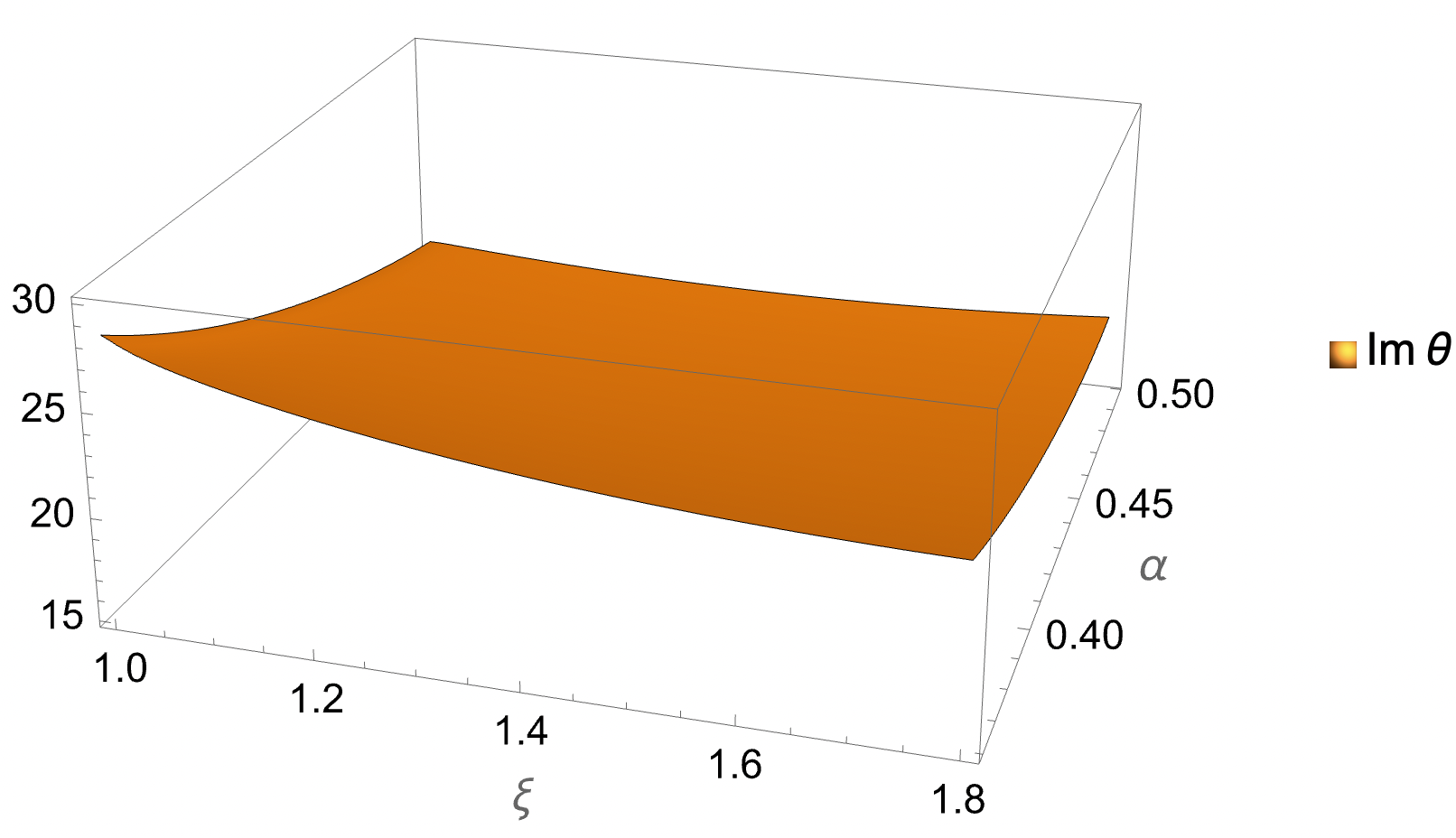}
    \caption{Critical exponents varying $\xi$ and $\alpha$ for $\zeta = 1$.}
    \label{fig:FP-xi-kappa-critical}
\end{figure}

Finally, we can study the dependence of the fixed point and critical exponents on the full parameter space spanned by $\xi$ and $\alpha$, shown in Figs.~\ref{fig:FP-xi-kappa} and~\ref{fig:FP-xi-kappa-critical}.
The overall picture is similar to the case $\zeta = \frac{1}{2}$, except that the dependence of the fixed points on $\xi$ is very weak.

\section{Discussion}
\label{sec:discussion}

We have derived the non-perturbative RG flow for Lorentzian quantum gravity in de~Sitter space in the Einstein--Hilbert truncation. As the main result, we showed evidence for the realization of the UV completion of gravity in de~Sitter space, the positively curved cosmological Einstein spacetime in accordance with cosmological observations in the early and late time universe. Notably, our findings in the functional framework could be compared to the discrete counterpart of the Lorentzian version of asymptotic safety, namely causal dynamical triangulations (CDT)~\cite{Loll:2019rdj}. In CDT, a macroscopic de~Sitter universe with small quantum fluctuations emerges from the complete gravitational path integral, and the effective action governing its dynamics can be reconstructed using Monte Carlo data~\cite{Ambjorn:2007jv,Ambjorn:2008wc,Ambjorn:2024qoe}.

\begin{figure}[h]
	\centering
    \includegraphics[width=0.5\linewidth]{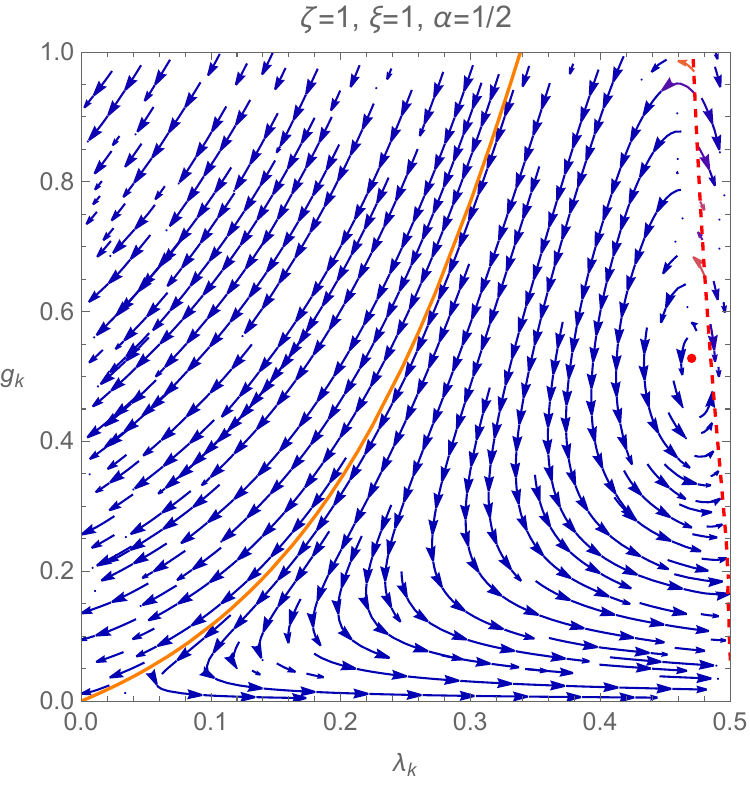}
	\caption{Phase diagram in gauge $\xi = \zeta = 1$ and with $\alpha = \frac{1}{2}$, with the separatrix in orange, the locus of divergences of the anomalous dimension in a dashed line, and the UV fixed point in red.} 
    \label{fig:phase-diagram-F-H-alpha-12}
\end{figure}

As an important technical advancement, for the first time we used a fully covariant Lorentzian RG flow~\cite{DAngelo:2023wje}, which explicitly displays a dependence on the quantum state. As the analogue of the Minkowski vacuum, we used the unique de~Sitter invariant Hadamard state for the graviton and ghost fields, for which we needed to determine the graviton and ghost propagators for massive gravity in de~Sitter space in a general gauge with two parameters $\xi$ and $\zeta$~\cite{DFF2024b}. We choose a regulator adapted to the gauge in such a way that the d'Alembertian is dressed as $\nabla^2 \to \nabla^2 - k^2$, and $k^2$ acts as an effective mass. While this results in IR-finite propagators in contrast to the IR divergences typical for massless fields in de~Sitter~\cite{Higuchi:2002sc,Higuchi:2010xt,Higuchi:2011vw,Miao:2011fc,Mora:2012zi,Morrison:2013rqa,Frob:2016hkx,Ferrero:2021lhd}, the Higuchi bound~\cite{Higuchi:1986py,Higuchi:1989gz} for massive spin-2 fields restricts the flow to a small cosmological constant $\lambda_k \leq \frac{1}{2}$. Moreover, a new parameter $\alpha$ appears due to the removal of the UV divergences via the covariant Hadamard normal ordering and which parametrizes finite renormalizations. A typical flow diagram is shown in Fig.~\ref{fig:phase-diagram-F-H-alpha-12}, which despite all the technical differences is qualitatively very similar to the well-known Euclidean results for the Einstein--Hilbert truncation~\cite{Reuter:1996cp,Reuter:2001ag}.
Unfortunately, the UV fixed point is often in close proximity to the asymptote $\lambda_k = \frac{1}{2}$, which introduces spurious effects in the numerical analysis.

We have investigated the dependence of the phase diagram, non-trivial fixed points and critical exponents on the gauge parameters $\xi$ and $\zeta$ and on the Hadamard parameter $\alpha$. We have shown the existence of non-trivial UV fixed points and scaling solutions for a wide range of parameters, corroborating the asymptotic safety scenario also in Lorentzian signature. As one would expect, the numerical values for the non-trivial fixed point exhibit gauge and scheme dependence. In turn, it is interesting to notice that the critical exponents, and in particular their real part, appear remarkably close to the Euclidean results~\cite{Reuter:1996cp,Reuter:2001ag} for the most common gauge choices. In fact, it might even be possible to adjust the Hadamard parameter $\alpha$ in such a way (depending on the gauge parameters $\xi$ and $\zeta$) that the critical exponents become at least approximately scheme- and gauge-independent. Clearly, a scheme- and gauge-independent RG flow would provide a powerful tool to investigate the asymptotic safety scenario in Lorentzian quantum gravity. As a step in this direction, in future works we would like to incorporate ideas of the essential RG scheme, restricting the flow to essential couplings and imposing the quantum equations of motion $\frac{\delta \Gamma_k}{\delta \phi} = 0$~\cite{Baldazzi:2021orb,Baldazzi:2021ydj,Baldazzi:2023pep,Knorr:2022ilz,Knorr:2023usb,Falls:2024noj}. This on-shell flow is then in principle gauge-independent (although truncations inevitably introduce such a dependence), and we could compare our results with those derived from other Lorentzian flow equations~\cite{Fehre:2021eob,Banerjee:2022xvi,Banerjee:2024tap,Thiemann:2024vjx,Ferrero:2024rvi}.

The main issue in our analysis is the lack of IR-complete RG trajectories. Complete trajectories would be a fundamental step both from a conceptual and a pragmatic point of view. Unfortunately, trajectories emanating from the UV fixed point are often interrupted by the asymptote at $\lambda_k = \frac{1}{2}$ (due to the Higuchi bound that is effective for large RG scale $k$), which heavily influences the numerical analysis. However, since Higuchi's analysis applies to a single non-interacting spin-2 field derived from the Einstein--Hilbert action with Fierz--Pauli mass term, it is possible that including more degrees of freedom in the form of matter fields could change this bound and thus the asymptote. Global flows with IR and UV non-trivial fixed points have been found in the Euclidean context, particularly in the presence of matter systems~\cite{Christiansen:2014raa,Korver:2024sam}. Another possibility is to work with a higher truncation of the effective average action, which also would change the effective dynamics of the spin-2 graviton such that the Higuchi bound could be overcome. One could furthermore consider different background such as anti-de~Sitter space (whose Euclidean analogue is hyperbolic space~\cite{Falls:2016msz}) or more general cosmological backgrounds, which apart from a cosmological constant necessarily involve a scalar field to source their evolution. It is also natural to ask whether the choice of a different state would improve the numerical evaluation of the flow. Here, we worked in the unique Hadamard de~Sitter-invariant state, the Bunch--Davies vacuum. Indeed, while the UV structure of the propagator is fixed by the Hadamard condition, the choice of the state influences its long-distance properties, due to the modification of its smooth part, and this could improve the IR sector of the FRGE.

Our work paves the way for a number of investigations in a cosmological setting. First, embedding this analysis into a gauge-invariant framework by means of relational observables would enable the study of the scaling behavior of the Mukhanov--Sasaki variable and other gauge-invariant observables~\cite{Dittrich:2005kc,Giesel:2007wi,Giesel:2007wk,Giesel:2012rb,Frob:2017lnt,Frob:2017gyj,Giesel:2018opa,Frob:2021ore,Giddings:2022hba,Goeller:2022rsx,Baldazzi:2021fye,Frob:2023awn,Ferrero:2024rvi}. Second, if in future works it will be possible to identify an IR fixed point, it could be compared with cosmological correlators, and in particular with the statistical properties of the CMBR distribution or, at later times, of the large scale structures~\cite{Bernardeau:2001qr,Friedrich:2015nga}.

\begin{acknowledgments}
The authors are grateful to Nicola Pinamonti for useful discussions on the topic of this paper. RF and MBF are pleased to thank Nicola Pinamonti and the Universit{\`a} di Genova, DIMA, for their kind hospitality. RF is grateful for the hospitality of Perimeter Institute where part of this work was carried out.

ED is partially supported by the European Union (ERC StG FermiMath, grant agreement no. 101040991), and partially supported by GNFM-INdAM.

Research at Perimeter Institute is supported in part by the Government of Canada through the Department of Innovation, Science and Economic Development and by the Province of Ontario through the Ministry of Colleges and Universities. This work was supported by a grant from the Simons Foundation (grant no. 1034867, Dittrich).

MBF is supported by the Deutsche Forschungsgemeinschaft (DFG, German Research Foundation) --- project no. 396692871 within the Emmy Noether grant CA1850/1-1.
\end{acknowledgments}

\bibliography{literature}

\end{document}